\documentclass[epsf,here,12pt]{article}
\usepackage{mypennames}
\usepackage{epsfig}  
\begin{document}

\newcommand{\rhobar}{\overline{\rho}}
\newcommand{\etabar}{\overline{\eta}}
\newcommand{\epsilonk}{\left | \epsilon_K \right |}
\newcommand{\vubovcb}{\left | \frac{V_{ub}}{V_{cb}} \right |}
\newcommand{\vtdovts}{\left | \frac{V_{td}}{V_{ts}} \right |}
\newcommand{\pr}{{\rm P.R.}}
\newcommand{\Ds}{{\rm D}_s^+}
\newcommand{\Dp}{{\rm D}^+}
\newcommand{\Do}{{\rm D}^0}
\newcommand{\piss}{\pi^{\ast \ast}}
\newcommand{\pis}{\pi^{\ast}}
\newcommand{\bbar}{\overline{b}}
\newcommand{\cbar}{\overline{c}}
\newcommand{\Dstar}{{\rm D}^{\ast}}
\newcommand{\Dstars}{{\rm D}^{\ast +}_s}
\newcommand{\Dstaro}{{\rm D}^{\ast 0}}
\newcommand{\Dstarp}{{\rm D}^{\ast +}}
\newcommand{\Dstarstar}{{\rm D}^{\ast \ast}}
\newcommand{\pistar}{\pi^{\ast}}
\newcommand{\pisstar}{\pi^{\ast \ast}}
\newcommand{\Dbar}{\overline{{\rm D}}}
\newcommand{\Bbar}{\overline{{\rm B}}}
\newcommand{\Bsbar}{\overline{{\rm B}^0_s}}
\newcommand{\Lcbar}{\overline{\Lambda^+_c}}
\newcommand{\Bstar}{{\rm B}^{*}}
\newcommand{\Bstarstar}{{\rm B}^{**}}
\newcommand{\nubar}{\overline{\nu_{\ell}}}
\newcommand{\tautaubar}{\tau \overline{\tau}}
\newcommand{\Vcb}{\left | {\rm V}_{cb} \right |}
\newcommand{\Vub}{\left | {\rm V}_{ub} \right |}
\newcommand{\Vtd}{\left | {\rm V}_{td} \right |}
\newcommand{\Vts}{\left | {\rm V}_{ts} \right |}
\newcommand{\fleisher}{\frac{BR({\rm B}^0~(\overline{{\rm B}^0}) \rightarrow \pi^{\pm} {\rm K}^{\mp})}
{BR({\rm B}^{\pm} \rightarrow \pi^{\pm} {\rm K}^0)}}
\newcommand{\bptre}{\rm b^{+}_{3}}
\newcommand{\bp}{\rm b^{+}_{1}}
\newcommand{\bo}{\rm b^0}
\newcommand{\bos}{\rm b^0_s}
\newcommand{\bss}{\rm b^s_s}
\newcommand{\qq}{\rm q \overline{q}}
\newcommand{\cc}{\rm c \overline{c}}
\newcommand{\BsDmX}{{B_{s}^{0}} \rightarrow D \mu X}
\newcommand{\BsDsm}{{B_{s}^{0}} \rightarrow D_{s} \mu X}
\newcommand{\BsDsX}{{B_{s}^{0}} \rightarrow D_{s} X}
\newcommand{\BDsX}{B \rightarrow D_{s} X}
\newcommand{\BDomX}{B \rightarrow D^{0} \mu X}
\newcommand{\BDpmX}{B \rightarrow D^{+} \mu X}
\newcommand{\Dsfmn}{D_{s} \rightarrow \phi \mu \nu}
\newcommand{\Dsfipi}{D_{s} \rightarrow \phi \pi}
\newcommand{\DsfX}{D_{s} \rightarrow \phi X}
\newcommand{\DpfX}{D^{+} \rightarrow \phi X}
\newcommand{\DofX}{D^{0} \rightarrow \phi X}
\newcommand{\DfX}{D \rightarrow \phi X}
\newcommand{\DsD}{B \rightarrow D_{s} D}
\newcommand{\DsmX}{D_{s} \rightarrow \mu X}
\newcommand{\DmX}{D \rightarrow \mu X}
\newcommand{\Zbb}{Z^{0} \rightarrow \rm b \overline{b}}
\newcommand{\Zcc}{Z^{0} \rightarrow \rm c \overline{c}}
\newcommand{\Rbb}{\frac{\Gamma_{Z^0 \rightarrow \rm b \overline{b}}}
{\Gamma_{Z^0 \rightarrow Hadrons}}}
\newcommand{\Rcc}{\frac{\Gamma_{Z^0 \rightarrow \rm c \overline{c}}}
{\Gamma_{Z^0 \rightarrow Hadrons}}}
\newcommand{\bb}{\rm b \overline{b}}
\newcommand{\str}{\rm s \overline{s}}
\newcommand{\Bs}{\rm{B^0_s}}
\newcommand{\Bsb}{\overline{\rm{B^0_s}}}
\newcommand{\Bsh}{\rm{B^{heavy}_s}}
\newcommand{\Bsl}{\rm{B^{light}_s}}
\newcommand{\Bssh}{\rm{B^{short}_s}}
\newcommand{\Bslg}{\rm{B^{long}_s}}
\newcommand{\Gsh}{\Gamma^{{\rm heavy}}_{{\rm B^0_s}}}
\newcommand{\Gsl}{\Gamma^{{\rm light}}_{{\rm B^0_s}}}
\newcommand{\Gss}{\Gamma^{{\rm short}}_{{\rm B^0_s}}}
\newcommand{\Gslg}{\Gamma^{{\rm long}}_{{\rm B^0_s}}}
\newcommand{\Gs}{\Gamma_{{\rm B^0_s}}}
\newcommand{\Gd}{\Gamma_{{\rm B^0_d}}}
\newcommand{\Msh}{{\rm m}^{{\rm heavy}}_{{\rm B^0_s}}}
\newcommand{\Msl}{{\rm m}^{{\rm light}}_{{\rm B^0_s}}}
\newcommand{\Bp}{\rm{B^{+}}}
\newcommand{\Bm}{\rm{B^{-}}}
\newcommand{\Bo}{\rm{B^{0}}}
\newcommand{\Bd}{\rm{B^{0}_{d}}}
\newcommand{\Bdb}{\overline{\rm{B^{0}_{d}}}}
\newcommand{\Lb}{\Lambda^0_b}
\newcommand{\Lbb}{\overline{\Lambda^0_b}}
\newcommand{\Kstar}{\rm{K^{\star 0}}}
\newcommand{\phim}{\rm{\phi}}
\newcommand{\Dsp}{\mbox{D}_s^+}
\newcommand{\Dsm}{\mbox{D}_s^-}
\newcommand{\Dn}{\mbox{D}^0}
\newcommand{\Dsb}{\overline{\mbox{D}_s}}
\newcommand{\Dm}{\mbox{D}^-}
\newcommand{\Dnb}{\overline{\mbox{D}^0}}
\newcommand{\Lc}{\Lambda_c^+}
\newcommand{\Lcb}{\overline{\Lambda_c}}
\newcommand{\Dstarm}{\mbox{D}^{\ast -}}
\newcommand{\Dsstarp}{\mbox{D}_s^{\ast +}}
\newcommand{\Dsstar}{\mbox{D}^{\ast \ast}}
\newcommand{\Km}{\mbox{K}^-}
\newcommand{\Pb}{P_{b-baryon}}
\newcommand{\KKpi}{\rm{ K K \pi }}
\newcommand{\GeV}{\rm{GeV}}
\newcommand{\MeV}{\rm{MeV}}
\newcommand{\nb}{\rm{nb}}
\newcommand{\Zzero}{{\rm Z}}
\newcommand{\MZ}{\rm{M_Z}}
\newcommand{\MW}{\rm{M_W}}
\newcommand{\GF}{\rm{G_F}}
\newcommand{\Gm}{\rm{G_{\mu}}}
\newcommand{\MH}{\rm{M_H}}
\newcommand{\MT}{\rm{m_{top}}}
\newcommand{\GZ}{\Gamma_{\rm Z}}
\newcommand{\Afb}{\rm{A_{FB}}}
\newcommand{\Afbs}{\rm{A_{FB}^{s}}}
\newcommand{\sigmaf}{\sigma_{\rm{F}}}
\newcommand{\sigmab}{\sigma_{\rm{B}}}
\newcommand{\NF}{\rm{N_{F}}}
\newcommand{\NB}{\rm{N_{B}}}
\newcommand{\Nnu}{\rm{N_{\nu}}}
\newcommand{\RZ}{\rm{R_Z}}
\newcommand{\rhob}{\rho_{eff}}
\newcommand{\Gammanz}{\rm{\Gamma_{Z}^{new}}}
\newcommand{\Gammani}{\rm{\Gamma_{inv}^{new}}}
\newcommand{\Gammasz}{\rm{\Gamma_{Z}^{SM}}}
\newcommand{\Gammasi}{\rm{\Gamma_{inv}^{SM}}}
\newcommand{\Gammaxz}{\rm{\Gamma_{Z}^{exp}}}
\newcommand{\Gammaxi}{\rm{\Gamma_{inv}^{exp}}}
\newcommand{\rhoZ}{\rho_{\rm Z}}
\newcommand{\thw}{\theta_{\rm W}}
\newcommand{\swsq}{\sin^2\!\thw}
\newcommand{\swsqmsb}{\sin^2\!\theta_{\rm W}^{\overline{\rm MS}}}
\newcommand{\swsqbar}{\sin^2\!\overline{\theta}_{\rm W}}
\newcommand{\cwsqbar}{\cos^2\!\overline{\theta}_{\rm W}}
\newcommand{\swsqb}{\sin^2\!\theta^{eff}_{\rm W}}
\newcommand{\ee}{{e^+e^-}}
\newcommand{\eeX}{{e^+e^-X}}
\newcommand{\gaga}{{\gamma\gamma}}
\newcommand{\mumu}{\ifmmode {\mu^+\mu^-} \else ${\mu^+\mu^-} $ \fi}
\newcommand{\eeg}{{e^+e^-\gamma}}
\newcommand{\mumug}{{\mu^+\mu^-\gamma}}
\newcommand{\tautau}{{\tau^+\tau^-}}
\newcommand{\qqb}{{q\overline{q}}}
\newcommand{\eegg}{e^+e^-\rightarrow \gamma\gamma}
\newcommand{\eeggg}{e^+e^-\rightarrow \gamma\gamma\gamma}
\newcommand{\eeee}{e^+e^-\rightarrow e^+e^-}
\newcommand{\eeeeee}{e^+e^-\rightarrow e^+e^-e^+e^-}
\newcommand{\eeeeg}{e^+e^-\rightarrow e^+e^-(\gamma)}
\newcommand{\eeeegg}{e^+e^-\rightarrow e^+e^-\gamma\gamma}
\newcommand{\eeeg}{e^+e^-\rightarrow (e^+)e^-\gamma}
\newcommand{\eemumu}{e^+e^-\rightarrow \mu^+\mu^-}
\newcommand{\eetautau}{e^+e^-\rightarrow \tau^+\tau^-}
\newcommand{\eehad}{e^+e^-\rightarrow {\rm hadrons}}
\newcommand{\eettg}{e^+e^-\rightarrow \tau^+\tau^-\gamma}
\newcommand{\eell}{e^+e^-\rightarrow l^+l^-}
\newcommand{\Ztopig}{{\rm Z}^0\rightarrow \pi^0\gamma}
\newcommand{\Ztogg}{{\rm Z}^0\rightarrow \gamma\gamma}
\newcommand{\Ztoee}{{\rm Z}^0\rightarrow e^+e^-}
\newcommand{\Ztoggg}{{\rm Z}^0\rightarrow \gamma\gamma\gamma}
\newcommand{\Ztomumu}{{\rm Z}^0\rightarrow \mu^+\mu^-}
\newcommand{\Ztotautau}{{\rm Z}^0\rightarrow \tau^+\tau^-}
\newcommand{\Ztoll}{{\rm Z}^0\rightarrow l^+l^-}
\newcommand{\Ztocc}{{\rm Z^0\rightarrow c \overline c}}
\newcommand{\Lamp}{\Lambda_{+}}
\newcommand{\Lamm}{\Lambda_{-}}
\newcommand{\Pt}{\rm P_{t}}
\newcommand{\Gee}{\Gamma_{ee}}
\newcommand{\Gpig}{\Gamma_{\pi^0\gamma}}
\newcommand{\Ggg}{\Gamma_{\gamma\gamma}}
\newcommand{\Gggg}{\Gamma_{\gamma\gamma\gamma}}
\newcommand{\Gmumu}{\Gamma_{\mu\mu}}
\newcommand{\Gtautau}{\Gamma_{\tau\tau}}
\newcommand{\Ginv}{\Gamma_{\rm inv}}
\newcommand{\Ghad}{\Gamma_{\rm had}}
\newcommand{\Gnu}{\Gamma_{\nu}}
\newcommand{\GnuSM}{\Gamma_{\nu}^{\rm SM}}
\newcommand{\Gll}{\Gamma_{l^+l^-}}
\newcommand{\Gff}{\Gamma_{f\overline{f}}}
\newcommand{\Gtot}{\Gamma_{\rm tot}}
\newcommand{\Rb}{\mbox{R}_b}
\newcommand{\Rc}{\mbox{R}_c}
\newcommand{\al}{a_l}
\newcommand{\vl}{v_l}
\newcommand{\af}{a_f}
\newcommand{\vf}{v_f}
\newcommand{\ael}{a_e}
\newcommand{\ve}{v_e}
\newcommand{\amu}{a_\mu}
\newcommand{\vmu}{v_\mu}
\newcommand{\atau}{a_\tau}
\newcommand{\vtau}{v_\tau}
\newcommand{\ahatl}{\hat{a}_l}
\newcommand{\vhatl}{\hat{v}_l}
\newcommand{\ahate}{\hat{a}_e}
\newcommand{\vhate}{\hat{v}_e}
\newcommand{\ahatmu}{\hat{a}_\mu}
\newcommand{\vhatmu}{\hat{v}_\mu}
\newcommand{\ahattau}{\hat{a}_\tau}
\newcommand{\vhattau}{\hat{v}_\tau}
\newcommand{\vtildel}{\tilde{\rm v}_l}
\newcommand{\avsq}{\ahatl^2\vhatl^2}
\newcommand{\Ahatl}{\hat{A}_l}
\newcommand{\Vhatl}{\hat{V}_l}
\newcommand{\Afer}{A_f}
\newcommand{\Ael}{A_e}
\newcommand{\Aferb}{\overline{A_f}}
\newcommand{\Aelb}{\overline{A_e}}
\newcommand{\AVsq}{\Ahatl^2\Vhatl^2}
\newcommand{\Iwk}{I_{3l}}
\newcommand{\Qch}{|Q_{l}|}
\newcommand{\roots}{\sqrt{s}}
\newcommand{\pT}{p_{\rm T}}
\newcommand{\mt}{m_t}
\newcommand{\Rechi}{{\rm Re} \left\{ \chi (s) \right\}}
\newcommand{\up}{^}
\newcommand{\abscosthe}{|cos\theta|}
\newcommand{\dsum}{\Sigma |d_\circ|}
\newcommand{\zsum}{\Sigma z_\circ}
\newcommand{\sint}{\mbox{$\sin\theta$}}
\newcommand{\cost}{\mbox{$\cos\theta$}}
\newcommand{\mcost}{|\cos\theta|}
\newcommand{\epair}{\mbox{$e^{+}e^{-}$}}
\newcommand{\mupair}{\mbox{$\mu^{+}\mu^{-}$}}
\newcommand{\taupair}{\mbox{$\tau^{+}\tau^{-}$}}
\newcommand{\gamgam}{\mbox{$e^{+}e^{-}\rightarrow e^{+}e^{-}\mu^{+}\mu^{-}$}}
\newcommand{\fullskip}{\vskip 16cm}
\newcommand{\halfskip}{\vskip  8cm}
\newcommand{\quarskip}{\vskip  6cm}
\newcommand{\abitskip}{\vskip 0.5cm}
\newcommand{\ba}{\begin{array}}
\newcommand{\ea}{\end{array}}
\newcommand{\bc}{\begin{center}}
\newcommand{\ec}{\end{center}}
\newcommand{\be}{\begin{eqnarray}}
\newcommand{\eeq}{\end{eqnarray}}
\newcommand{\bes}{\begin{eqnarray*}}
\newcommand{\ees}{\end{eqnarray*}}
\def\etal{{\it et al.}}
\newcommand{\Kz}{\ifmmode {\rm K^0_s} \else ${\rm K^0_s} $ \fi}
\newcommand{\Zz}{\ifmmode {\rm Z} \else ${\rm Z } $ \fi}
\newcommand{\qqbar}{\ifmmode {\rm q\overline{q}} \else ${\rm q\overline{q}} $ \fi}
\newcommand{\ccbar}{\ifmmode {\rm c\overline{c}} \else ${\rm c\overline{c}} $ \fi}
\newcommand{\bbbar}{\ifmmode {\rm b\overline{b}} \else ${\rm b\overline{b}} $ \fi}
\newcommand{\xxbar}{\ifmmode {\rm x\overline{x}} \else ${\rm x\overline{x}} $ \fi}
\newcommand{\rphi}{\ifmmode {\rm R\phi} \else ${\rm R\phi} $ \fi}
\def\vcb{$\left | {\rm V}_{cb} \right |$}
\def\fw{${\cal F}(w)$}
\def\fone{${\cal F}_{D^{*}}(1)$}
\def\fvcb{${\cal F}_{D^{*}}(1)|{\rm}V_{cb}|$}
\def\btods{$ \Bdb \rightarrow {\rm D}^{*+}\ell^-{\overline \nu_{\ell}}$}
\def\Btau{$ \Bdb \rightarrow {\rm D}^{*+}\tau^-{\overline \nu_{\tau}}$}
\def\Bxc{$ \Bdb\rightarrow {\rm D}^{*+} {\rm X}_{\overline{c}}$}
\def\btodss{$ \Bdb\rightarrow {\rm D}^{**+}\ell^-{\overline \nu_{\ell}}$}
\newcommand {\bl}    {{\rm BR}({b \rightarrow \ell})}
\newcommand {\cl}    {{\rm BR}({c \rightarrow \overline{\ell}})}
\newcommand {\bcbl}   {{\rm BR}({b \rightarrow \overline{c} \rightarrow \ell})}
\newcommand {\bcl}   {{\rm BR}({b \rightarrow c \rightarrow \overline{\ell}})}
\newcommand {\btaul}  {{\rm BR}({b \rightarrow \tau \rightarrow \ell})}
\newcommand {\bpsill} {{\rm BR}({b \rightarrow {\rm J}/\psi\rightarrow \ell^+\ell^-})}
\newcommand {\glcc}   {\rm{g \rightarrow c \overline c}}
\newcommand {\glbb}   {\rm{g \rightarrow b \overline b}}
\def\dsp{${\rm D}^{*+}$}
\def\bbar{$\overline{{\rm B}^0_d}$}
%
%
\newcommand{\taubav}{1.564 \pm 0.014} 
\newcommand{\taubd}{1.562 \pm 0.029} 
\newcommand{\taubp}{1.656 \pm 0.025} 
\newcommand{\taubs}{1.464 \pm 0.057} 
\newcommand{\taulb}{1.229 ^{+0.081}_{-0.079}} 
\newcommand{\tauxib}{1.39 ^{+0.34}_{-0.28}} 
\newcommand{\taubbar}{1.208^{+0.051}_{-0.050}} 
\newcommand{\taubpovertaubd}{1.065 \pm 0.023} 
\newcommand{\dmdw}{0.470\pm 0.018} 
\newcommand{\dmdwerr}{0.482\pm 0.016}  
\newcommand{\dmdx}{0.476\pm 0.016} 
\newcommand{\dmdxev}{3.13 \pm 0.11} 
\newcommand{\chix}{0.177\pm 0.008} 
\newcommand{\chicls}{0.180\pm 0.009} 
%
\newcommand{\fbsdir}{(11.2^{+3.1}_{-2.7})\%} 
\newcommand{\fbudir}{(41.4 \pm 1.6)\%} 
\newcommand{\flbdir}{(9.8^{+3.8}_{-2.5})\%} 
\newcommand{\fxidir}{(1.1^{+0.6}_{-0.5})\%} 
\newcommand{\fbardir}{(12.0^{+3.9}_{-2.7})\%} 
\newcommand{\fbarspec}{(10.2 \pm 2.8)\%} 
\newcommand{\fbaravg}{(11.0^{+2.3}_{-1.9})\%} 
\newcommand{\fsa}{(9.4 \pm 2.2)\%} 
\newcommand{\fbara}{(10.1 \pm 1.8)\%} 
\newcommand{\fua}{(40.3 \pm 1.2)\%} 
\newcommand{\rhosbara}{-0.31} 
\newcommand{\rhosua}{-0.68} 
\newcommand{\rhobarua}{-0.48} 
\newcommand{\fbsmix}{(10.3\pm 1.5)\%} 
\newcommand{\fbs}{(10.0\pm 1.2)\%} 
\newcommand{\fbar}{(9.9\pm 1.7)\%} 
\newcommand{\fbd}{(40.1\pm 1.0)\%} 
\newcommand{\rhosbar}{-0.03} 
\newcommand{\rhosd}{-0.57} 
\newcommand{\rhobard}{-0.81} 
\newcommand{\dmslim}{14.3}  
\newcommand{\dmsb}{12.1}    
\newcommand{\dmssen}{14.5}  
\newcommand{\ldmslim}{11.5} 
\newcommand{\ldmsb}{10.8}   
\newcommand{\ldmssen}{12.9} 
%
\newcommand{\vcbinc}{(40.76 \pm 0.41({\rm exp.}) \pm 2.04({\rm theo.})) \times 10^{-3}} 
\newcommand{\favcb}{(33.8 \pm 0.9({\rm stat.}) \pm 1.9({\rm syst.})) \times 10^{-3}} 
\newcommand{\rhobb}{1.01 \pm  0.09 \pm 0.19} 
\newcommand{\vcbexc}{(38.4 \pm 1.0({\rm stat.}) \pm 2.1({\rm syst.}) \pm 2.2({\rm theo.}))\times 10^{-3}} 
\newcommand{\vcbavg}{(40.2 \pm 1.9) \times 10^{-3}} 

%

\setlength{\textheight}{24.0cm}
\setlength{\topmargin}{-0.5cm}
\setlength{\textwidth}{15.0cm}
\newcommand{\dmd}{\ensuremath{\Delta m_d}}
\newcommand{\dms}{\ensuremath{\Delta m_s}}
\newcommand{\Bb}{\mbox{$b$-baryon}}
\newcommand{\chib}{\ensuremath{\overline \chi}}
\newcommand{\chid}{\ensuremath{\chi_d}}
\newcommand{\chis}{\ensuremath{\chi_s}}
\newcommand{\Yfs}{\ensuremath{{\Upsilon({\rm 4S})}}}
\newcommand{\fd}{f_{\PdB}}
\newcommand{\fs}{f_{\PsB}}
\newcommand{\fu}{f_{\PBp}}
\newcommand{\fb}{f_{\mbox{\scriptsize \Bb}}}
\newcommand{\gs}{g_{\PsB}}
\newcommand{\gd}{g_{\PdB}}
\newcommand{\gu}{g_{\PBp}}
\newcommand{\gb}{g_{\mbox{\scriptsize \Bb}}}
\newcommand{\ips}{${\mathrm{ps}}^{-1}$}
\newcommand{\blankvalue}{\multicolumn{2}{c|}{ }}
\newcommand{\Brr}[2]{\ensuremath{\mathrm{BR}(#1 \rightarrow #2)}}
\newcommand{\mffbs}{\Brr{\overline b}{\PsBz}}
\newcommand{\mbsdslX}{\Brr{\PsBz}{\PsDm \ell^+\nu_{\ell} X}}
\newcommand{\prodbs}{$\mffbs \cdot \mbsdslX$}
\newcommand{\bsdslX}{\Brr{\PsBz}{\PsDm \ell^+ \nu_{\ell} \mathrm{X}}}
\newcommand{\prodlb}{\Brr{b}{\PbgLz}~
  \Brr{\PbgLz}{\PcgLp \ell^-\overline {\nu_{\ell}} \mathrm{X}}}
\newcommand{\prodxb}{\Brr{b}{\PbgX}~
  \Brr{\PbgXm}{\PgXm \ell^-\overline {\nu_{\ell}} \mathrm{X}}}
\newcommand{\brpkpi}{\Brr{\PcgLp}{\Pp \PKm \Pgpp}}
\newcommand{\mbXlnu}{\Brr{\overline b}{\ell^+\nu_{\ell}{\mathrm{X}}}}
\newcommand{\mBXlnu}{\Brr{\PB}{\ell^+ \nu_{\ell}{\mathrm{X}}}}
\newcommand{\mups}{\Upsilon(4\mathrm{S})}
\newcommand{\bzdlnu}{\Brr{\PdBz}{\PDm \ell^+ \nu_{\ell}}}
\newcommand{\bzdstlnu}{\Brr{\PdBz}{\PDstm\ell^+ \nu_{\ell}}}
\newcommand{\bpdlnu}{\Brr{\PBp}{\PaDz \ell^+ \nu_{\ell}}}
\newcommand{\bpdstlnu}{\Brr{\PBp}{\PDstz \ell^+ \nu_{\ell}}}
\newcommand{\mbdstpilnu}{\Brr{\PB}{\PDst \Pgp \ell^+ \nu_{\ell}}}
\newcommand{\mbdpilnu}{\Brr{\PB}{\PaD \Pgp \ell^+ \nu_{\ell}}}
\newcommand{\brphipi}{\Brr{\PsDm}{\phi \Pgpm}}
\newcommand{\mysection}[1]{\section{\boldmath #1}}
\newcommand{\mysubsection}[1]{\subsection[#1]{\boldmath #1}}
\newcommand{\mysubsubsection}[1]{\subsubsection[#1]{\boldmath #1}}
\newcommand{\dgs}{\Delta \Gamma_{\Bs}}
\newcommand{\dgbs}{\Delta \Gamma_{\rm \Bs}/\Gamma_{\rm \Bs}}
\newcommand{\tbs}{\tau_{\Bs}}
\newcommand{\tbd}{\tau_{\Bd}}
\newcommand{\tbssemi}{\tau_{\rm B^{semi.}_s}}
\newcommand{\tbsshort}{\tau_{\Bssh}}
\newcommand{\tbsdh}{\tau_{\rm B^{D_s-had.}_s}}
\newcommand{\tbspsi}{\tau_{\rm B^{{\rm J}/\psi \phi}_s}}
\newcommand{\tbsinc}{\tau_{\rm B^{incl.}_s}}

\begin{titlepage}

\pagenumbering{arabic}
\vspace*{-0.9cm}
\begin{center}
{\large EUROPEAN ORGANIZATION FOR NUCLEAR RESEARCH}
\end{center}
\vspace*{1.1cm}
\begin{tabular*}{15.cm}{l@{\extracolsep{\fill}}r}
{  } & 
SLAC-PUB-8492 \\
&
CERN-EP-2000-096
\\
& 
19 March 2000\\
\\
&\\
\end{tabular*}
\vspace*{1.cm}
\begin{center}
\Large 
{\bf Combined results on {\boldmath $b$}-hadron production rates, lifetimes,
oscillations and semileptonic decays} 
\vspace*{2.cm}
\\
\normalsize {    {\bf 
 ALEPH, CDF, DELPHI, L3, OPAL, SLD }}\\
\vskip 0.5truecm
\footnotesize{Prepared 
\footnote{The members of the working groups involved in 
this activity are:
D. Abbaneo, 
J. Alcaraz, V. Andreev, E. Barberio, M. Battaglia,
S. Blyth, G. Boix, C. Bourdarios,
 M. Calvi, P. Checchia, P. Coyle,
L. Di Ciaccio,  P. Gagnon, R. Hawkings, O. Hayes, P. Henrard, T. Hessing, 
M. Jimack, I.J. Kroll, O. Leroy, D. Lucchesi, M. Margoni, S. Mele, 
H.G. Moser, F. Muheim, F. Palla, D. Pallin, F. Parodi, M. Paulini, E. Piotto,
P. Privitera, Ph. Rosnet, 
P. Roudeau, D. Rousseau, O. Schneider, Ch. Schwick,
C. Shepherd-Themistocleous,
 F. Simonetto, P. Spagnolo, A. Stocchi, D. Su, T. Usher, C. Weiser, P. Wells, 
B. Wicklund and S. Willocq.}
from Contributions to the 1999 Summer conferences.} \\
\end{center}

\vskip 2.5truecm

\begin{abstract}
\noindent
 Combined results on $b$-hadron lifetimes,
$b$-hadron production rates,
$\Bd-\Bdb$ and $\Bs-\Bsb$ oscillations,
the decay width difference 
between 
the mass eigenstates of the $\Bs-\Bsb$ system,
and 
the values of the CKM matrix elements $\Vcb$ 
and $\Vub$ are
obtained from published and preliminary measurements
available in Summer 99 from the ALEPH, CDF, DELPHI, L3, OPAL and SLD
Collaborations.

\end{abstract}

\vskip 2.0truecm
\vspace{\fill}
\begin{center}
\end{center}
\noindent

\vspace{\fill}
\end{titlepage}

\setcounter{page}{1}    
\tableofcontents

\setcounter{footnote}{0}

\newpage
\section {Introduction}
\label{sec:intro}
Accurate determinations of $b$-hadron decay properties provide constraints
on the values of the elements of the Cabibbo-Kobayashi-Maskawa (CKM) matrix
\cite{ckm}. The $\Vcb$ and $\Vub$ elements can be obtained from semileptonic
decay rates into charmed and non-charmed hadrons, and measurements
of the oscillation frequency in $\Bd-\Bdb$ and $\Bs-\Bsb$ systems give
access to $\Vtd$ and $\Vts$.

Elements of the CKM matrix govern weak transitions between quarks.
Experimental results are obtained from processes involving $b$-hadrons. Effects
from strong interactions have thus to be controlled; and
 $b$-hadrons are also
a good laboratory in this respect. 
Lifetime differences between the different
weakly decaying hadrons 
can be related to interactions between
the heavy quark and the light quark system inside the hadron. The polarization
of $\Lb$ baryons produced by $b$-quarks of known polarization,
 emitted from $\Zz$ decays at LEP or SLC,
indicates how polarization is transmitted from the 
heavy quark to the baryon(s) in the hadronization process.
Rates and decay properties of excited $\Dstarstar$ 
states\footnote{The notation $\Dstarstar$ includes all 
charm mesons and non-resonant charmed final states
which are not simply D or $\Dstar$ mesons.} produced in
$b$-hadron semileptonic decays are needed to obtain accurate
determinations of $\Vcb$. 
Decays of $\Dstarstar$ states are also an important source
of background in other channels, and their properties have to be monitored.
Finally, it is mandatory to measure
the production rates
of the different weakly decaying $b$-hadrons produced during
the hadronization of $b$-quarks created in high energy collisions,
because all these states have different properties so the study
of any one of them requires control of the background from the others.

Results obtained on $b$-hadron lifetimes,
$b$-hadron production rates,
$\Bd-\Bdb$ and $\Bs-\Bsb$ oscillations,
and 
the values of the CKM matrix elements $\Vcb$ 
and $\Vub$ made available 
during Summer 1999 are presented here. A limit on the decay width difference 
of 
the mass eigenstates of the $\Bs-\Bsb$ system is also given.
These quantities have been obtained by averaging
    published and preliminary measurements released publicly
    by the ALEPH, CDF, DELPHI, L3, OPAL and SLD experiments.
Whenever possible, the input
    parameters used in the various analyses have been adjusted
    to common values, and all known correlations have been
    taken into account.
Close contacts have been 
established between representatives from the experiments and members of 
the different working
groups in charge of the averages, to ensure that the data are prepared in a form 
suitable for combinations.
Working group activities are coordinated by a steering 
group\footnote{The present members of the Heavy Flavour Steering Group are: 
D. Abbaneo, 
J. Alcaraz, E. Barberio, M. Battaglia, S. Blyth, 
D. Su, P. Gagnon, R. Hawkings, S. Mele, 
D. Pallin, M. Paulini, P. Roudeau, O. Schneider,
 F. Simonetto, A. Stocchi, P. Wells, B. Wicklund and S. Willocq.}. 

Section \ref{sec:systgen} presents the values of the common input parameters 
that contribute to the systematic
uncertainties given in this note.
Published results obtained by other experiments,
mainly operating at the $\Yfs$ resonance, have been included 
to obtain most accurate values.
In particular, it contains studies 
on 
some characteristics of $\Dstarstar$
mesons
in semileptonic $b$-decays, and on the $\Lb$ polarization.
As some measured quantities are needed
for the evaluation
of others, an iterative procedure
has been adopted to obtain stable results.


Section \ref{sec:Atau} describes the averaging of the  $b$-hadron lifetime
 measurements.
The combined values
are compared with expectations from theory. The inclusive  
$b$-hadron and $\Bdb$ meson\footnote{Throughout the paper charge
 conjugate states are implicitly included unless stated otherwise.}
 lifetimes are needed in the 
determination of $\Vub$ and $\Vcb$ in order to convert
measured branching fractions into partial widths that can then be
compared with theory.

In Section \ref{sec:boscill}, oscillations of neutral B mesons are studied.
Also the production rates of the different $b$-hadrons in jets induced by a 
$b$-quark are determined using direct measurements 
together with the constraints provided by B mixing.
The $\Bdb$ production rate is an important
input for the $\Vcb$ measurement using 
$\Bdb \rightarrow \Dstarp \ell^- \overline{\nu_{\ell}}$
decays, and the sensitivity
to $\Bs$-$\Bsb$ oscillations depends on the $\Bsb$ production rate.

In Section \ref{sec:deltag}, a limit on the decay width difference
between mass eigenstates of the  $\Bs$-$\Bsb$ system is given.
As this average is obtained for the first time, more details
on the adopted procedure have been given than for lifetimes
or oscillations measurements.

The determination of $\Vcb$ and $\Vub$ presented in Sections \ref{sec:vcb}
and \ref{sec:vub} includes only LEP results.
These averages are also released for the first time. The
determination of $\Vub$ is based on a novel technique using
the lepton momentum and the mass of the hadronic system.
The accuracy of these results (especially on $\Vcb$) depends mostly
on theoretical uncertainties. Appendix \ref{appendixC} gives details on the
theoretical inputs used.

A summary of all results obtained
by the different working groups is given  in Section \ref{sec:conclusion}.
In addition, Appendices \ref{appendixA}-\ref{appendixAc} contain,
respectively, the individual measurements of the production rates 
of narrow $\Dstarstar$ states, $\Lb$ polarization, $b$-hadron
lifetimes, and direct measurements of $b$-meson and $b$-baryon
production rates, that have been used in the
present averages.

\mysection{Common input parameters}
\label{sec:systgen}
The $b$-hadron properties used as common input parameters in these averages
are given in Table \ref{tab:gensys}. 
Most of the quantities 
have been taken from results obtained
by the LEPEWWG \cite{HFLEPEW, EWWG} or quoted by the PDG \cite{PDG98}.
The others, which concern the production rates and decay properties
of $\Dstarstar$ mesons in $b$-hadron semileptonic decays and
the value of the $\Lb$ polarization, are explained later in this section.

\begin{table}[ht!]
  \begin{center}
    \begin{minipage}{\linewidth}
    \renewcommand{\thempfootnote}{\fnsymbol{mpfootnote}}
    \addtocounter{mpfootnote}{1}
    \begin{tabular}{|l| l | l @{\,$\pm$\,}l | c |}
      \hline
      Quantity & Symbol &  \multicolumn{2}{c|}{Value}    
      & Reference \\
      \hline
      Fraction of $b$ events  & ${\rm R}_b$      & 0.21643  & 0.00073 
      & \cite{HFLEPEW} \\
      Fraction of $c$ events  & ${\rm R}_c$      & 0.1694  & 0.0038 
      & \cite{HFLEPEW} \\
      Beam energy fraction  & $<x_E>$      & 0.702  & 0.008 
      & \cite{EWWG} \\
     $b$-hadron sl. BR  & BR$(b \rightarrow \ell^- \overline{\nu_{\ell}}X)$ & 0.1058  & 0.0018       & \cite{HFLEPEW} \\
      Cascade $b$ sl. decay (r.s.) & BR$(b \rightarrow \overline{c}\rightarrow \ell^-\overline{\nu_{\ell}}X)$ & 0.0162  & $^{0.0044}_{0.0036}$       &  \cite{EWWG} \\
      Cascade $b$ sl. decay (w.s.) & BR$(b \rightarrow c \rightarrow \ell^+\nu_{\ell}X)$ & 0.0807  & 0.0025       &  \cite{HFLEPEW} \\
      $c$-hadron sl. BR  & BR$(c \rightarrow \ell^+\nu_{\ell}X)$ & 0.0985  & 0.0032       &  \cite{HFLEPEW} \\
      $b$ quarks from gluons & P$(g \rightarrow b\overline{b})$ &  0.00251  & 0.00063       &  \cite{HFLEPEW} \\
       $c$ quarks from gluons & P$(g \rightarrow c\overline{c})$ & 0.0319  & 0.0046       &  \cite{HFLEPEW} \\
      $b$ decay charged mult.  & $n_{ch}^b$      & 4.955  & 0.062 
      & \cite{EWWG} \\
      $b$-hadron mixing & \chib\         & 0.1186   & 0.0043
      & \cite{HFLEPEW}      \\
      \PdBz\ mixing from \Yfs
                         & $\chid(\Yfs)$        & 0.156   & 0.024 
      & \cite{PDG98, argchi, cleochi} \\
\hline
$\Dstarstar$ in sl. $b$ decays & ${\rm BR}(\Bdb \rightarrow \rm D^{\ast \ast +} \ell^- \overline{\nu_{\ell}})$ &  0.0304  & 0.0044 &  Sect. \ref{sec:dssa} \\ 
      $\Dstarp$ in $\Dstarstar$ sl. $b$ decays & BR(B$^- \rightarrow \Dstarp \pi^- \ell^-\overline{\nu_{\ell}}$)  & 0.0125  & 0.0016
      & Sect. \ref{sec:dssb}  \\
      $\Dstarp$ in $\tau$ sl. $b$ decays  & BR$(\Bdb \rightarrow \Dstarp
\tau^- \overline{\nu_{\tau}}X)$ & 0.0127  & 0.0021 
      &  Sect. \ref{sec:ajous1}  \\
      $\Dstarp$ in double charm & BR$(b \rightarrow \Dstarp
X_{\overline{c}}( \rightarrow \ell^-X))$ & 0.008  & 0.003 
      &  Sect. \ref{sec:ajous1}  \\
      $\Lb$ polarization  & ${\cal P}(\Lb)$ &-0.45   & $^{0.19}_{0.17}$
      &  Sect. \ref{sec:dssd}  \\
      \hline
      $\tau$ in sl. $b$ decays & BR$(b \rightarrow \tau^- \overline{\nu_{\tau}}X)$ & 0.026  & 0.004       &\cite{PDG98}\\
      J/$\psi$ in $b$ decays& BR$(b \rightarrow {\rm J}/\psi X $ & 0.0116  & 0.0010       &\cite{PDG98}\\
      $\Do$ branching fraction
   & BR($\Do \rightarrow {\rm K}^- \pi^+$) & 0.0385   & 0.0009 
      & \cite{PDG98} \\
      $\Dp$ branching fraction
   & BR($\Dp \rightarrow {\rm K}^- \pi^+ \pi^+$) & 0.091   & 0.006 
      & \cite{PDG98} \\
      $\Dsp$ branching fraction
   & BR($\Dsp \rightarrow \phi \pi^+$) & 0.036   & 0.009 
      & \cite{PDG98} \\
      $\Lc$ branching fraction
   & BR($\Lc \rightarrow {\rm p} {\rm K}^- \pi^+$) & 0.050   & 0.013 
      & \cite{PDG98} \\
      $\Dstarp$ branching fraction
   & BR($\Dstarp \rightarrow \Do \pi^+$) & 0.683   & 0.014 
      & \cite{PDG98} \\
\hline
    \end{tabular}
  \end{minipage}
  \end{center}
    \caption{{\it Common set of input parameters used for the 
derivation of the various measurements presented in this paper.
The first set of results has been taken from those obtained
by the LEPEWWG \cite{HFLEPEW, EWWG} or quoted by the PDG \cite{PDG98},
the second set corresponds to averages obtained in the present report and,
for the sake of completeness, in the last set,
branching fractions in charm decays needed for
the production rates of these particles are listed as well.
The abbreviated notations sl., r.s. and. w.s. correspond, respectively,
to semileptonic, right sign and wrong sign.}
    \label{tab:gensys}}
\end{table}

Note the following:
\begin{itemize}
\item {\it ${\rm R}_b,~{\rm R}_c$ and $<x_E>$}

These values apply only to $b$-hadrons produced in $\Zz$ decays.
${\rm R}_b$ and ${\rm R}_c$ are the respective branching fractions
of the $\Zz$ boson into $b \overline{b}$ and $c \overline{c}$ pairs
in hadronic events, $<x_E>$ is the mean fraction of the beam energy  
taken by a weakly decaying $b$-hadron.

\item {\it shape of the $b$-quark fragmentation function.}


The value of $<x_E>$ obtained assuming the Peterson function 
\cite{ref:peters}, is given
in Table \ref{tab:gensys}. To evaluate the corresponding systematic 
uncertainty, parameter(s) governing $b$-quark fragmentation functions,
taken from other models, have been varied in accordance with the uncertainty
quoted for $<x_E>$. Fragmentation functions taken from two
models \cite{ref:colsop,ref:kart} have been chosen to estimate
the systematic uncertainties coming from the shape of the function.
These models typically yield results on either side of those obtained
using the Peterson function
 which is commonly used by the experiments.
For analyses which are rather insensitive to the fragmentation uncertainty,
it is considered adequate to use only the Peterson model and to inflate
the uncertainty on $<x_E>$ to $\pm 0.02$.

\item {\it the inclusive semileptonic branching fraction of $b$-hadrons.}

The average LEP value for 
${\rm BR} (b \rightarrow \ell^- \overline{\nu_{\ell}} X) = (10.58 \pm 0.07({\rm stat.}) 
\pm 0.17({\rm syst.}))\% $ is taken from the global LEPEWWG fit
which combines the heavy flavour measurements performed 
 at the $\Zz$ \cite{HFLEPEW}.

 The largest contribution to the systematic error
comes from the uncertainty on the semileptonic decay model. This
error is reduced from $\pm 0.084 \times 10^{-2}$ to 
$\pm 0.065 \times 10^{-2}$
when the fit is performed including the forward-backward asymmetry measurements
in addition to all heavy flavour measurements performed at the $\Zz$.
This happens because
only asymmetry measurements
obtained using leptons depend on the semileptonic
decay model, measurements using a lifetime tag
combined with jet-charge or D-meson reconstruction do not.
To achieve consistency between these measurements, the fit
effectively constrains the size of the error attributed to
the semileptonic decay model, thus reducing the 
corresponding systematic error 
on the semileptonic branching fraction. 
Including asymmetry measurements gives an
inclusive $b$-hadron semileptonic branching fraction  of
$(10.62 \pm 0.17)\%$.
To be conservative, we have not
included the asymmetry measurements in the fit to extract
the semileptonic branching fraction used in the present averages.      
The average value was obtained 
with the average $\rm B^0-\overline{B^0}$ mixing parameter 
and  $ {\rm R}_b$ value
given in Table \ref{tab:gensys}.  


In the absence of direct measurements of the semileptonic branching
fractions for the different B meson states, it has been assumed, when 
needed in the following analyses, that all $b$-hadron semileptonic widths  
are equal. This hypothesis is strictly valid
for $\Bm$ and $\Bdb \rightarrow c \ell^- \overline{\nu_{\ell}}$ decays
because of isospin invariance originating from the $b \rightarrow c$ transition,
which is $\Delta {\rm I}=0$ (similar considerations apply also
to ${\rm D}^{0,+} \rightarrow s \ell^+ \nu_{\ell}$ decays). 
It is not valid in
$\Bm~{\rm and}~\Bdb \rightarrow u \ell^- \overline{\nu_{\ell}}$ decays, but
the induced difference between $\Bm$ and $\Bdb$ total semileptonic
decay rates can be neglected.
It has been assumed valid also
for $\Bsb$ mesons and $b$-baryons, but for $b$-baryons an uncertainty of 15$\%$ has been added,
estimated by comparing the lifetime ratios and semileptonic branching 
fraction 
ratios for $b$-mesons and $b$-baryons \cite{ref:pauline}.
Exclusive semileptonic branching fraction averages,
given in the following for the $\Bdb$ meson, have been obtained 
using this hypothesis.
Results for other $b$-hadron flavours
can be obtained using, in addition, the corresponding lifetime ratios.
The latter are obtained from $b$-hadron lifetimes given in 
Section \ref{sec:Atau}.

\item {\it gluon splitting to heavy quarks.}

The quantities P$(g \rightarrow c\overline{c})$ and 
P$(g \rightarrow b\overline{b})$ are defined as the ratios
$\frac{{\rm BR}(\Zz \rightarrow q \overline{q} g,~g \rightarrow Q \overline{Q})}{{\rm BR}(\Zz \rightarrow {\rm hadrons})}$ in which, respectively, $Q$
is a $c$ or a $b$ quark.

\item {\it $b$-hadron decay multiplicity.}

The value given in Table \ref{tab:gensys} is an average of 
DELPHI \cite{multDELPHI} and OPAL
\cite{multOPAL} measurements which does not
include charged decay products from the long lived particles
${\rm K}^0_s$ and $\Lambda^0$.

\mysubsection{Inclusive $\Dstarstar$ production rate in $b$-hadron semileptonic decays}
\label{sec:dssa}

The inclusive $b$-hadron semileptonic branching fraction into 
$\Dstarstar$  mesons has been 
measured in three different ways:
\begin{itemize}
\item[-]by subtracting the contributions of 
$\Bdb \rightarrow (\Dp + \Dstarp) \ell^- \overline{\nu_{\ell}}$ from the total
semileptonic branching fraction of $\Bdb$ mesons, yielding:
\begin{eqnarray}
{\rm BR}(\Bdb \rightarrow \rm D^{\ast \ast +} \ell^- \overline{\nu_{\ell}})& = &
(3.93 \pm 0.44 )\%
\label{eq:dsstarinc}
\end{eqnarray}
using the inclusive semileptonic branching 
fraction given in Table \ref{tab:gensys} and published values for the exclusive rates 
\cite{PDG98}.
\item[-]from a measurement at the $\Upsilon(4S)$ of the rate of final states
with a $\Dstarp$ and using models to account for the other channels
\cite{DssARGUS}:
\begin{eqnarray}
{\rm BR}(\Bdb \rightarrow \rm D^{\ast \ast +} \ell^- \overline{\nu_{\ell}}) &= &
(2.7 \pm 0.7 )\% 
\end{eqnarray}
\item[-]from an inclusive measurement of final states in which a D or a 
$\Dstarp$ meson is accompanied by a charged hadron and assuming
that non-strange $\Dstarstar$ decay channels involve only ${\rm D} \pi$ and
$\Dstar \pi$ final states:
\begin{eqnarray}
\nonumber {\rm BR}(\Bdb \rightarrow \rm D^{\ast \ast +} \ell^- \overline{\nu_{\ell}}) & = &
(2.16 \pm 0.30 \pm 0.30) \%~ \cite{DssALEPH}\\
     & =& (3.40 \pm 0.52 \pm 0.32) \%~\cite{DssDELPHI}
\end{eqnarray}
\end{itemize}
As the $\chi^2$ of these four measurements
is equal to three per degree of freedom, the uncertainty on the weighted
average has been multiplied by $\sqrt{3}$,
giving:
\begin{equation}
{\rm BR}(\Bdb \rightarrow \rm D^{\ast \ast +} \ell^- \overline{\nu_{\ell}}) ~=~ (3.04 \pm 0.44 ) \% 
\label{eq:ajous5}
\end{equation}

\mysubsection{$\Dstarstar$ decays to $\Dstar$ mesons in semileptonic $b$-decays}
\label{sec:dssb}

Semileptonic decays to excited charm states which subsequently decay 
to a $\Dstarp$\ are a source of correlated (physics) background in studies
of $\Bdb$ meson properties.
It is appropriate to express the different measurements in terms of the
parameter ${\rm b}^{\ast\ast}$, defined as the branching fraction 
of $\Bdb$
semileptonic decays involving $\Dstarstar$ final states in which a 
$\Dstar$, charged or neutral, is produced:
\begin{equation}
{\rm b}^{\ast\ast}={\rm BR}(\Bdb \rightarrow \rm D^{\ast \ast +} \ell^- \overline{\nu_{\ell}})
\times BR(D^{\ast \ast +} \rightarrow \Dstar X) 
\end{equation}
 This is to differentiate
such $\Dstarstar$ decays from those into a D meson directly.
Throughout this section 
it is assumed that decays of 
$\Dstarstar$
mesons involve at most one pion (or one kaon for strange states).

The following measurements have been 
interpreted in terms of the quantity ${\rm b}^{\ast\ast}$ and the
production fractions ($f_{{\rm B}_i}$) and lifetimes ($\tau({\rm B}_i)$)
of the different types of weakly decaying $b$-hadrons:
\begin{itemize}
\item[-] semi-inclusive measurements of semileptonic decays in which
a $\Dstarp$ and a charged pion have been isolated:
\begin{eqnarray}
\nonumber {\rm BR}(b \rightarrow \Dstarp \pi^- {\rm X} \ell^- \overline{\nu_{\ell}}) & = &
(4.73 \pm 0.77 \pm 0.55)\times 10^{-3}~\cite{DssALEPH} \\
\nonumber     & =& (4.8 \pm 0.9 \pm 0.5) \times 10^{-3}~\cite{DssDELPHI}\\
     & =& \fu\frac{2}{3}{\rm b}^{\ast\ast}\frac{\tau(\Bm)}{\tau(\Bdb)}
\end{eqnarray}
\item[-] the ARGUS measurement \cite{DssARGUS}:
\begin{eqnarray}
\nonumber {\rm BR}(\Bdb \rightarrow \rm D^{\ast \ast +} \ell^- \overline{\nu_{\ell}}) &= &
(2.7 \pm 0.7 )\% \\
     & = &\frac{{\rm b}^{\ast\ast}}{0.77}
\end{eqnarray}
where the value of 0.77 corresponds to the modelling used for 
$\Dstarstar$ decays
in that analysis;
\item[-] the inclusive production rate of charged $\Dstar$ mesons
in semileptonic $b$ decays:
\begin{eqnarray}
\nonumber {\rm BR}(b \rightarrow \Dstarp {\rm X} \ell^- \overline{\nu_{\ell}}) & = &
(2.75 \pm 0.17 \pm 0.16)\%~\cite{DssDELPHI} \\
\nonumber     & =& (2.86 \pm 0.18 \pm 0.23) \%~\cite{DssOPAL}\\
& =& \fd {\rm b}^{\ast} + \fd \frac{1}{3}{\rm b}^{\ast\ast}+
\fu\frac{2}{3}{\rm b}^{\ast\ast}\frac{\tau(\Bm)}{\tau(\Bdb)}
\label{ajoute}\\
\nonumber & & +\fs\frac{\alpha}{2}{\rm b}^{\ast\ast}\frac{\tau(\Bsb)}{\tau(\Bdb)}
\end{eqnarray}
where the quantity ${\rm b}^{\ast}$ is the exclusive 
semileptonic branching fraction:
\begin{equation}
{\rm b}^{\ast}={\rm BR}(\Bdb \rightarrow \rm \Dstarp \ell^- \overline{\nu_{\ell}})  = (4.64 \pm 0.25) \% 
\end{equation}
which has been obtained by averaging results quoted by the PDG
\cite{PDG98} for corresponding decays of $\Bdb$ and $\Bm$ mesons, and
the scaling factor $\alpha=0.75\pm0.25$ has been
introduced in (\ref{ajoute}) to account for a possible SU(3) flavour violation
when comparing $\rm D^{**+}_s \rightarrow \Dstarp K^0$ and
$\rm D^{**0} \rightarrow \Dstarp \pi^-$ decays.
\end{itemize}
In the above, the
production fractions ($f_{{\rm B}_i}$) and lifetimes ($\tau({\rm B}_i)$)
are taken from Sections \ref{sec:boscill} and \ref{sec:Atau} respectively.

These measurements form a coherent set of results yielding:
\begin{equation}
{\rm b}^{\ast\ast}={\rm BR}(\Bdb \rightarrow \rm D^{\ast \ast +} \ell^- \overline{\nu_{\ell}})
\times BR(D^{\ast \ast +} \rightarrow \Dstar X)  = (1.75 \pm 0.21 \pm 0.08 )
 \% 
\label{eq:ajoute1}
\end{equation}

From this result, and using the same hypotheses, it is possible to derive 
other quantities of interest
in several analyses presented in this paper such as:
\begin{eqnarray}
 {\rm BR}(\rm B^- \rightarrow \rm D^{\ast \ast 0} (\rightarrow \Dstarp \pi^-) \ell^-\overline{\nu_{\ell}})~ = ~\frac{2}{3}b^{\ast\ast}\frac{\tau(\Bm)}{\tau(\Bdb)} ~=~
(1.25 \pm 0.16) \% \label{eq:ajoute2}\\
 {\rm BR}(\Bdb ~\rightarrow~ \Dstarp \pi^0 \ell^-\overline{\nu_{\ell}})~ = ~
\frac{1}{3}b^{\ast\ast}~=~
(0.58 \pm 0.08) \% \\
 {\rm BR}(\Bsb \rightarrow \Dstarp {\rm K}^0 \ell^-\overline{\nu_{\ell}})~ = ~
\frac{1}{2}b^{\ast\ast}\frac{\tau(\Bsb)}{\tau(\Bdb)}\alpha~=~
(0.61  \pm 0.22 ) \%. 
\end{eqnarray}


It is also possible to determine the fraction of $\Bdb$ semileptonic 
decays which contain a $\Dstarp$:

\begin{equation}
\frac{{\rm BR}(\Bdb \rightarrow \Dstarp \ell^- \overline{\nu_{\ell}})+
{\rm BR}(\Bdb \rightarrow \Dstarp \pi \ell^- \overline{\nu_{\ell}})}
{{\rm BR}(\Bdb\rightarrow  \ell^- \overline{\nu_{\ell}} {\rm X})}
  = 0.49 \pm 0.03 
\label{eq:fracdstar}
\end{equation}

and the fraction of $b$-quark jets in which the $\Dstarp$ comes from a 
$\Dstarstar$ decay:
\begin{equation}
\frac{{\rm BR}(b \rightarrow \Dstarp \pi \ell^- \overline{\nu_{\ell}})}
{{\rm BR}(b\rightarrow \Dstarp \ell^- \overline{\nu_{\ell}})
+{\rm BR}(b \rightarrow \Dstarp \pi \ell^- \overline{\nu_{\ell}})}
  = 0.28 \pm 0.04
\label{eq:ajoute3}
\end{equation}

As these results (\ref{eq:ajoute1}) to (\ref{eq:ajoute3}) are highly correlated
they are represented in Table \ref{tab:gensys} by a single entry, 
chosen to be result
(\ref{eq:ajoute2}).

\mysubsection{Other semileptonic decays to $\Dstarp$ mesons}
\label{sec:ajous1}
Other mechanisms giving a $\Dstarp$ accompanied by a lepton
are important because they are further sources of background
for the exclusive channel $\Bdb \rightarrow \Dstarp \ell^- \overline{\nu_{\ell}}$.

\mysubsubsection{Charged $\Dstar$ production in semileptonic decays of 
$b$-hadrons involving $\tau$ leptons}

The value of  BR$(\Bdb \rightarrow \Dstarp \tau^- \overline{\nu_{\tau}} {\rm X})$ 
quoted in Table \ref{tab:gensys} is 
obtained from the average measured value of 
\mbox{BR($b \rightarrow \tau \overline{\nu_{\tau}} {\rm X}$)} taken from 
the PDG \cite{PDG98} and
assuming that a $\Dstarp$\ is produced in $(50 \pm 10) \%$ of the cases,
as measured in semileptonic decays involving light leptons. 
The uncertainty on this last number has been increased, as compared to 
the value
obtained in Equation (\ref{eq:fracdstar}), to account for a possible
different behaviour of decays when a $\tau$ is produced
instead of a light lepton because of differences in the masses
involved in the final state.

\mysubsubsection{Charged $\Dstar$ production in double charm semileptonic
$( b \rightarrow \overline{c} \rightarrow \ell)$ decays}

The double charm production rate in $b$-decays was measured by the ALEPH 
\cite{twocharma} and 
CLEO \cite{twocharmc} Collaborations, isolating the contributions of different
D meson species. From these measurements the inclusive semileptonic
branching fraction, involving a wrong sign charmed meson, has been obtained~\cite{EWWG}:
\begin{equation}
{\rm BR}(b \rightarrow \overline{c} \rightarrow {\ell})~=~
0.0162^{+0.0044}_{-0.0036}.
\end{equation}


 The product  
\mbox{$\rm BR( b \rightarrow \Dstarp X_{\overline{c}} Y) \times 
BR( X_{\overline{c}} \rightarrow \ell^-X$)}
quoted in Table \ref{tab:gensys} 
is then determined assuming again that a $\Dstarp$\ is produced in 
$(50 \pm 10) \%$ of the cases.

\mysubsection {Production of narrow $\Dstarstar$ states in $b$-hadron
semileptonic decays}
\label{sec:dssc}

A modelling of $b$-hadron semileptonic decays requires detailed 
rate measurements of the different produced states which can be resonant 
or non-resonant 
${\rm D}^{(*)}n\pi$ systems\footnote{In the following the study has been 
limited to $n$=1.},
 each having characteristic 
decay properties. As an example, there are four orbitally excited states
with L=1. They can be grouped in two pairs according to the
value of the spin of the light system, $j={\rm L} \pm 1/2$ (L=1).

States with $j=3/2$ can have ${\rm J}^{\rm P}=1^+$ and $2^+$.
The $1^+$ state decays only through ${\rm D}^*\pi$, and the $2^+$
through ${\rm D}\pi$ or ${\rm D}^*\pi$.
Parity and angular momentum conservation imply that in the $2^+$ the
${\rm D}^*$ and $\pi$ are in a D wave but allow both S and D waves
in the $1^+$. However, if the heavy quark spin is assumed to decouple, conservation of $j$(~=~3/2) forbids S waves even in the $1^+$.
An important D wave component, and the fact that the
masses of these states are not far from threshold, imply that the
$j=3/2$ states are narrow. These states have been observed with a typical
width of 20 $\MeV/c^2$, in accordance with the expectation.

On the contrary, $j=1/2$ states can have ${\rm J}^{\rm P}=0^+$ and $1^+$, so
they are expected to decay mainly through an S wave and to be broad resonances
with typical widths of several hundred $\MeV/c^2$. The first experimental
evidence for broad $1^+$ states has been obtained by CLEO \cite{ref:broaddss}.

At present, information on the composition 
of $b$-hadron semileptonic decays
involving $\Dstarstar$ is rather scarce.
ALEPH \cite{DssALEPH} and CLEO \cite{ref:cleodss}
have reported evidence for the 
production of narrow resonant states (${\rm D}_1$ and ${\rm D}_2^*$), 
which can be summarised as follows:
\begin{eqnarray}
{\rm BR}(\overline{{\rm B}} \rightarrow {\rm D}_1 \ell^-\overline{\nu_{\ell}} ) &=& (0.63 \pm 0.11) \% \label{eq:darate}\\
{\rm BR}(\overline{{\rm B}} \rightarrow {\rm D}_2^* \ell^-\overline{\nu_{\ell}} ) &=& (0.23 \pm 0.09)\%~{\rm or}~<0.4\% ~{\rm at~the~95\%~CL} 
\label{eq:dbrate} \\
{\rm R}^{**} = \frac{{\rm BR}(\overline{{\rm B}} \rightarrow {\rm D}^*_2 \ell^-\overline{\nu_{\ell}})}
{{\rm BR}(\overline{{\rm B}} \rightarrow {\rm D}_1 \ell^-\overline{\nu_{\ell}})} &=& 0.37 \pm 0.16 ~{\rm or}~<0.6 ~{\rm at~the~95\%~CL} \label{eq:Rss}
\end{eqnarray}
where ${\rm R}^{**}$ is the ratio between the production rates 
of ${\rm D}^*_2$ and
${\rm D}_1$ in $b$-meson semileptonic decays.
The original measurements are listed in Appendix \ref{appendixA}.
Absolute rates for the ${\rm D}_1$ and ${\rm D}_2^*$ mesons 
have been obtained
assuming that the former decays always into $\Dstar \pi$ and the latter
decays into ${\rm D} \pi$ in $(71 \pm 7)\%$ of the cases, as measured
by ARGUS \cite{argusn} and CLEO \cite{cleon}.

It is of interest to note that these results are very different from
naive expectations from HQET
in which ${\rm R}^{**} \simeq 1.6$ \cite{Pene}.
This implies large $1/m_c$ corrections in these models, as was noted
for instance in \cite{ref:ligeti}. 
These corrections not only reduce the expected value
for ${\rm R}^{**}$, they also enhance the expected 
production rate of $j=1/2$ states which can become of similar importance
as the $j=3/2$ rate. 

Comparing Equations (\ref{eq:darate}-\ref{eq:dbrate}) and (\ref{eq:ajous5}),
it can be deduced that
narrow $\Dstarstar$ states account for less than one third
of the total $\Dstarstar$ rate.
If, in addition, model expectations for the production rates of
$j=1/2$ states are used, it can be concluded that non resonant
production of ${\rm D}^{(*)}\pi$ states is as important as, or
even dominates over, the production of resonant states. As a consequence,
models like ISGW \cite{isgw,isgw2}, which contain only resonant charmed states,
do not provide a complete description of $b$-hadron semileptonic decays.



\mysubsection{$\Lb$ polarization}
\label{sec:dssd}

Even for unpolarized ${\rm e}^{\pm}$ beams, the $b$-quarks produced
in $\Zz$ decays are highly polarized: ${\cal P}_b=-0.94$.
Hard gluon emission and quark mass effects 
are expected to change ${\cal P}_b$ by only $3\%$
\cite{ref:popol}. However, during the hadronization process of the 
heavy quark, part or all of the initial $b$-quark polarization may be lost 
by the final weakly decaying $b$-hadron state. The $b$-mesons always 
decay finally to spin zero  pseudoscalar states, which do not retain
any polarization information. In contrast, $\Lb$ baryons
are expected to carry the initial $b$-quark polarization since the
light quarks are arranged, according to the constituent quark model, 
in a spin-0 and isospin-0 singlet. 
However, $b$-quark
fragmentation
into intermediate $\Sigma_b^{(*)}$ states can lead to a 
depolarization of the heavy quark
\cite{ref:popolb}.

The $\Lambda \ell$ final state is used to select event samples enriched
with $\Lb$ semileptonic decays, and the polarization is measured using
the average values of the lepton and neutrino energies. Samples
of events enriched in $b$-mesons, which carry no polarization information,
are used to calibrate the measurements.

Measurements from ALEPH \cite{ref:alpol}, DELPHI \cite{ref:delpol} and
OPAL \cite{ref:oppol}, collected in Appendix \ref{appendixAb},
 have been averaged, assuming
systematic uncertainties to be correlated, yielding:
\begin{equation}
{\cal P}(\Lb)~=~-0.45^{+0.17}_{-0.15}\pm 0.08
\end{equation}
where the uncertainty is still dominated by the statistics.
The following sources of systematic uncertainties, common to all
measurements, have been considered:
\begin{itemize}
\item $b$-quark fragmentation,
\item $\Lambda$ production in semileptonic and inclusive decays of the $\Lc$,
\item $\Lc$ polarization,
\item theoretical uncertainties in the modelling of polarized $\Lb$
semileptonic decays.
\end{itemize}

\end{itemize}

\mysection{Averages of $b$-hadron lifetimes}
\label{sec:Atau}

Best estimates for the various $b$-hadron lifetimes, for the ratio of the 
$\Bp$ and $\Bd$ lifetimes, and for the average lifetime of a sample
of $b$-hadrons produced in $b$-jets have been obtained by the B lifetime
working group\footnote{The present members of the B lifetime working group are:
J. Alcaraz, L. Di Ciaccio, T. Hessing, I.J. Kroll, H.G. Moser and
C. Shepherd-Themistocleous.}.
Details on the procedure used to combine the different measurements
can be found in  
\cite{ref:lifetimenote}.


The measurements used in the averages presented here are  
listed in Appendix \ref{appendixB}.
Possible biases can originate from the
averaging procedure. They depend on the statistics and time resolution
of each measurement and in the way systematic uncertainties have been
included. These effects have been studied in detail using simulations
and have been measured to be at a level 
which can be neglected as compared to the statistical uncertainty.

\mysubsection{Dominant sources of systematic uncertainties}
 Dominant sources of systematic uncertainties leading to correlations
between measurements are briefly reviewed (more details can be 
found in \cite{ref:lifetimenote}). 
These are the background estimation, the evaluation of the $b$-hadron momentum, 
and the decay length reconstruction.

 The background can be due either to physics processes leading
to a final state similar to that used to tag the signal, or to accidental 
combinations of tracks which simulate the decay of interest.
When ``physics'' background is present, the experimental uncertainties on the
branching fractions of the background processes and on the lifetimes of the
background particles lead to a systematic error which is correlated between
different experiments. When the background is combinatorial, the amount 
and/or the lifetime of the background ``particles'' is normally extracted 
from the data using the sidebands of mass distributions
 or wrong sign combinations. 
In this case the related systematic uncertainty is usually not correlated 
between experiments. But, in the measurement of the $b$-baryon lifetime
using $\Lambda \ell$ correlations, the amount of accidental background is 
obtained from the wrong sign combinations and a correction, common to all 
analyses, has to be applied
to this number to take into account the production asymmetry of accidental
$\Lambda \ell$ pairs.

\begin{figure}[tb!]
\begin{center}
\begin{tabular}{cc}
\mbox{\epsfxsize8.0cm \epsfysize10.0cm\epsffile{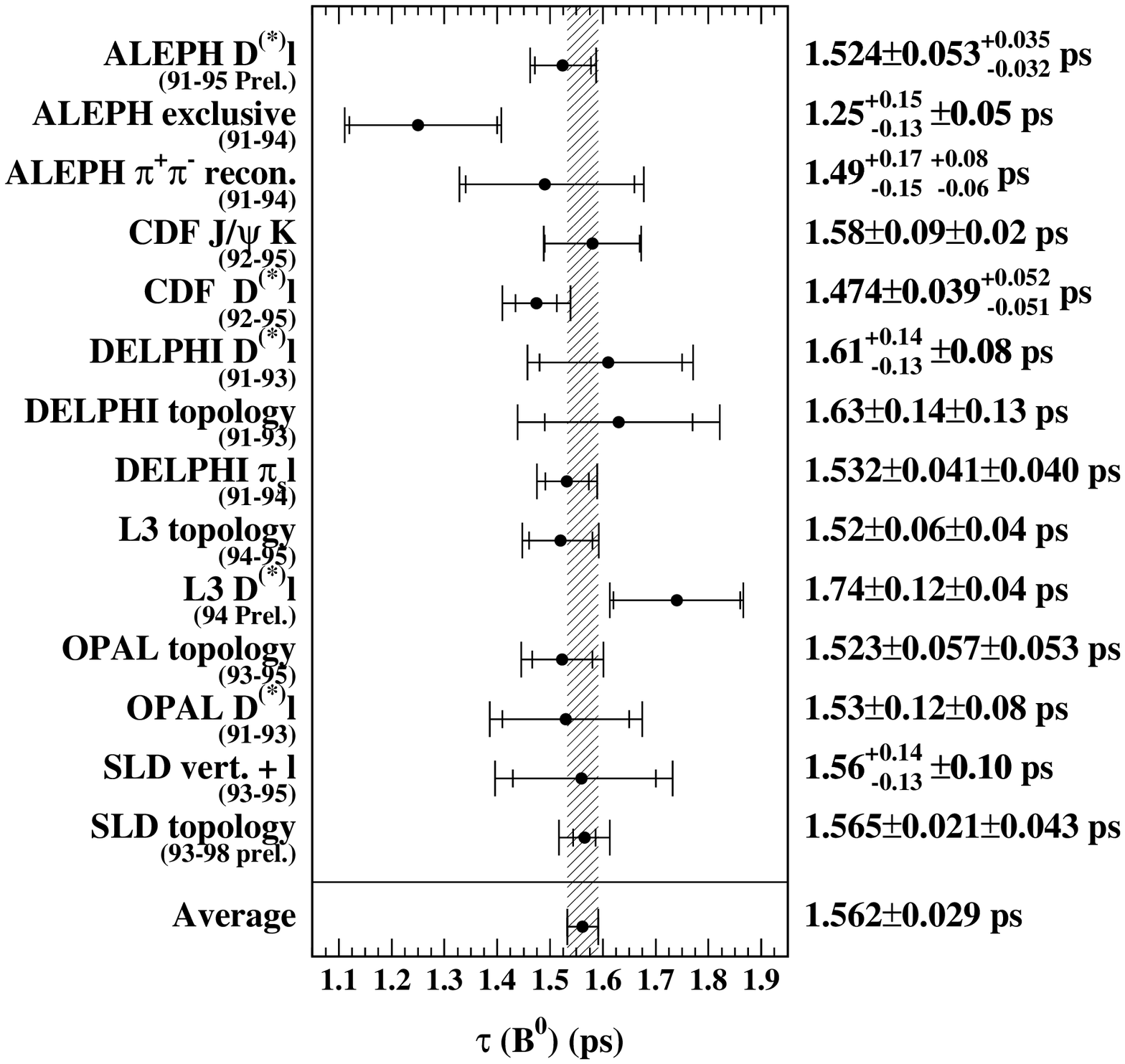}} &
\mbox{\epsfxsize8.0cm \epsfysize10.0cm\epsffile{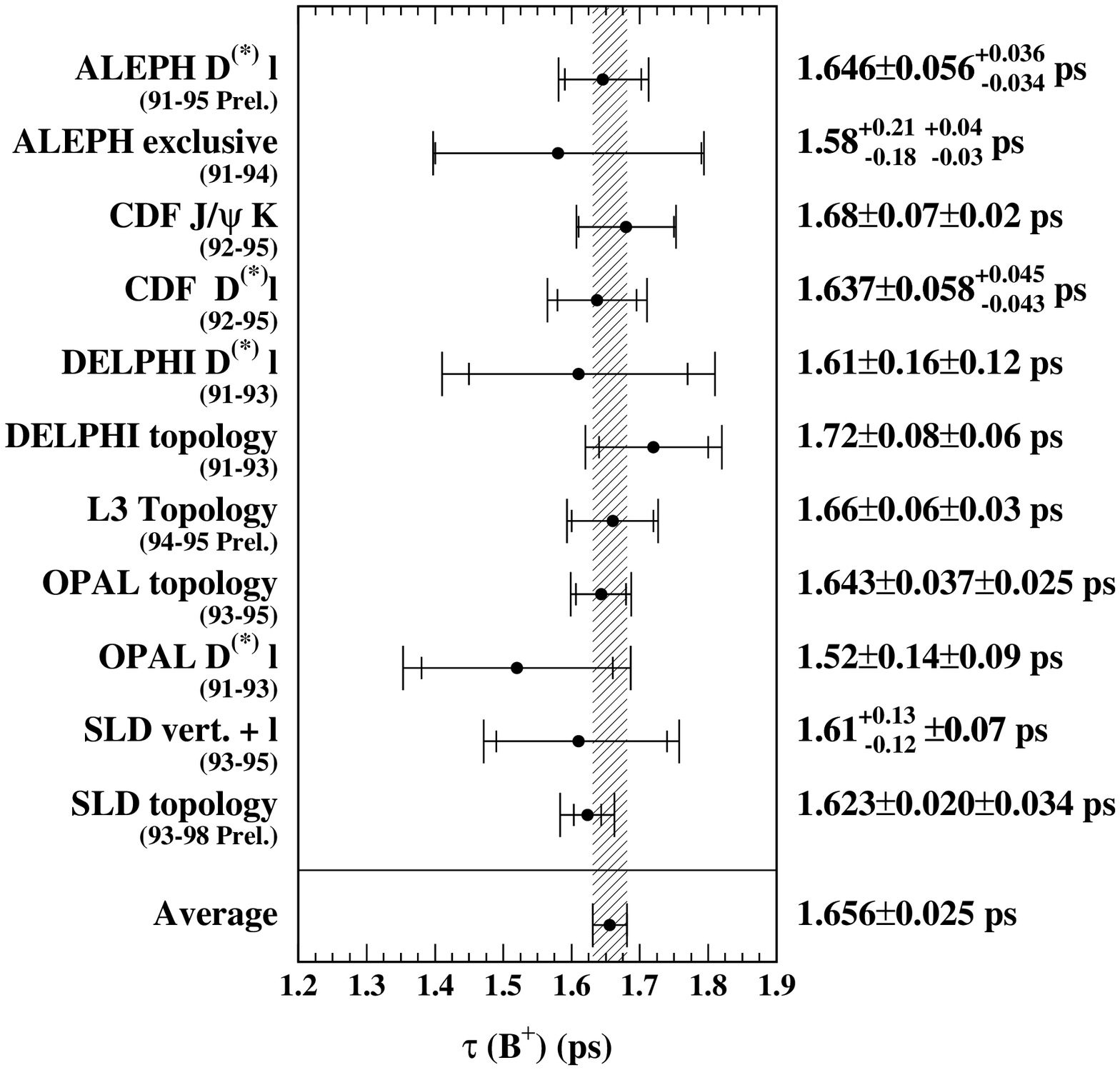}} \\
\end{tabular}
\caption{{\it Left: $\Bd$ lifetime measurements.
 Right: $\Bp$ lifetime measurements.
Hatched areas correspond to present averages of the measurements given in the 
corresponding figures. Internal error bars correspond to statistical
uncertainties and full error bars include systematics.}
\label{fig:tbdbp}}
\end{center}

\end{figure}

Most of the exclusive B lifetime measurements are based on the reconstruction
of B decay length and momentum. In most analyses the B particles are 
only partially reconstructed, and their energies are estimated from 
the energies of the detected decay products. 
In all cases, systematic uncertainties have been evaluated 
for the following effects:
\begin{itemize}
\item determination of the $b$-quark fragmentation function, 
\item branching fractions of B and charmed hadrons,
\item $b$-hadron masses,
\item $b$-baryon polarization and
\item modelling of neutral hadronic energy.
\end{itemize}
Finally there are uncertainties correlated within an experiment.
They are due to primary and secondary vertex
 reconstruction procedures, detector resolution, tracking errors, 
B flight direction reconstruction and detector alignment.

\begin{figure}[tb!]
\begin{center}
\begin{tabular}{cc}
\mbox{\epsfxsize8.0cm \epsfysize10.0cm\epsffile{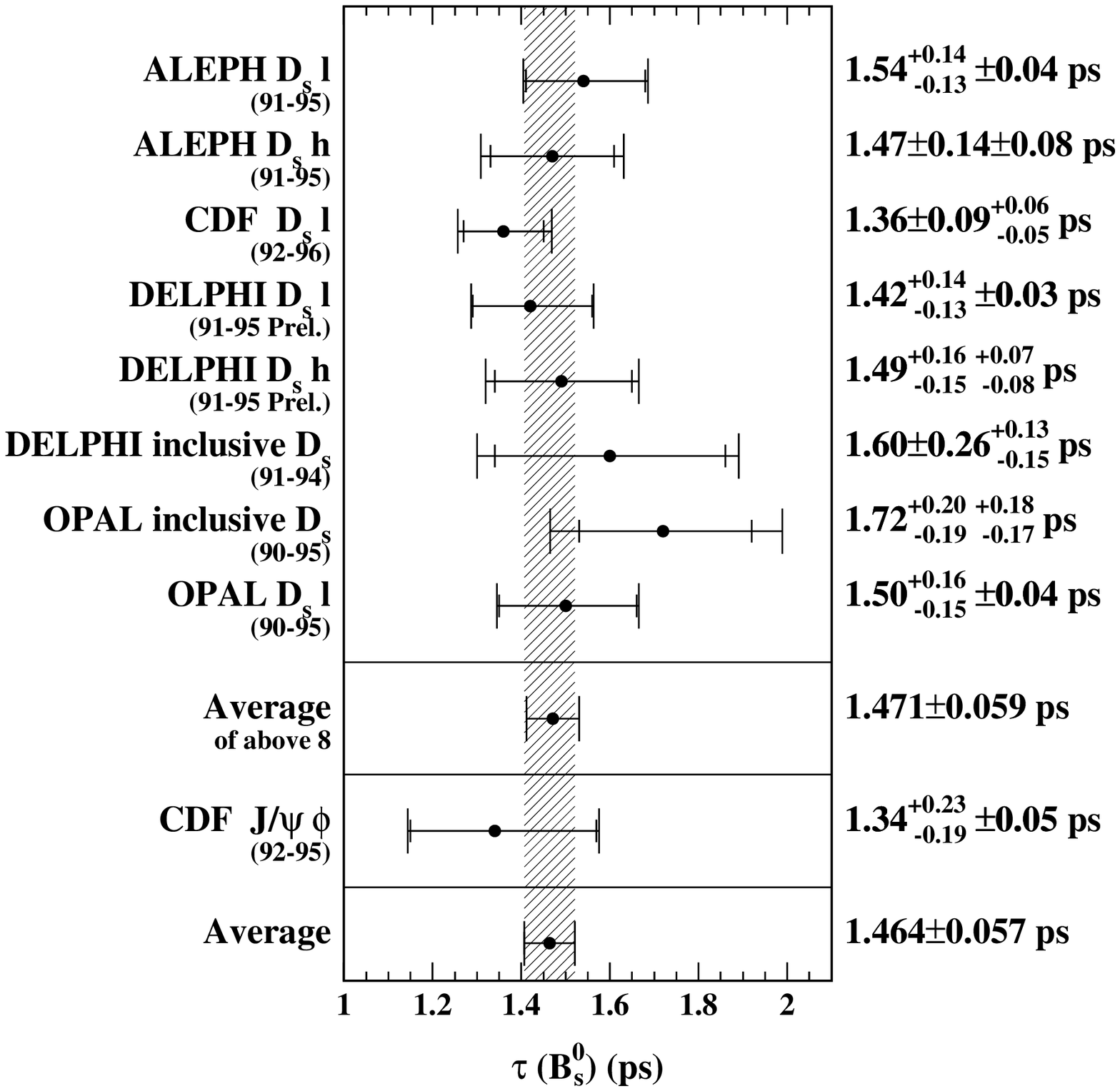}} &
\mbox{\epsfxsize8.0cm \epsfysize10.0cm\epsffile{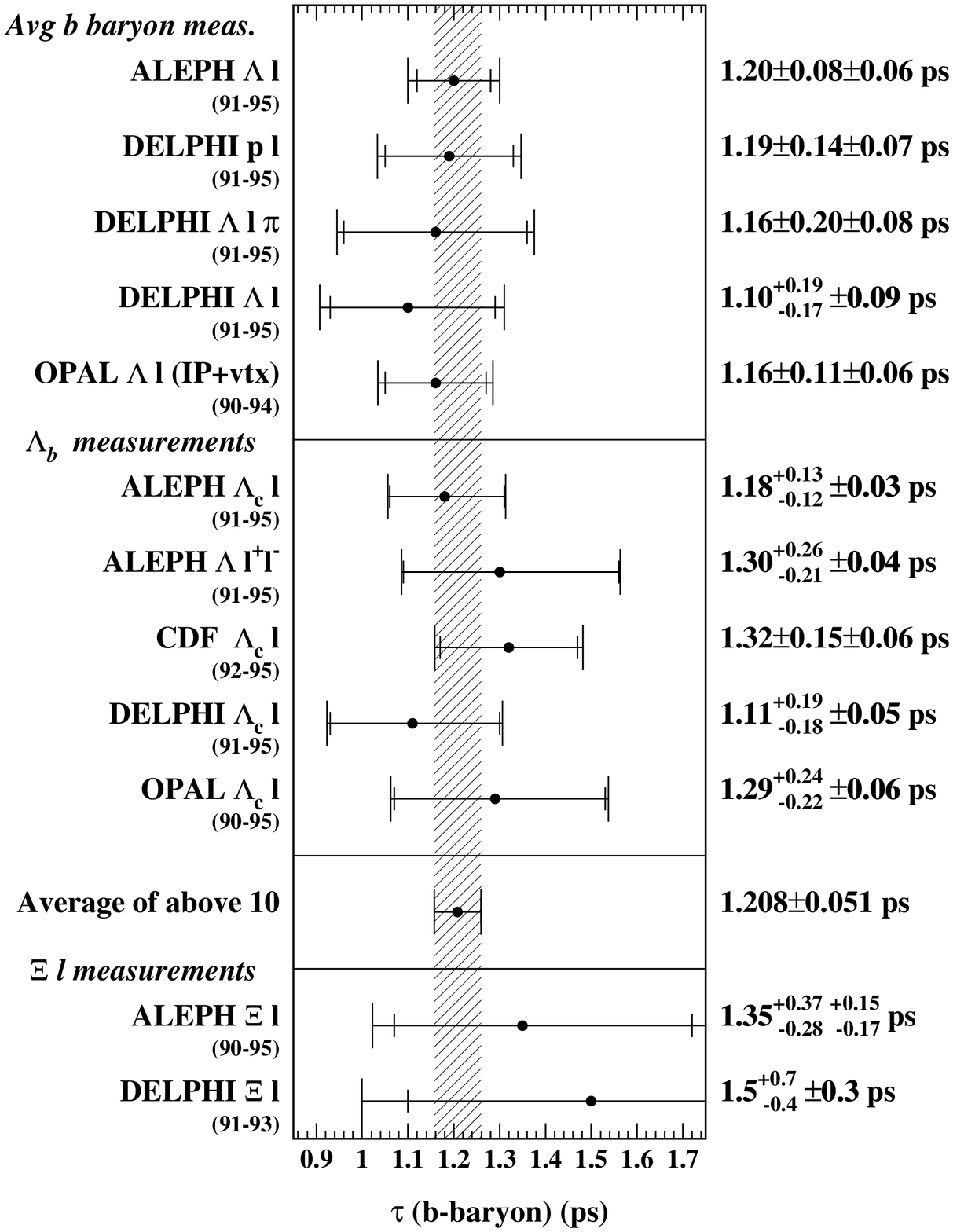}} \\
\end{tabular}
\caption{{\it Left: $\Bs$ lifetime measurements, 
Right: $b$-baryon lifetime measurements.
Hatched areas correspond to present averages of the measurements given in the 
corresponding figures. Internal error bars correspond to statistical
uncertainties and full error bars include 
systematics.} \label{fig:tbsbb}}
\end{center}

\end{figure}

\mysubsection{Measurements of $\Bd$ and $\Bp$ lifetimes}
\label{sec:taubd}
Apart from the measurements by CDF using large samples of exclusive 
$\Bd \rightarrow {\rm J}/\psi K^{(*)0}$ and $\Bp \rightarrow {\rm J}/\psi K^+$
decays, the most precise measurements of $\Bd$ and $\Bp$ lifetimes
originate from two classes of partially reconstructed decays.
In the first class the decay ${\rm B} \rightarrow \overline{{\rm D}^{(*)}} \ell^+ \nu_{\ell} {\rm X}$
is used in which the charge of the charmed meson distinguishes between
neutral and charged B mesons. In the second class the charge attached to
the B decay vertex is used to achieve this separation.

The following sources of correlated systematic uncertainties have been 
considered:
background composition (includes $\Dstarstar$ branching fraction uncertainties
as obtained in Section \ref{sec:systgen}),
momentum estimation,
lifetimes of $\Bs$ and $b$-baryons (as obtained in Sections \ref{sec:taubs}
and \ref{sec:taulb}),
and fractions of $\Bs$ and $b$-baryons
 produced in $\Zz$ decays (as measured in Section \ref{sec:results}).

The world average lifetimes of $\Bd$ and $\Bp$ mesons are:
\begin{eqnarray}
 \tau(\Bd)& = & (\taubd)~{\rm ps} \\
 \tau(\Bp)& = & (\taubp)~{\rm ps}  .
\end{eqnarray}
The several measurements and the average values are shown
in Figure \ref{fig:tbdbp}.

\mysubsection{$\Bs$ lifetime measurements}
\label{sec:taubs}

The most precise measurements of the $\Bs$ lifetime
originate from partially reconstructed decays in which
a ${\rm D}^-_s$ meson has been completely reconstructed
(see Figure \ref{fig:tbsbb}-left).

The following sources of correlated systematic uncertainties have been 
considered:
average $b$-hadron lifetime used for backgrounds,
$\Bs$ decay multiplicity,
and branching fractions of B and charmed hadrons.

 
The world average lifetime of $\Bs$ mesons is equal to:
\begin{eqnarray}
 \tau(\Bs)& = & (\taubs)~{\rm ps} 
\end{eqnarray}

This result has been obtained neglecting a possible difference, $\dgs$,
between the
decay widths of the two mass eigenstates of the $\Bs-\Bsb$ system. 
The measurement of $\dgs$ is explained in Section \ref{sec:deltag}. 
The several measurements and the average value are shown
in Figure \ref{fig:tbsbb}-left.

\mysubsection{$b$-baryon lifetime measurements}
\label{sec:taulb}

The most precise measurements of the $b$-baryon lifetime
originate from two classes of partially reconstructed decays
(see Figure \ref{fig:tbsbb}-right).
In the first class, decays with an exclusively reconstructed $\Lc$ baryon
and a lepton of opposite charge are used.
In the second class, more inclusive final states with a baryon
(${\rm p},~\overline{{\rm p}},~\Lambda~{\rm or}~\overline{\Lambda}$) and a 
lepton have been used.

The following sources of correlated systematic uncertainties have been 
considered:
experimental time resolution within a given experiment, $b$-quark
fragmentation distribution into weakly decaying $b$-baryons,
$\Lb$ polarization,
decay model,
and evaluation of the $b$-baryon purity in the selected event samples.
As the measured $b$-hadron lifetime is proportional to the assumed 
$b$-hadron mass,
the central values of the masses are scaled to m($\Lambda_b$) = ($5624 \pm 9$) 
$\MeV/c^2$ and
m($b$-baryon) = ($5670 \pm 100$) $\MeV/c^2$, before computing the averages.
Uncertainties related to the decay model include 
mostly assumptions on the fraction of $n$-body decays.
 To be conservative it is 
assumed
that they are correlated whenever given as an error.
Furthermore, in computing the average, results have been corrected 
for the effect of the measured value of the $\Lb$ polarization 
 and it has been assumed
that $b$-baryons have the same fragmentation distribution as all
$b$-hadrons (Section \ref{sec:systgen}).  
    
The world average lifetime of $b$-baryons is then:
\begin{eqnarray}
 \tau(b-{\rm baryon})& = & (\taubbar)~{\rm ps}
\end{eqnarray}
This value and the single measurements  are given
in Figure \ref{fig:tbsbb}-right.

Keeping only $\Lambda^{\pm}_c \ell^{\mp}$ 
and $\Lambda \ell^- \ell^+$ final states, as representative of 
the $\Lb$ baryon, the following lifetime is obtained:
\begin{eqnarray}
 \tau(\Lb)& = & (\taulb)~{\rm ps}
\end{eqnarray}

Averaging the measurements based on the $\Xi^{\mp} \ell^{\mp}$
final states gives a lifetime value for a sample of events 
containing $\Xi_b^0$ and $\Xi_b^-$ baryons:
\begin{eqnarray}
 \tau(\Xi_b)& = & (\tauxib)~{\rm ps}
\end{eqnarray}

\begin{figure}[tb!]
\begin{center}
\begin{tabular}{cc}
\mbox{\epsfxsize8.0cm \epsfysize10.0cm\epsffile{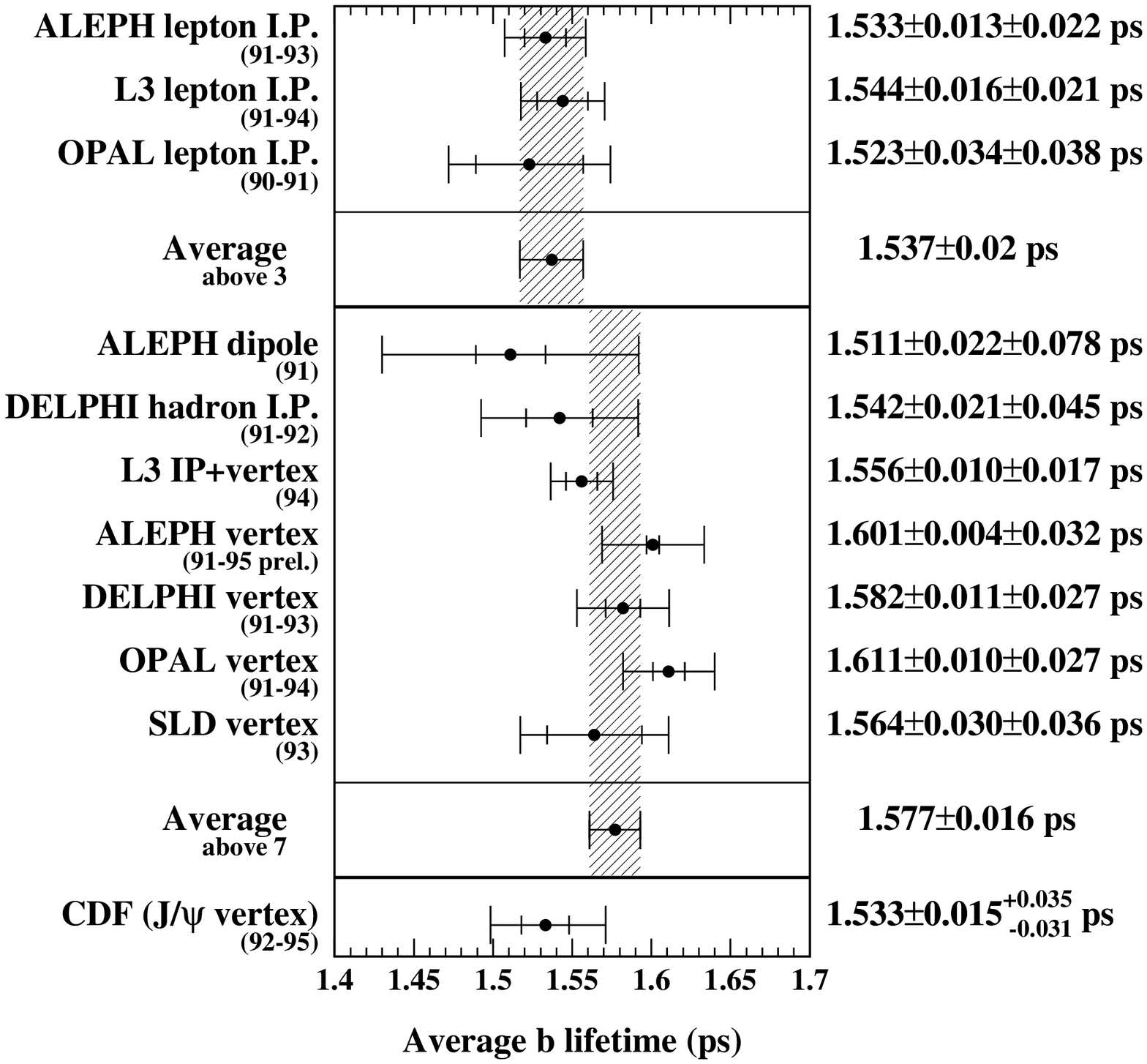}} &
\mbox{\epsfxsize8.0cm \epsfysize10.0cm\epsffile{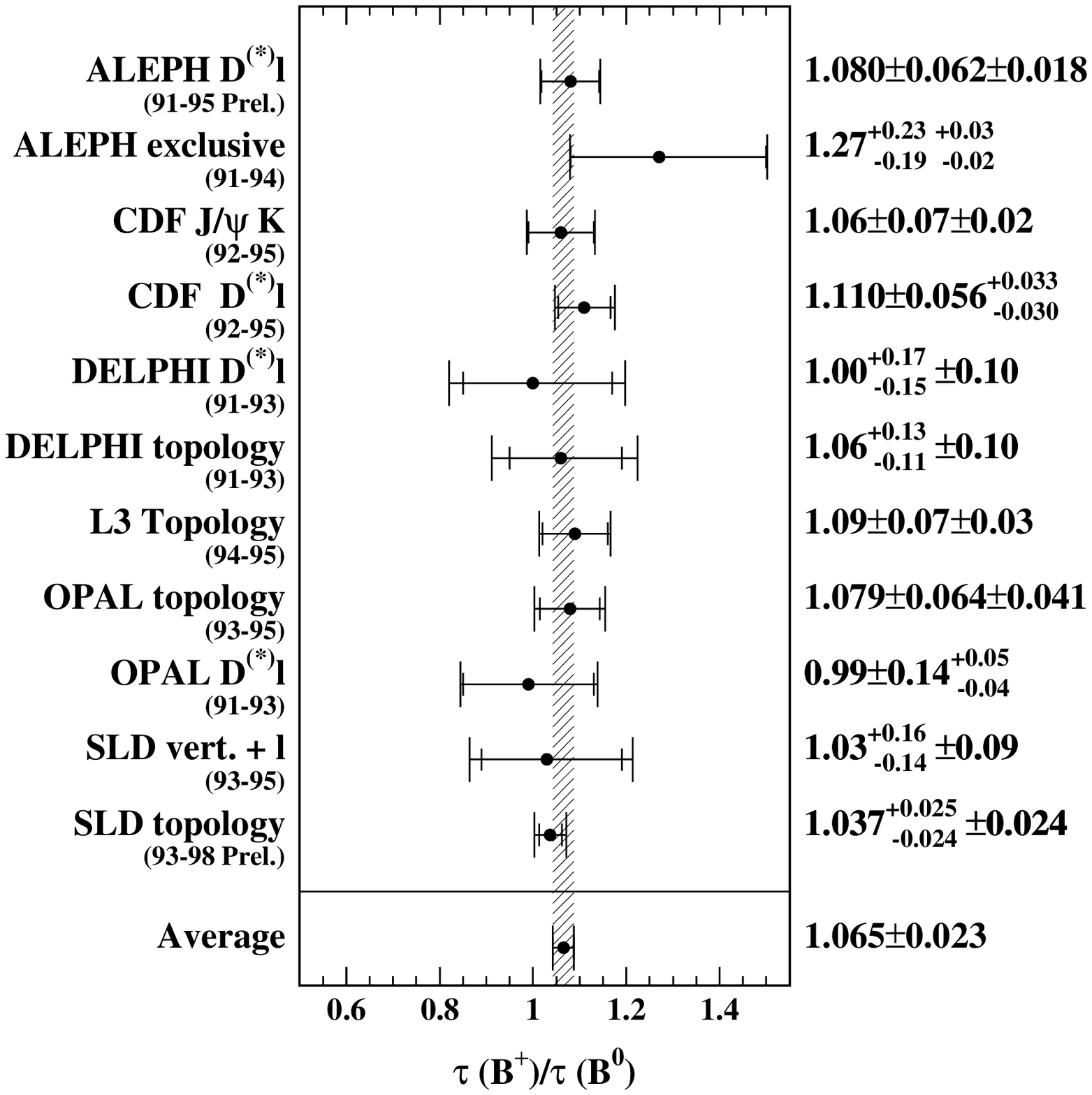}} \\

\end{tabular}
\caption{{\it Left: Measurements of the average $b$-hadron lifetime.
 Right: Ratio between the $\Bm$ and $\Bdb$ lifetimes.
Hatched areas correspond to present averages of the measurements given in the 
corresponding figures. Internal error bars correspond to statistical
uncertainties and full error bars include 
systematics.}  \label{fig:tbhad}}
\end{center}
\end{figure}

\mysubsection{Average $b$-hadron lifetime}
\label{sec:tauav}

Available results have been divided into three different sets and a 
separate average has been computed for each set
(see Figure \ref{fig:tbhad}-left):
\begin{itemize}
\item [(1)]measurements at LEP and SLD which accept any $b$-hadron decay:
\begin{equation}
\tau_b^{incl.}~=~(1.577 \pm 0.016)~{\rm ps}
\end{equation}

 
\item [(2)]measurements at LEP based on the identification of a lepton
      from a $b$-hadron decay:
\begin{equation}
\tau_b^{sl}~=~(1.537 \pm 0.020)~{\rm ps}
\end{equation}

\item [(3)]measurement at CDF based on the identification of a J/$\psi$ 
from a $b$-hadron decay: 
\begin{equation}
\tau_b^{{\rm J}/\psi}~=~(1.533^{+0.038}_{-0.034})~{\rm ps}
\end{equation}
\end{itemize}

The reason for this division is that, since the lifetimes of the individual 
$b$-hadron species are different,
a meaningful average lifetime 
can only be computed for samples in which the composition is the
same.

The following sources of correlated systematic uncertainties have been 
considered when evaluating the averages:
$b$- and $c$-quark fragmentation and decay models,
BR($b \rightarrow \ell$),
BR($b \rightarrow$ c $\rightarrow \overline{\ell}$),
$\tau_c$,
BR(c $\rightarrow \overline{\ell}$)
and the charged track multiplicity in $b$-hadron decays
(see Table \ref{tab:gensys}).

The measurement in Set 1 is the average 
$b$-hadron lifetime for a sample of weakly decaying $b$-hadrons 
produced in $\Zz$ decays.
\begin{equation}
\tau_b~=~\fd \tau(\Bd)+\fu\tau(\Bp)+\fs\tau(\Bs)+\fb\tau({\it b}-{\rm baryon})
\end{equation}


The measurements of Set 2 are based on a sample 
enriched with $\Bp$ 
with respect to other production
fractions, and thus 
are expected to correspond to a higher average lifetime.
In practice this difference is expected to be small.
Assuming that all $b$-hadrons have the
same semileptonic partial width,
the average lifetimes in the inclusive and semileptonic samples
can be related using the measured values of the production rates 
(Section \ref{sec:boscill})
and
lifetimes of the different types of $b$-hadrons, giving:
\begin{equation}
\frac{\tau_b^{sl}}{\tau_b}~=~1.007 \pm 0.005.
\label{eq:taurapp}
\end{equation}

Finally the CDF ${\rm J}/\psi$ measurement probably corresponds
to a different composition in $b$-hadron species 
compared to any of
the previous sets of measurements.

As a consequence, the average $b$-lifetime, $\tau_b$, is obtained by combining
the values of $\tau_b^{incl.}$
and $\tau_b^{sl}$ only, and neglecting the small correction expected in
Equation (\ref{eq:taurapp}):
\begin{equation}
\tau_b~=~(\taubav)~{\rm ps}.
\label{eq:taub}
\end{equation}
 
%
The difference between the values of $\tau_b^{incl.}$ and $\tau_b^{sl}$
amounts to less than two standard deviations.

\mysubsection{$b$-hadron lifetime ratios}
\label{sec:tauth}

Ratios of other $b$-hadron lifetimes to the $\Bd$ lifetime
have also been determined, using the previous averages of the 
individual lifetimes
and the direct measurements of the ratio between $\Bp$ and $\Bd$ lifetimes
shown in Figure \ref{fig:tbhad}-right. The results are given
in Table \ref{tab:lifetimefrac}.

\begin{table}[ht!]
  \begin{center}
    \begin{tabular}{|c| c |}
      \hline
      Lifetime ratio & Measured value \\
      \hline
     $\tau(\Bp)/\tau(\Bd)$ & $\taubpovertaubd$ \\
     $\tau(\Bs)/\tau(\Bd)$ & $0.937 \pm0.040$ \\
     $\tau(b-{\rm baryon})/\tau(\Bd)$ & $0.773 \pm0.036$ \\
\hline
    \end{tabular}
  \end{center}
    \caption{\it {Ratios of $b$-hadron lifetimes relative to the $\Bd$ 
lifetime.}    \label{tab:lifetimefrac}}

\end{table}

It can be seen that
the predictions \cite{ref:bigilife} for the meson lifetimes
based on factorization,
given in Equation (\ref{eq:predt}),
which had encountered theoretical criticism \cite{ref:sacrelife},
are on the mark. A recent lattice study \cite{ref:latlife}
finds a result quite consistent with \cite{ref:bigilife}.

\begin{equation}
\frac{\tau(\Bp)}{\tau(\Bd)}~=~1 + 0.05 \times \frac{f_B^2}{(200~\MeV)^2},~
\frac{\tau(\Bs)}{\tau(\Bd)}~=~1 \pm {\cal O}(1\%),~
\frac{\tau(\Lb)}{\tau(\Bd)}~=~0.90-0.95.
\label{eq:predt}
\end{equation}
However a large discrepancy is observed for baryons.
It remains to be clarified if this problem is related only to the validity
of the quark models used to determine the parameters in the case of baryons, 
as was
suggested in reference \cite{ref:bigilife}, or to more basic
defects in the whole theoretical picture.
Lattice evaluations
for $b$-baryons are on the way \cite{ref:latlifeb}.

\mysection{$\Bd$ and $\Bs$ oscillations and
  $b$-hadron production fractions}
\label{sec:boscill}

The four LEP Collaborations and CDF and SLD have published or 
otherwise released
measurements of \dmd~\cite{A:dmd,C:dmd,D:dmd,L:dmd,O:dmd,S:dmd}
and lower limits for \dms~\cite{A:dms,C:dms,D:dms,O:dms,S:dms}.
Combined results, as well as estimates of $b$-hadron fractions, have been 
prepared by the B oscillation working 
group\footnote{The present members of the B oscillation working group are:
V. Andreev, E. Barberio, 
G. Boix, C. Bourdarios, P. Checchia, O. Hayes, R. Hawkings, M. Jimack, O. Leroy, S. Mele,
 H-G. Moser, F. Parodi, M. Paulini,
P. Privitera, 
P. Roudeau, O. Schneider, A. Stocchi, T. Usher, C. Weiser
and S. Willocq.}.  

The estimates of the $b$-fractions, described in Section \ref{sec:results},
are important inputs needed for the combination of the $\Bd$ and $\Bs$
oscillation results. 
The procedure used to combine \dmd\ results is explained in
Section \ref{sec:method}, followed by a description of the
common systematic uncertainties in the analyses, and the result for
the mean value of \dmd.
Then combined results on the \PsBz\ oscillation amplitude,
as well as an overall lower limit on \dms\, are presented
in Section \ref{sec:dmsosc}.
More details on these procedures can be found in reference \cite{ref:nimosc}.

\mysubsection{Measurements of $b$-hadron production rates in $b$-jets}
\label{sec:results}

In this analysis, the relative production rates
of the different types of weakly-decaying $b$-hadrons are
assumed to be similar in $b$-jets originating from $\Zz$ decays
and in high transverse momentum $b$-jets in ${\rm p}\overline{{\rm p}}$
collisions at 1.8 TeV.
This hypothesis can be justified by considering the last steps of 
jet hadronization to be a non-perturbative QCD process occurring
at a scale of order $\Lambda_{{\rm QCD}}$.

Direct information on production rates is available
from measurements of branching fraction products using channels
with characteristic signatures. All the measurements
used in the present analysis are listed in Appendix \ref{appendixAc}.

 At CDF and LEP, the $\Bsb$ production rate has been evaluated using
events with a $\Ds$ accompanied by a lepton of opposite sign in 
the final state (Table \ref{tab:sfrac}).
 Since the rate of these events is given by
$\fs \times \bsdslX$, it is necessary to evaluate \bsdslX.
This has been done by assuming, from SU(3) symmetry, that
the partial semileptonic decay widths into D, ${\rm D}^*$ 
and $\Dstarstar$ final states
are the same for all B mesons.
A lower value for the $\Bs$ semileptonic branching fraction
with a $\Dsm$ in the final state is obtained by assuming
that all $\overline{{\rm D}^{**}_s}$ states decay into a non-strange D meson.
The branching fraction corresponding to
$\Bs \rightarrow {\rm D}_s^{**-} \ell^+ \nu_{\ell}$ is obtained
using Equation (\ref{eq:dsstarinc}) multiplied by the lifetime ratio
$\tau(\Bs)/\tau(\Bd)$. This corresponds to a lower limit
on $\Dsm$ production because the possibility that 
${\rm D}_s^{**-}$ states decay into $\Dsm \overline{{\rm K}}{\rm K}$
or $\Dsm \eta^{(')}$ has
been neglected.
In addition, assuming that only the measured ${\rm D}^{(*)} \pi$
final state in $\Bd$ or $\Bp$ semileptonic decays corresponds
to ${\rm D}^{(*)} {\rm K}$ transitions for the $\Bs$, an upper value
can be obtained for the $\Bs$ decay of interest using the relation:
\begin{equation}
\Gamma(\Bd \rightarrow \overline{{\rm D}^{(*)0}} \pi^- \ell^+ \nu_{\ell})
=\frac{4}{3} \Gamma(\Bs \rightarrow \overline{{\rm D}^{(*)0}} {\rm K}^- \ell^+ \nu_{\ell})
\end{equation}
based on isospin symmetry. In practice, the lower and upper values are 
compatible within uncertainties.
This allows $\fs=\fbsdir$ to be extracted from the LEP measurements.
Using the same assumptions, the CDF Collaboration has measured the 
ratio $\fs/(\fu+\fd)=0.213\pm0.068$
from final states with an electron and a charm meson 
($\Do,~\Dp,~\Dstarp,~\Dsp$) 
\cite{A:flamcdf}\footnote{CDF has published a 
second measurement of the same ratio,
using double semileptonic B decays with $\phi \ell$ and ${\rm K}^* \ell$
final states as
characteristic signatures for $\Bs$ and non-strange B mesons 
\cite{ref:cdfrates}. At present,
this result has not been included in the average  
for $b$-hadron production fractions because of
difficulties in properly correlating certain assumptions in that analysis
with the other measurements.}. 

In a similar way, the fraction of $b$-baryons is estimated 
from the measured production rates
of $\Lc \ell^-$ \cite{AD:prodlamb, AD:prodlamb2} and $\Xi^- \ell^-$ 
\cite{AD:cascb, AD:cascb2} final states (Table \ref{tab:bfrac})
yielding, respectively,
$f_{\Lb}=\flbdir$ and $f_{\Xi_b^-}=\fxidir$. 
The value for BR($\Lb \rightarrow \Lc {\rm X} \ell^- \overline{\nu_{\ell}}$)
has been obtained considering that there could be one 
$\Lc$ produced in every decay or, at the other extreme, 
that all excited charmed
baryons, assumed to be produced with a rate similar to $\Dstarstar$ final 
states in B mesons, 
decay into a D meson and a non-charmed baryon.
Similar considerations
have been applied to $\Xi_b$ semileptonic decays.
The semileptonic decay width 
$\Gamma(\Xi_b \rightarrow \Xi_c {\rm X} \ell^- \overline{\nu_{\ell}}$)
has been taken to be equal to
$\Gamma(\Lb \rightarrow \Lc {\rm X} \ell^- \overline{\nu_{\ell}}$)
and it has been assumed that, within present uncertainties,
BR$(\Xi_c^0 \rightarrow \Xi^- {\rm X})$ = 
BR$(\Lc \rightarrow \Lambda^0 {\rm X})$.
The total $b$-baryon
production rate is then: $\fb=\fbardir$, assuming the same
production rates for $\Xi_b^0$ and $\Xi_b^-$ baryons.
This is then averaged with a direct measurement of $\fb=\fbarspec$
from the number
of protons in $b$-events \cite{A:flamdir}.
Finally, the CDF Collaboration
has measured the production rate of $b$-baryons relative
to non-strange B mesons,
$\fb/(\fu+\fd)=0.118\pm0.042$, using $\Lc e^-$ final states 
\cite{A:flamcdf}.

The fraction of $\Bp$ mesons has also been measured \cite{ref:delphibplus}
 using the charge of a large
sample of inclusively reconstructed secondary vertices: 
$\fu=\fbudir$ (Table \ref{tab:bpfrac}).

All these measurements have been combined to obtain average production rates
for $\Bs$, $b$-baryons, $\Bp$ and $\Bd$ \footnote{The production
of ${\rm B}_c$ mesons and of other weakly decaying
states made of several heavy quarks has been neglected.}, imposing that:
\begin{equation}
\fu + \fd + \fs +\fb~=~1 
\label{eq:unitary}
\end{equation}
and 
\begin{equation}
\fu = \fd.
\label{eq:equal}
\end{equation}
The results obtained are given in Table \ref{tab:ratesstepa}.
Correlated systematics between the different measurements, 
coming mainly from
the poorly measured branching fractions of $\Ds$ and $\Lc$ charmed hadrons,
have been taken into account.
It may be noted that constraint (\ref{eq:equal}) is expected to be true for
$b$-mesons even though it is known to be wrong for $c$-mesons.
In $b$-jets, and also in $c$-jets, isospin invariance of strong 
interactions predicts
similar production rates of mesons
in which the heavy quark is associated to a $\overline{u}$ or
$\overline{d}$ antiquark. 
Strong and electromagnetic decays of these states result in different rates
for weakly decaying mesons with a $u$ or a $d$ flavour
because the $\Dstar \rightarrow {\rm D} \pi$ decays occur
very close to threshold, and the threshold prevents the transition
$\Dstarp \rightarrow \Dp \pi^-$. However
for $\Bd$ and 
$\Bp$ mesons, no asymmetry is expected: $\Bstar$ mesons decay 
electromagnetically, leaving the flavour of the light spectator quark
in the $b$-meson unchanged;
non-strange $\Bstarstar$ mesons decaying through strong interactions and having
masses away from threshold will not induce any asymmetry either;
${\rm B}^{**0}_s$ mesons decay also by strong interactions into
${\rm B}^{0(*)}_d \overline{{\rm K}^0}$ or 
${\rm B}^{+(*)} {\rm K}^-$ with equal probabilities,
although a possible tiny difference
between these two rates can be expected for decays of narrow states occurring
very close to threshold, because of the mass difference between ${\rm K}^0$
and ${\rm K}^+$ mesons\footnote{The mass difference between $\Bd$ and $\Bp$
mesons is compatible with zero within $\pm 0.3~\MeV/c^2$.}. 

\begin{table}[t]
\begin{center}
\begin{tabular}{|l|c|}
\hline
 $b$-hadron fractions  & correlation coefficients \\ \hline
 $\fs$ = $\fsa$ & \\
 $\fb$ = $\fbara$ &$\rho(\fs,\fb)=\rhosbara$ \\
$\fd = \fu$ = $\fua$ & $\rho(\fd,\fs)=\rhosua$,~$\rho(\fd,\fb)=\rhobarua$\\
\hline
\end{tabular}
\end{center}
\caption{{\it Average values of $b$-hadron production rates
and their correlation coefficients obtained from direct measurements.}
\label{tab:ratesstepa}} 
\end{table}


Additional information on the production rates
can be obtained from measurements of the time-integrated
mixing probability of $b$-hadrons. For an unbiased sample of
semileptonic $b$-hadron decays in $b$-jets, with fractions
$\gd$ and $\gs$ of $\Bd$ and $\Bs$ mesons, this mixing
probability is equal to:
\begin{equation}
\chib= \gs~\chis + \gd ~\chid
\label{eq:chibar}
\end{equation}
where $\chid$
is the time-integrated mixing probability for $\Bd$ mesons 
(see Section \ref{sec:method})
and $\chis$=1/2 is the corresponding quantity for $\Bs$ 
mesons\footnote{The assumption $\chis =\frac{1}{2}$ can be justified
  by the existence of limits on $\dms$ obtained from $\rm D_s$-lepton 
analyses, which have negligible dependence on $\fs$.}.
As already mentioned in Section \ref{sec:systgen}, the semileptonic width
is assumed to be the same for all $b$-hadron species, implying
$g_i = f_i ~{\mbox R}_i$, where ${\mbox R}_i=\frac{\tau_i}{\tau_b}$
are the lifetime ratios. This leads to the relation:
\begin{equation}
\fs~=~\frac{1}{{\mbox R}_s}~
\frac{(1+r)~\chib-(1-\fb ~{\mbox R}_{b-{\rm baryon}})~ \chid}
{(1+r)~ \chis - \chid}
\end{equation}
where $r={\mbox R}_u/{\mbox R}_d = \tau(\Bp)/\tau(\Bd)$.
This is used to extract another determination of $\fs$
from the $b$-baryon fraction of Table \ref{tab:ratesstepa},
the lifetime ratio averages of Table \ref{tab:lifetimefrac},
the world average value of $\chid$ from Equation (\ref{eq:dmdbb}) 
of Section \ref{sec:method}, and the $\chib$ average from the 
LEPEWWG (see Table \ref{tab:gensys}). This new estimate of
$\fs$,
%
$\fs^{\rm mixing}= \fbsmix$, is then combined
with the $b$-hadron rates from direct measurements, 
taking into account correlations\footnote{
There is a small statistical correlation between 
$\chid$ and $\chib$, arising from the fact that a few $\dmd$ analyses at LEP 
are based on the same samples of dilepton events as the ones used to extract 
$\chib$. This correlation is ignored, with a negligible effect on the final result.} and imposing the conditions (\ref{eq:unitary}) and (\ref{eq:equal}).
The final $b$-hadron fractions are displayed in Table \ref{tab:rates}.

\begin{table}[t]
\begin{center}
\begin{tabular}{|l|c|}
\hline
 $b$-hadron fractions  & correlation coefficients \\ \hline
 $\fs$ = $\fbs$ & \\
 $\fb$ = $\fbar$ &$\rho(\fs,\fb)=\rhosbar$ \\
$\fd = \fu$ = $\fbd$ & $\rho(\fd,\fs)=\rhosd$,~$\rho(\fd,\fb)=\rhobard$\\
\hline
\end{tabular}
\end{center}
\caption{{\it Average values of $b$-hadron production rates
and their correlation coefficients obtained from direct measurements 
and including constraints from results on B mixing.} \label{tab:rates} }
\end{table}


%


\begin{figure}[htp]
  \begin{center}
    \leavevmode
    \mbox{\epsfxsize16.0cm \epsfysize22.0cm
\epsffile{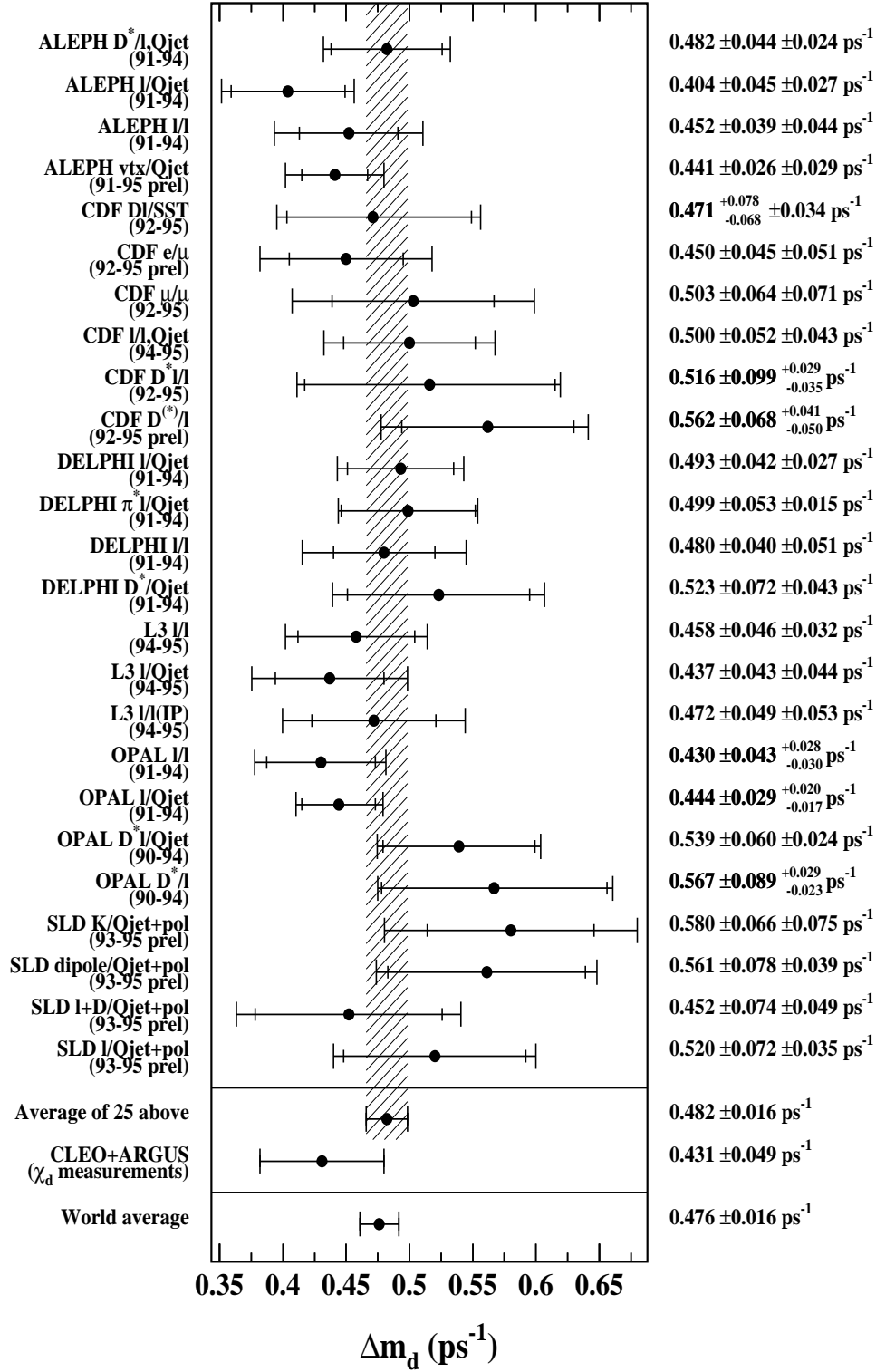}}
    \caption{{\it Individual and combined measurements of $\dmd$ at LEP, CDF
      and SLD. 
Note that the
      individual measurements are as quoted in the original
      publications, but their average includes the effects of adjustments
      to a common set of input parameters.
The world average also includes $\chid$ measurements
\cite{PDG98} performed
      at the \Yfs\ resonance. }\label{fig:dmdvalues}}

  \end{center}
\end{figure}

The production rate for $b \overline{s}$ states 
is expected to be 
rather similar to the probability ($\gamma_s \sim 12\%$)
for creating an $s  \overline{s}$ pair
during the hadronization process. But, because of their masses,
non strange $\Bstarstar$ mesons are not expected to decay into
${\rm B}^{0(*)}_s \overline{{\rm K}}$, whereas strange $\Bstarstar$ 
mesons are expected to decay mainly into non-strange weakly 
decaying $b$-mesons.
As a result, the fraction of $\Bs$ in $b$-jets is expected to be given by
$\fs \simeq \gamma_s~(1 -{\rm P}(b \rightarrow \Bstarstar)) \simeq 10\%$,
as observed.

\mysubsection{Combination method for $\dmd$}
\label{sec:method}

The measurements of \dmd\ are now quite precise and it is
important that the correlated systematic uncertainties are correctly
handled. Furthermore, many results depend on physics parameters for which
different values were used in the original analyses. 
Before being combined, the measurements
of \dmd\ are therefore adjusted on the basis of a common consistent set of
input values.

For each input parameter the \dmd\ measurement and the corresponding
systematic uncertainty are linearly rescaled in accordance with
the difference between the originally used parameter value and error, and
the new common values.
This is done for the systematic uncertainties
originating from the $b$-hadron lifetimes, the
$b$-hadron fractions and the mixing parameters
$\chib$ and $\chid$. The
statistical and systematic uncertainties of each individual
measurement are symmetrized. An alternative approach using
asymmetric uncertainties (when quoted as such) would
produce negligible differences in the combined result.

The combination procedure makes a common fit of \dmd\ and the common input
parameters.
It is assumed that \dmd\ may be expressed as a function
$X(Y_1 \dots Y_{N_{sys}})$ with a weak dependence on the systematic
sources $Y_\alpha$. Expanding $X$ then gives:
\begin{equation}
  X(Y_1 \dots Y_{N_{sys}}) \approx
  X^0 + \sum_{\alpha=1}^{N_{sys}} Y_\alpha \sigma_\alpha^{syst}
\end{equation}
where the quantities $\sigma_\alpha^{syst}$ are the 
correlated errors on \dmd\
from systematic sources $\alpha$, and $Y_\alpha = (z_\alpha -
z^{fit}_\alpha)/\delta_\alpha$. In this last expression,
 $z_\alpha$ and $z^{fit}_\alpha$
are the input and fitted values of the systematic parameter $\alpha$, and
$\delta_\alpha$ is the variation used to calculate
$\sigma_\alpha^{syst}$.

The following $\chi^2$ is then constructed:
\begin{equation}
  \chi^2 = \sum_{i=1}^{N}
  \left(\frac{X_i - 
      \sum_{\alpha=1}^{N_{sys}} Y_\alpha \sigma_{i \alpha}^{syst} - X^0}
    {\sigma_i^{uncor}} \right)^2
      +\sum_{\alpha=1}^{N_{sys}} Y_\alpha^2,
\label{eq:compl}
\end{equation}
where $X_i$ are the measurements of \dmd\ and 
$\sigma_i^{uncor}$ the quadratic sum of the statistical and uncorrelated 
systematic uncertainties on $X_i$. This $\chi^2$ is minimized 
with respect to the parameters $X^0$ and $Y_\alpha$; the result for 
$\dmd$ is taken as the value of $X^0$ at the minimum (and the 
values of $Y_\alpha$ at the minimum are ignored).
This method gives the
same results as a $\chi^2$ minimization with inversion of a global
correlation matrix. As several measurements also have a statistical
correlation, 
the first sum in Equation (\ref{eq:compl}) is then
    generalized to handle an $N \times N$ error matrix describing the
    statistical and uncorrelated systematic uncertainties.

The following sources of systematic uncertainties, common to analyses
from different experiments, have been considered:
\begin{itemize}
\item $b$-lifetime measurements (Section \ref{sec:Atau}).
Different measurements use the $b$-hadron lifetimes in different ways as
input: some use the actual lifetimes, and others use ratios of
lifetimes. As this leads to complicated correlations between the
analyses, the following procedure is adopted. For each measurement a
single combined value of all $b$-lifetime-related systematic
uncertainties is computed. This systematic is then treated as fully
correlated with all other such numbers from the other
measurements. Tests show that this procedure gives a 
negligible bias and results in a conservative evaluation
of the uncertainty.

\item $b$-quark fragmentation (Section \ref{sec:systgen}).
In some analyses the
$b$-hadron momentum dependence is treated through the variation of the
$\epsilon$ parameter in the Peterson fragmentation function, and in others
through the mean scaled $b$-hadron energy $<x_E>$. In such a case,
the corresponding systematic uncertainties are treated as fully correlated, 
without attempting to adjust the individual results to a common value
of $\epsilon$ or $<x_E>$.

\item fraction of cascade decays (Section \ref{sec:systgen}),
related mainly to the value
of BR$(b \rightarrow c \rightarrow \ell^+\nu_{\ell}X)$.

\item fraction of $\Bp$ mesons remaining in $\Bd$ enriched samples
for analyses using $\Dstar$ mesons. This depends on decay branching
fractions and on the mistag probability.

\item production rate of $\Dstarstar$ mesons in $b$-hadron semileptonic decays
(Section \ref{sec:systgen}).

\item $b$-hadron fractions (Section \ref{sec:results}).


\item $\Do$ lifetime (\cite{PDG98}).

\end{itemize}

Common systematics due to purely experiment-dependent 
factors (i.e.\ common to different results in a particular
experiment) have not been included in the above list, 
but nonetheless are treated
as correlated in the fit.

Following this procedure, a
combined value:
\begin{eqnarray}
   \begin{array}{lll}
\dmd^{{\rm LEP+CDF+SLD}}& = &(\dmdwerr)~{\rm ps}^{-1}
   \end{array}
\label{eq:dmda}
\end{eqnarray}
is obtained. 
Using the 
relation\footnote{Equation (\ref{eq:chid}) assumes that there is no decay width
difference in the $\Bd-\Bdb$ system.}:
\begin{equation}
\chid = \frac{1}{2}~\frac{(\dmd~\tau_{B_d})^2}{(\dmd ~\tau_{B_d})^2 +1}
\label{eq:chid}
\end{equation}
and the $\Bd$ lifetime of Section \ref{sec:Atau},
the above $\dmd$ result can be converted to:
\begin{eqnarray}
   \begin{array}{lll}
\chid^{{\rm LEP+CDF+SLD}}& = & \chicls. 
   \end{array}
\label{eq:dmdb}
\end{eqnarray}
Averaging with the time-integrated mixing
results obtained by ARGUS and CLEO at the $\Upsilon$(4S)
(see Table \ref{tab:gensys}), yields finally:
\begin{eqnarray}
   \begin{array}{lll}
\chid^{{\rm LEP+CDF+SLD}+\Yfs}& = &\chix, 
   \end{array}
\label{eq:dmdbb}
\end{eqnarray}
or equivalently
\begin{eqnarray}
   \begin{array}{lllll}
\dmd^{{\rm LEP+CDF+SLD}+\Yfs}& = &(\dmdx)~{\rm ps}^{-1}& =& (\dmdxev)~10^{-4}~{\rm eV}/c^2.
   \end{array}
\label{eq:dmdc}
\end{eqnarray}

The individual measurements
    of $\dmd$ and their combined value are shown in 
Figure~\ref{fig:dmdvalues}.

The determination of $\dmd$ as described above and 
of the $b$-hadron fractions described in
    Section \ref{sec:results} cannot be performed sequentially: the
    values of the fractions are needed to perform the
    $\dmd$ fit, and the best estimates of the fractions can
    be obtained only once the final $\dmd$ average is known.
    This circular dependence has been handled by including
    the calculation of the fractions in the $\dmd$ fitting
    procedure, in such a way that the final results quoted for
    $\dmd$ and the $b$-hadron fractions form a consistent set.

\mysubsection{Combination method for $\Bs$ oscillation amplitudes
         and derived limits on $\dms$}
\label{sec:dmsosc}

\begin{figure}[htb]
  \begin{center}
    \leavevmode
    \mbox{\epsfxsize12.0cm \epsfysize12.0cm
\epsffile{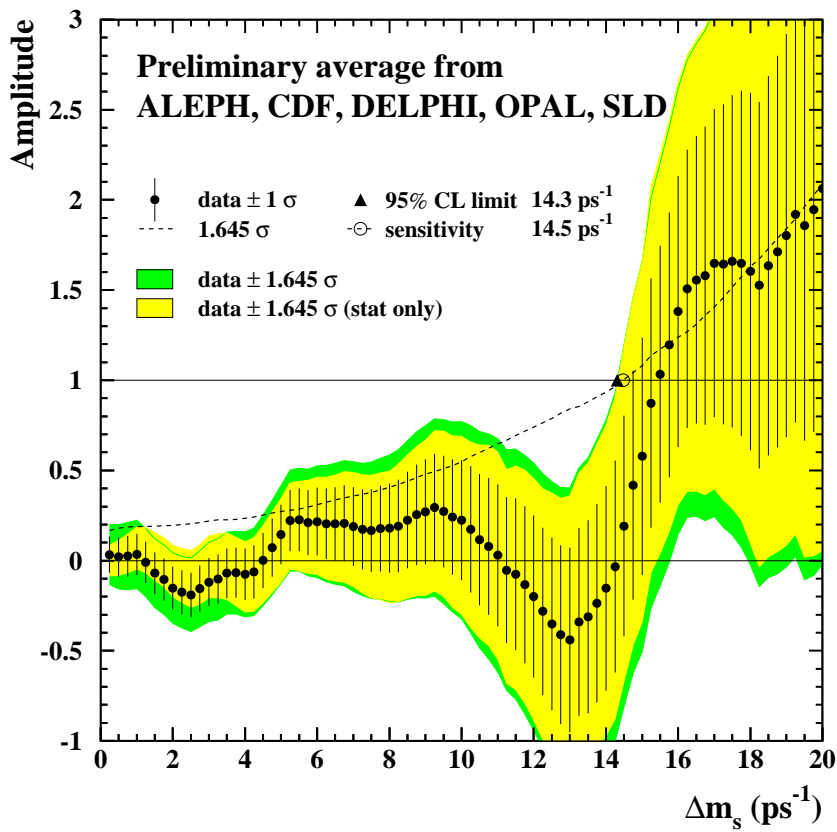}}
    \caption{{\it Combined \PsBz\ oscillation amplitude as a
      function of $\dms$.  A 95\% CL lower limit on $\dms$ of \dmslim\
      \ips\ is derived from this spectrum.
Points with error bars are the fitted amplitude values and 
corresponding uncertainties, including systematics. The dark shaded area
is obtained by multiplying these uncertainties
by 1.645 such that the integral of the probability distribution,
 assumed to be Gaussian, is equal to 5$\%$ above this range.
The clear shaded area is obtained in a similar way but including only 
statistical uncertainties in the fit; as a consequence, central values
of the two regions do not necessarily coincide.
}\label{fig:dmsampl}}

  \end{center}
\end{figure}

No experiment has yet directly observed \PsBz\ oscillations, so the task
of a combination procedure is to calculate an overall limit 
or to quantify the evidence for a signal from the
information provided by each measurement. This is done using the
amplitude method of reference~\cite{AmpMeth}. At each value of $\dms$, in
the range of interest, an amplitude is measured in each analysis,
where the expected value of the amplitude is unity at the
true frequency.
An overall limit on $\dms$ is then inferred from the combined 
amplitude spectrum by excluding regions of $\dms$ where the amplitude
is incompatible with unity.

Studies with Monte Carlo simulation show that the measured amplitude
has a Gaussian distribution around its expected value, which is zero for
frequencies much lower than the true one, and unity at the true frequency.
In addition, the error on the amplitude depends on the statistical
power of the analysis but not on the true value of the oscillation
frequency. Therefore, if at a given frequency ($\Delta m_s$) a measured
amplitude ($\mu$) with error $\sigma$ is found, this value of the frequency
can be excluded at the~95\% C.L. if the probability of measuring an amplitude
equal or smaller than $\mu$, for a true amplitude of unity, is smaller than 5\%,
that is $ \int_{-\infty}^{\mu} G(A-1, \sigma) dA  < 0.05$, or equivalently
$\mu + 1.645 \sigma <1$ where $G$ is the Gaussian function.

Before combining the measured amplitudes, the individual central
       values and systematic uncertainties are adjusted, using the same
       procedure as for $\dmd$ measurements, to common values of the $b$-hadron
       fractions (Table \ref{tab:rates}), 
$b$-hadron lifetimes (Section \ref{sec:Atau}), and $\dmd$
       (Equation (\ref{eq:dmdc})).
       The adjustment to a common value of $\fs$ is performed first,
       and needs a special treatment for certain analyses,
       as explained below. Although the final value of $\fs$ is
       calculated under the assumption that $\Bs$ mixing is maximal, studies
       indicate that the effect on the combined $\dms$ result is negligible.

The statistical uncertainty on the measured amplitude is
    expected to be inversely proportional to the $\Bs$ purity
    of the analysed sample \cite{AmpMeth}. This purity is of the order
    of $\fs$ for inclusive analyses. For analyses where a
    full or partial $\Bs$ reconstruction is performed, the purity
    is much less dependent on the assumed value of $\fs$.
    Therefore, the statistical uncertainties on the amplitudes
    measured in inclusive analyses have been multiplied by
$\fs^{\rm used}/\fs^{\rm new}$, where
    $\fs^{\rm used}$ is the $\Bs$ fraction assumed in the
    corresponding analysis, and $\fs^{\rm new}$ is the $\Bs$
    fraction from Table \ref{tab:rates}. This correction is
    also applied on the
    central values of the amplitude measured in inclusive
    analyses, after having checked that the relative uncertainty
    on the amplitude is essentially independent of $\fs$ for
    large enough values of $\dms$.

The amplitudes (measured at each
value of $\dms$) are averaged using the same procedure as for 
the combination of $\dmd$ results.
In addition to systematics which are correlated within the individual 
experiments, the following sources of correlated systematic uncertainties 
have been taken into account:
\begin{itemize}
\item $b$-hadron lifetime measurements,
\item $b$-quark fragmentation,
\item direct and cascade semileptonic branching fractions of $b$-hadrons,
\item $b$-hadron fractions,
\item $\dmd$ measurements,
\item $\Delta \Gs$ measurements.
\end{itemize}

\begin{figure}[t]
  \begin{center}
    \leavevmode
    \mbox{\epsfxsize12.0cm \epsfysize12.0cm
\epsffile{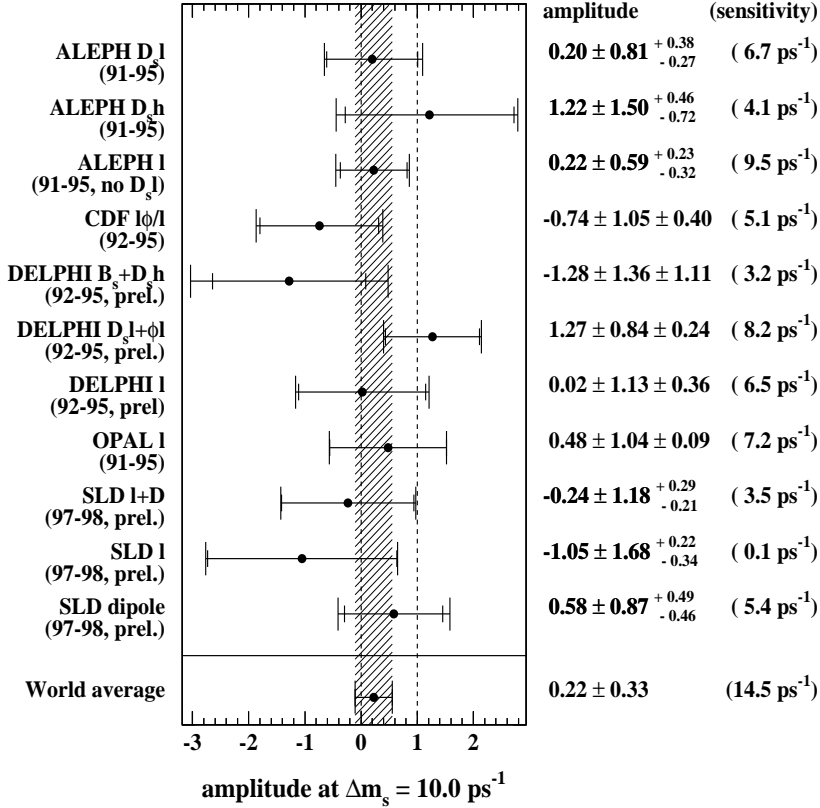}}
    \caption{{\it Measurements of the \PsBz\ oscillation
      amplitude. 
      The amplitudes are given at $\dms=10$ \ips, along with
      the relevant statistical and systematic errors. The
      sensitivities are given within parentheses in the column on the right.
      Note that the individual measurements are as quoted in the original
      publications, but the average, corresponding to the hatched area,
 includes the effects of adjustments
      to a common set of input parameters. The dashed lines correspond to 
amplitude values of 0 and 1.} \label{fig:dmsexpts}}

  \end{center}
\end{figure}

Data using the amplitude method have been provided by
ALEPH, DELPHI, OPAL (for their lepton-jet analysis 
only\footnote{The OPAL Collaboration has other results on searches 
for \PsBz\ oscillations 
which are not included in this combination 
(see reference~\cite{O:dms} for details).}),
CDF and SLD.
The
combined amplitude spectrum is shown in
Figure~\ref{fig:dmsampl}.  All values of \dms\ below
\dmslim\ \ips\ are excluded at the~95\% CL.
The combined expected sensitivity
to \dms, defined as the largest value of \dms\
that would have been excluded if the combined amplitudes were zero at
all values of \dms, is \dmssen\ \ips. 
In Figure~\ref{fig:dmsexpts} the values of the amplitude at 
$\dms =10$ \ips\,
from the various analyses, are shown. In addition, the sensitivities
are given.



Positive measured amplitudes are found for frequency values near and
above the sensitivity limit. Some values deviate from zero by more than
twice the total estimated error.
The likelihood profile obtained from the amplitude spectrum, using the 
prescription of \cite{AmpMeth}, shows a minimum at 16.3 \ips. The likelihood
value at this minimum is 1.9 units below the asymptotic likelihood
value for $\dms \rightarrow \infty$. Because the measurements at different frequencies
are correlated, it is not possible to calculate analytically the probability 
that,
in the explored frequency range, a fluctuation as or more unlikely occurs in 
a sample where the true frequency is far beyond the sensitivity. Therefore, 
an estimation procedure
based
on fast Monte Carlo experiments has been developed \cite{ref:dega}: the above
probability is found to be $(10 \pm 2)\%$, where the quoted uncertainty is due
to
the assumptions and approximations adopted in the fast Monte Carlo generation.

\mysection{Limit on the decay width difference for mass eigenstates
in the $\Bs$-$\Bsb$ system}
\label{sec:deltag} 

The CKM picture of weak charge-changing transitions predicts that 
the $\Bs$ and its 
charge conjugate mix. This 
results in new states $\Bsh$ and $\Bsl$, with masses $\Msh$ and 
$\Msl$
and (probably) different widths $\Gsh$ and $\Gsl$. 

Neglecting CP violation, the mass eigenstates are also CP eigenstates, the 
$\Bsl$ being 
CP even and $\Bsh$ being CP odd. The decay of a $\Bsb$ meson via the quark 
subprocess $b (\overline{s})\rightarrow c \overline{c} s (\overline{s})$ gives rise to predominantly 
CP-even eigenstates, thus the CP-even eigenstate should have the greater decay rate 
and hence the shorter lifetime.  For convenience of notation, in the following
we therefore substitute 
$\Gsl \equiv \Gss$ and $\Gsh \equiv \Gslg$, and 
define $\Gs=1/\tbs=(\Gslg+\Gss)/2$ and 
$\Delta \Gs = \Gss-\Gslg$.

Theoretical calculations \cite{Beneke2} of the ratio $\dgbs$ at 
next-to-leading order give:
\begin{equation}
\frac{\Delta \Gs}{\Gs} = \left({\frac{{\cal F}_{\Bs}}{210~\MeV}}\right)^2 [0.006 {\cal B}_b(m_b)+0.150 {\cal B}_s(m_b)-0.063]
\end{equation} 
where ${\cal F}_{\Bs}$ is the $\Bs$ decay constant and
where ${\cal B}_b(m_b)$ and 
${\cal B}_s(m_b)$ are bag parameters. 
Using the values from recent lattice calculations 
\cite{Hashimoto} of ${\cal B}_b(m_b)=0.80\pm0.15$,  
${\cal B}_s(m_b)=1.19\pm0.02\pm0.20$ 
and ${\cal F}_{\Bs}=(245\pm30)~\MeV$  yields 
$\dgbs =0.16\pm0.03\pm0.04$, where the uncertainties are from ${\cal F}_{\Bs}$
 and 
${\cal B}_s$ respectively. Care should be taken here as the values for 
the bag constants are preliminary 
and are correctly normalised at one loop only, 
so there may be some additional systematic 
uncertainty to be included. 

The width difference and the mass difference are correlated 
($\Delta \Gs/\dms =\frac{3}{2}\pi \frac{m_b^2}{m_t^2}$ to first approximation
\cite{ref:dgnaif}),
thus offering the 
possibility of measuring $\dms$ via the lifetime difference rather than the 
oscillation frequency. This could be particularly important if the oscillation 
frequency is too fast to be measured with the present experimental 
proper time resolution. In addition, if $\Delta \Gs$ does turn out to be 
sizable, the 
observation of CP violation and the measurement of CKM phases from untagged 
$\Bs$ samples 
can be imagined \cite{Dunietz}. 

The existing experimental constraints on the width difference and the 
combination of these constraints 
are reported here\footnote{The present members of the $\Delta \Gs$ working group are:
P. Coyle, D. Lucchesi, S. Mele, F. Parodi and P. Spagnolo.}.


\mysubsection{Experimental constraints on $\Delta \Gs/\Gs$}
Experimental information on $\Delta \Gs$ can be extracted by studying the 
proper time distribution of data samples enriched in $\Bs$ mesons. 
An alternative method based on  
measuring the branching fraction ${\rm B}_s \rightarrow{\rm D} _s^{(*)+}{\rm D} _s^{(*)-}$ has also
been proposed  recently \cite{ALEPH-phiphi}. 
The available results are summarised in 
Table \ref{tab:dgammat}. 
The values of the limit on $\dgs/\Gs$ quoted in the last column of this 
table have been obtained by the working group.

Methods based on double exponential lifetime fits to samples containing a 
mixture of CP eigenstates have a quadratic sensitivity to $\Delta \Gs$ 
(inclusive, semileptonic, $\Ds$-hadron), whereas methods based on isolating
a single CP eigenstate have a linear dependence on $\Delta \Gs$ ($\phi\phi$,
${\rm J}/\psi\phi$). The latter are therefore, in principle, more sensitive to
$\Delta \Gs$; but they tend to suffer from reduced statistics.

In order to obtain an improved limit on $\Delta \Gs$,
the results based on fits to the proper time distributions 
are used to apply a constraint on the allowed 
range of $1/\Gs$. The world average $\Bs$ lifetime is not used, as its meaning 
is not clear if
$\Delta \Gs$ is non-zero.
Instead, it is chosen to constrain $1/\Gs$
 to the world average $\tbd$ lifetime 
($1/\Gs \equiv 1/\Gd=\tbd= (\taubd)$~ps).
This is well motivated theoretically, as 
the total widths of the $\Bs$ and $\Bd$ mesons
are expected to be 
equal within less than one percent~\cite{ref:bigilife,Beneke}
and $\Delta \Gd$ is expected to be negligible. 
  
Further information on the various individual measurements is now given.

\begin{table}
\begin{center}

\begin{tabular}{|l|c|c|c|} 

\hline
Experiment & Selection        & Measurement            & $\Delta \Gs/\Gs$ \\ 
\hline

L3~\cite{L3B01}         & inclusive $b$-sample              &                               & $<0.67$         \\

DELPHI~\cite{DELBS0}     & $\Bsb\rightarrow \Dsp  \ell^- \overline{\nu_{\ell}} X$ & $\tbssemi=(1.42^{+0.14}_{-0.13}\pm0.03)$~ps  & $<0.46$ \\

OTHERS~\cite{ref:others}& $\Bsb \rightarrow \Dsp \ell^-  \overline{\nu_{\ell}} X$  & $\tbssemi=(1.46\pm{0.07})$~ps & $<0.30$ \\

ALEPH~\cite{ALEPH-phiphi}      & $\Bs \rightarrow\phi\phi X$      & 
${\rm BR}(\Bssh \rightarrow {\rm D}_s^{(*)+} {\rm D}_s^{(*)-}) =(23\pm10^{+19}_{-~9})\%$       & $0.26^{+0.30}_{-0.15}$ \\ 

ALEPH~\cite{ALEPH-phiphi}      & $\Bs \rightarrow\phi\phi X$      & $\tbsshort=(1.27\pm0.33\pm0.07)$~ps           & 
$0.45^{+0.80}_{-0.49}$ \\


DELPHI~\cite{dnewdsh}     & $\Bsb \rightarrow \Dsp$ hadron     &  $\tbsdh=(1.53^{+0.16}_{-0.15}\pm0.07)$~ps                          & $<0.69$         \\

CDF~\cite{CDFB01}        & $\Bs \rightarrow {\rm J}/\psi\phi$        & $\tbspsi=(1.34^{+0.23}_{-0.19}\pm0.05)$~ps & $0.33^{+0.45}_{-0.42}$ \\ 

\hline
\end{tabular}
\caption{{\it Experimental constraints on $\Delta \Gs/ \Gs$. The upper limits,
which have been obtained by the working group, are quoted at the~95~\%C.L.}
\label{tab:dgammat}}
\end{center}

\end{table}

\begin{itemize}

\item {\it L3 inclusive $b$-sample:}
in an unbiased inclusive B sample, all $\Bs $ decay modes are measured, 
including decays into CP eigenstates. An equal number of $\Bssh$ and $\Bslg$ 
mesons 
are therefore selected, and the proper time dependence of the $\Bs$ signal is 
given by:
\begin{equation} 
P_{incl}(t) = \frac{1}{2}(\Gslg \exp{(-\Gslg t)}+\Gss \exp{(-\Gss t)}).
\end{equation}
If the proper time dependence of this sample
is fitted assuming only a single exponential lifetime
then, using the definitions of $\Gs$ and $\Delta \Gs$
and assuming that $\Delta \Gs/\Gs$ is small,
the measured lifetime is
 given by:
\begin{equation} 
\tbsinc = \frac{1}{\Gs} \frac{1}{1- \left (\frac{\Delta \Gs}{2\Gs}\right )^2}.
\end{equation}

L3 effectively incorporates $P_{incl}(t)$ into the proper time fit of an inclusive $b$-sample and 
applies the constraint $1/\Gs=(1.49\pm0.06)$~ps. 
 
\item{\it DELPHI $\Bsb \rightarrow {\rm D}^+_s \ell^- \overline{\nu_{\ell}} X$:}
in a semileptonic selection, the ratio of short and long $\Bs $'s 
selected is proportional to the ratio of the decay widths and the proper time dependence of the 
signal is:
\begin{equation}  
 P_{semi}(t) =   \frac{\Gss~ \Gslg}{(\Gss+\Gslg)} 
(\exp{(-\Gss t)}+\exp{(-\Gslg t)}).
\label{eq:semi}
\end{equation}
If this sample is fitted assuming only a single exponential lifetime for the $\Bs $, then the 
measured lifetime is, always in the limit that $\Delta \Gs/\Gs$ is small, 
given by: 
\begin{equation}
\tbssemi = \frac{1}{\Gs} \frac{{1 + \left (\frac{\Delta \Gs}{ 2\Gs}\right )^2}}
{{1- \left (\frac{\Delta \Gs}{2\Gs}\right )^2}}.                  
\label{eqsemi}
\end{equation}
The single lifetime fit is thus more sensitive to the effects of 
$\Delta \Gs$ in the semileptonic than in the fully inclusive case.
Information on 
$\Delta\Gs$ is obtained by scanning the likelihood as a function the two parameters
$1/\Gs$ and $\Delta \Gs/\Gs$ and applying the $\tbd$ constraint.

\item {\it OTHERS:}
other analyses of the $\Bs $ semileptonic lifetime \cite{ref:others}
have not explicitly 
considered the possibility 
of a non-zero $\Delta \Gs$ value.
Nevertheless, the fact that the single exponential lifetime 
for this case 
(Equation~(\ref{eqsemi})) is sensitive to $\Delta \Gs$ allows information on 
$\Delta \Gs$ to be extracted. 
The average $\Bs $ semileptonic lifetime has been recalculated
excluding the DELPHI result just discussed, and information on 
$\Delta\Gs$ has been obtained by using Equation~(\ref{eqsemi}) and applying
the $\tbd$ constraint.
The validity of Equation~(\ref{eqsemi}) in the presence of background contributions has 
been verified using a toy Monte Carlo~\cite{Moser}.

\item {\it ALEPH $\Bs  \rightarrow \phi \phi X$ (counting method):}
only those $\Bs $ decays which are CP eigenstates can contribute to a width difference 
between the CP even and CP odd states. An analysis~\cite{Aleksan} of such decays 
shows that $\Bs  \rightarrow D_s^{(*)+} D_s^{(*)-}$ is by far the dominant
contribution and is almost 100\% CP even. Under this assumption,  
$\Delta \Gs = \Gss (\Bssh \rightarrow  {\rm D}_s^{(*)+} {\rm D}_s^{(*)-})$
where:
\begin{equation}
{\rm BR}(\Bssh \rightarrow {\rm D}_s^{(*)+} {\rm D}_s^{(*)-})= 
\frac{\Gss (\Bssh \rightarrow  {\rm D}_s^{(*)+} {\rm D}_s^{(*)-})} {\Gsh} = 
\frac{\Delta \Gs}{\Gs (1+\frac{\Delta \Gs}{2 \Gs}) }. 
\end{equation}
ALEPH~\cite{ALEPH-phiphi} has measured this branching fraction, and
this is the only constraint on $\Delta \Gs/\Gs$ which does not rely on 
a measurement of the average $\Bs$($\Bd$) lifetime.

\item {\it ALEPH $\Bs  \rightarrow \phi \phi X$ (lifetime method):}
as the decay $\Bs  \rightarrow D_s^{(*)+} D_s^{(*)-} \rightarrow \phi \phi X$, 
is  predominantly CP even,
the proper time dependence of the $\Bs $ component of the 
$\phi \phi$ sample is 
therefore just a simple exponential with the appropriate lifetime: 
\begin{equation} 
P_{single}(t)=\Gss \exp(-\Gss t).
\label{eq:single}
\end{equation}
This lifetime is related to $\dgs$ and $\Gs$ via the expression:
\begin{equation}  
\frac{\Delta \Gs}{\Gs}=2(\frac{1}{\Gs~\tbsshort}-1).  
\end{equation} 
ALEPH~\cite{ALEPH-phiphi} has measured 
the lifetime of $\phi\phi X$ events and extracted information on 
$\Delta \Gamma_s/\Gamma_s$ with the help of the world average
$\Bs$ lifetime obtained from semileptonic $\Bs$ decays.
The result listed in Table \ref{tab:dgammat} has been obtained 
by the working group with the assumption
$1/\Gs \equiv \tbd$.

\item {\it DELPHI inclusive ${\rm D}_s^+$:}
a fully inclusive $\Dsp$ selection is expected to have 
an increased CP-even content, as the $\Bs  \rightarrow {\rm D}_s^{(*)+} {\rm D}_s^{(*)-}$ 
contribution is enhanced by the selection criteria. 
If $f_{DsDs}$ is the fraction of ${\rm D}_s^{(*)+} {\rm D}_s^{(*)-}$ in the sample, 
then the proper time dependence is expected to be: 
\begin{equation} 
P(t)_{D_s-had.}= f_{DsDs}P_{single}(t)+ (1-f_{DsDs}) P_{semi}(t) 
\end{equation}
in which $ P_{semi}(t)$ and $P_{single}(t)$ are defined in Equations
(\ref{eq:semi}) and (\ref{eq:single}) respectively.
For the DELPHI analysis
a value $f_{DsDs}=(22\pm7)\%$ is estimated from simulation.  
Scanning the likelihood as a function $1/\Gs$ and $\Delta \Gs/\Gs$ and 
applying the $\tbd$ constraint yields an upper limit  
on $\Delta \Gs/\Gs$.

\item{\it CDF $\Bs  \rightarrow {\rm J}/\psi \phi$: } 
the final state $\Bs  \rightarrow {\rm J}/\psi \phi$ is thought to be predominantly CP even 
(i.e. measures mainly $\tbsshort$) \cite{Aleksan}.
An update \cite{CDF-angles} of the CDF measurement of the polarization 
in $\Bs  \rightarrow {\rm J}/\psi \phi$ decays measures the fraction of CP even in the final state 
to be $f_{short}=(79\pm19)\%$ and supports this expectation. 
For this case, the proper time dependence of the $\Bs $ component of the sample is: 
\begin{equation} 
P_{{\rm J}/\psi \phi}(t)=f_{short}P_{single}(\Gss, t)+(1-f_{short})P_{single}(\Gslg,t)
\end{equation}
where $P_{single}$ has been defined in Equation (\ref{eq:single}).
CDF measures the lifetime of ${\rm J}/\psi \phi$ events and information on
$\dgbs$ is obtained after applying the $1/\Gs \equiv \tbd$ constraint
and including the experimental uncertainty on $f_{short}$.

\end{itemize}

\mysubsection {Combined limit on $\dgs$}

\begin{figure}
\begin{center}
\resizebox{0.95\textwidth}{!}{\includegraphics{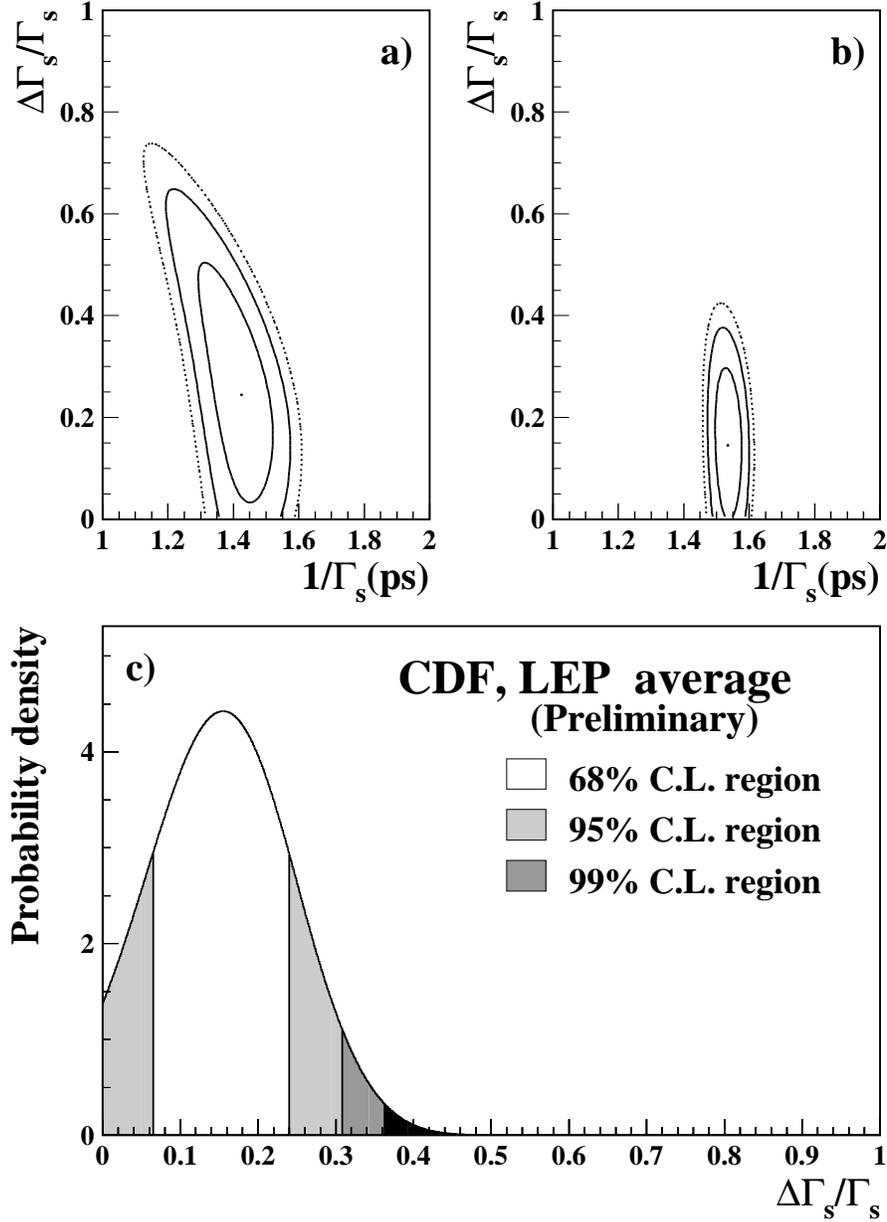}}
\end{center}
\caption[]{{ \it
          a) 68\%, 95\% and 99\% C.L. contours of the negative log-likelihood 
             distribution in the plane $1/\Gs$-$\dgbs$.
          b) Same as a) but with the constraint $1/\Gs \equiv\tau_{\Bd}$
supposed to be exact.
          c) Probability density distribution for $\dgbs$ after applying the constraint; 
             the three shaded regions show the limits at the 68\%, 95\% and 
99\% C.L. respectively.} \label{fig:dgplot}}

\end{figure}

In order to combine the analyses summarised in Table \ref{tab:dgammat},
the result of each analysis has been converted 
to a two-dimensional log-likelihood in the ($1/\Gs$,~$\dgbs$) plane. This log-likelihood has 
either been provided by each experiment or reconstructed from the measured lifetimes using 
the expected dependence of this quantity on $1/\Gs$ and $\dgbs$.
The latter procedure was necessary for the OTHERS and CDF entries
of Table \ref{tab:dgammat}.
The L3 analysis is not included in the
average as the two-dimensional likelihood was not provided and 
could not be reconstructed from
the available information. The $\tbd$ constraint is not applied
on $1/\Gs$ at this stage.
Systematic uncertainties are 
included in the individual log-likelihood distributions. 

The log-likelihood distributions  have been summed and 
the variation of the global negative log-likelihood function has been 
measured with respect to its minimum ($\Delta {\cal L}$). 
The 68\%, 95\% and 99\% C.L. contours of the combined negative 
log-likelihood are shown in Figure~\ref{fig:dgplot}a.
The corresponding limit on $\dgbs$ is:
\begin{eqnarray}
\nonumber   \dgbs & = & 0.24^{+0.16}_{-0.12}  \\
   \dgbs & < & 0.53~{\rm at~the~} 95\%~{\rm C.L.} 
\end{eqnarray}

An improved limit on $\dgbs$
can be obtained by applying the $\tbd = (\taubd)~{\rm ps}$ constraint. 
When expressed as 
a probability density, this constraint is:
\begin{eqnarray}
\begin{array}{lll}
  f_C(1/\Gs) & = & 
   \frac{1}{\sqrt{2\pi} \sigma_{\tbd}} \exp \left(-\frac{(1/\Gs-\tbd)^2} {2\sigma^2_{\tbd}} \right). 
\end{array}
\end{eqnarray}
The value of the uncertainty $\sigma_{\tbd}$ includes, eventually,
the theoretical uncertainty on the equality $1/\Gs=\tbd$,
expected to be of the order of 1$\%$ or less.
Using Bayes theorem, the probability density for $\dgbs$, 
with the constraint applied
\footnote{A flat, a priori probability density distribution has been assumed
for $\dgbs$.}, is obtained 
by convoluting $f_C(1/\Gs)$ with the 2-D probability density for $1/\Gs$ and $\dgbs$ 
(${\cal P}(\tbs,\dgbs)$), and normalizing the result to unity:
\begin{eqnarray}
   {\cal P}(\dgbs) = { {\int {\cal P}(1/\Gs,\dgbs) f_C(1/\Gs) d(1/\Gs) } \over 
    {\int {\cal P}(1/\Gs,\dgbs) f_C(1/\Gs) d(1/\Gs) d\dgbs } }  
\end{eqnarray}
where ${\cal P}(1/\Gs,\dgbs)$ is proportional to $\exp(-\Delta {\cal L})$. 

The two-dimensional log-likelihood obtained, after including the constraint,
supposed to be exact, is shown in 
Figure~\ref{fig:dgplot}b. The resulting probability density distribution for $\dgbs$ is 
shown in Figure~\ref{fig:dgplot}c. The corresponding limit on $\dgbs$ is:
\begin{eqnarray}
\nonumber   \dgbs & = & 0.16^{+0.08}_{-0.09}  \\
            \dgbs & < & 0.31~{\rm at~the}~ 95\%~{\rm C.L.} 
\end{eqnarray}

If an additional 2$\%$ uncertainty, assumed to be Gaussian distributed, 
is incorporated to account for the theory assumption $\tbd =1/\Gs$,
the effect on the result is small:
\begin{eqnarray}
\nonumber   \dgbs & = & 0.17^{+0.09}_{-0.10}  \\
  \dgbs & < & 0.32~{\rm at~ the}~95\%~{\rm C.L.} 
\end{eqnarray}

 

\mysection{Average of LEP $\Vcb$ measurements}
\label{sec:vcb}

Within the framework of the Standard Model of electroweak interactions, the
elements of the Cabibbo-Kobayashi-Maskawa mixing matrix
are free parameters, constrained only by the requirement that the matrix
be unitary. 
The Operator Product Expansion (OPE) and Heavy Quark Effective
Theory (HQET) provide means to determine \vcb\
with relatively small theoretical uncertainties, 
by studying the decay rates of
inclusive and exclusive semileptonic $b$-decays respectively.
Relevant branching fractions have to be determined experimentally.
Inputs from theory are needed to obtain the values of the matrix 
elements.

There are two methods to measure \vcb: the inclusive
method, which 
uses the semileptonic decay width of $b$-decays and the OPE; and the  exclusive
method, where \vcb\ is extracted by studying the exclusive 
\btods\ decay process using HQET.   
The  ${\rm B} \rightarrow {\rm D} \ell^- \overline{\nu_{\ell}}$ channel
has not been averaged to date.


In this note, both methods are used to determine values for $\Vcb$, which 
are then combined to produce a single 
average\footnote{The present members of the $\Vcb$ working group are: 
D. Abbaneo, E. Barberio,
S. Blyth, M. Calvi, P. Gagnon, R. Hawkings, M. Margoni, S. Mele, F. Muheim,
D. Rousseau and F. Simonetto.}.
 The semileptonic $b$-decay width,
determined by the LEP heavy flavour electroweak fit to ALEPH, DELPHI, L3
and OPAL data, is used to 
determine ${\rm BR}(b \rightarrow \ell^- \overline{\nu_{\ell}} {\rm X})$.
Results from ALEPH  \cite{ALEPH_vcb}\footnote{
Updated to use the parametrisation of \cite{CLN}, the Ligeti 
model \cite{ref:ligeti,ref:grinstein} for \btodss\ decays, and various updated inputs.},  
DELPHI \cite{DELPHI_vcb}\footnote{Updated to use 
the Ligeti model  \cite{ref:ligeti} for \btodss~ decays.} and 
OPAL \cite{OPAL_vcb} are used to perform a LEP \vcb\ average in the
\btods\ decay channel. These measurements are combined using
a method similar to that used by the B oscillations working group.

Theoretical input parameters needed to 
extract $\Vcb$ from actual measurements
are detailed in Appendix \ref{appendixC}. In both analyses, the validity
of the quark/hadron duality hypothesis has been assumed.

\mysubsection{Inclusive $\Vcb$ determination}
\label{sec:vcbinc}

In the inclusive method, the partial width for semileptonic 
B meson decays to charmed mesons is related to \vcb\ 
using the following expression (Appendix \ref{appendixC}):
\begin{eqnarray}
 \Vcb &=& 0.0411~  \sqrt{\frac{1.55}{0.105} \Gamma(b \rightarrow \ell^-  \overline{\nu_{\ell}}{\rm X}_c )} 
\left(1-0.024 \left(\frac{\mu_\pi^2-0.5}{0.2}\right)\right)\times  \nonumber \\ 
 & & (1 \pm 0.030(pert.) \pm 0.020(m_b) \pm 0.024(1/m_b^3)).\label{eq:incl}
\end{eqnarray}
where $\rm{X}_c$ represents all final states containing a charmed quark.
The rest of the expression, within parentheses, represents the
correction to the muon decay formalism, depending on the $b$- and $c$-quark
masses and on the strong coupling constant; $\mu_\pi^2$ is the average of the 
square of the $b$-quark momentum inside the $b$-hadron.
The three last contributions to the uncertainty correspond respectively,
to the uncertainties coming from the QCD perturbative expansion
(scale dependence, truncation at finite order, ..),
from the $b$-quark mass determination and from neglected terms of order 
$m_{b,c}^{-n}$, with $n \geq 3$, appearing in the O.P.E. formalism.

Experimentally, the semileptonic width of $b$-hadrons is determined from 
its semileptonic branching fraction and lifetime.
In $\Zz$ decays, a mixture of 
$\Bdb,~ \Bm, ~\Bsb$ and $b$-baryons is produced, such that the 
inclusive semileptonic branching fraction measured at LEP is an average over
the different hadrons produced: 
\begin{eqnarray}
{\rm BR}( b \rightarrow  \ell^- \overline{\nu_{\ell}}{\rm X}_c  )& =& 
       \fd \frac{\Gamma(\Bdb \rightarrow  \ell^- \overline{\nu_{\ell}}{\rm X}_c  )}{\Gamma(\Bdb)} +
       \fu \frac{\Gamma(\Bm \rightarrow \ell^- \overline{\nu_{\ell}} {\rm X}_c )}{\Gamma(\Bm)}  \nonumber \\
    & &  + \fs \frac{\Gamma(\Bsb \rightarrow  \ell^- \overline{\nu_{\ell}}{\rm X}_c  )}{\Gamma(\Bsb)} +
\fb \frac{\Gamma( b-{\rm baryon} \rightarrow  \ell^- \overline{\nu_{\ell}}{\rm X}_c  )}{\Gamma( b-{\rm baryon})}  
  \nonumber \\
   &  \simeq & \Gamma(b \rightarrow \ell^-  \overline{\nu_{\ell}}{\rm X}_c  ) ( \fd \tau_{B^0} + \fu \tau_{B^-} +
   \fs \tau_{B_s} + \fb\tau_{ b-{\rm baryon}})  \nonumber \\ 
  &  =& \Gamma(b \rightarrow  \ell^- \overline{\nu_{\ell}}{\rm X}_c ) \tau_b \label{eq:bri}
\end{eqnarray}
where 
$\tau_b $ is the average 
$b$-hadron lifetime.
Therefore the semileptonic width of $b$-hadrons can be obtained 
using the inclusive 
semileptonic branching fraction and the average  $b$-hadron lifetime.
 The two last equalities in Equation (\ref{eq:bri})
assume that all $b$-hadrons have the same semileptonic width.
This hypothesis may be incorrect for $b$-baryons.
Taking into account the present precision of LEP measurements of $b$-baryon 
semileptonic branching fractions and lifetimes, an estimate of
the correction 
to Equation (\ref{eq:bri}) is about 1.5\% (see Section \ref{sec:systgen}). 

The average LEP value for 
${\rm BR} (b \rightarrow \ell^- \overline{\nu_{\ell}}{\rm X}) = (10.58 \pm 0.07{\rm (stat.)} 
\pm 0.17{\rm (syst.)})\% $ is taken from the global fit
which combines the heavy flavour measurements performed 
at the $\Zz$ (see Section \ref{sec:systgen}).
The ${\rm BR}( b \rightarrow \ell^- \overline{\nu_{\ell}}{\rm X}_u )$ 
contribution is subtracted from  
${\rm BR} (b \rightarrow \ell^- \overline{\nu_{\ell}}{\rm X} )$,
 using the LEP average value 
given in Section \ref{sec:vub}.
For the average $b$-hadron lifetime, the world average value of
$\tau_b$
is used, as obtained in Equation (\ref{eq:taub}).

\mysubsubsection{Sources of systematic errors}
\label{subs:vcb}
The systematic errors assumed at present in the determination of 
the inclusive semileptonic decay width
can be grouped into the following categories:
\begin{itemize}
\item {\it errors related to the efficiency and purity of the $b$-tagging algorithm.}

 As a lifetime $b$-tag is involved, effects due to the uncertainties
in the sample composition in terms of different heavy hadrons, 
and uncertainties in the  heavy hadron lifetimes, are considered. 

\item 
{\it input parameters influencing the signal and background normalisation.}

The values of the production fractions: $\fd$, $ \fu$, $\fs$, 
$ \fb$ were taken from Section \ref{sec:results}.
The branching fractions $\bcbl$, $\btaul$, $\bpsill$, and the rates of gluon 
splitting P($\glcc$) and P($\glbb$) have been fixed to the values of \cite{HFLEPEW}. 
$\Rb$,  $\rm R_c$, $\bcl$ and $\cl$ are parameters of the LEPEWWG fit. Values 
for all these quantities are 
listed in Table~\ref{tab:gensys}.
\item  
{\it the average fraction of the beam energy carried by the weakly decaying 
$b$-hadron.} 

Different models have been considered for the shape of the 
fragmentation function and the free parameters of the models have
been determined 
from the data. 
\item 
{\it $\Lambda_b$ polarization.}

These effects on the lepton spectra have been 
included.
\item 
{\it semileptonic decay models.}

The average LEP value for 
${\rm BR} (b \rightarrow \ell^- \overline{\nu_{\ell}}{\rm X}) $ 
is taken from the global LEPEWWG fit
which combines the heavy flavour measurements performed 
 at the $\Zz$ \cite{HFLEPEW},
but removing the
forward-backward asymmetry measurements (see Section \ref{sec:systgen}). 

\item 
{\it detector specific items.}

These include: lepton efficiencies, misidentification 
probabilities, detector resolution effects, jet reconstruction,
etc.
\end{itemize}
The errors listed in the last item are uncorrelated among the different 
experiments. The others have been split into their uncorrelated and  
correlated parts.
The error on the average $b$-hadron lifetime is assumed to be uncorrelated
with the error on the semileptonic branching fraction.
The propagation of these errors to the error on \vcb\ is done assuming that 
they are Gaussian in the branching fraction and the lifetime, respectively.

\mysubsubsection{Inclusive $\Vcb$ average}

Using the  expression given in Equation (\ref{eq:incl}), the following value is 
obtained:
\begin{equation}
\Vcb^{incl.} = \vcbinc
\end{equation}
where the first error is experimental and the second is from theory.
The experimental contributions due to the semileptonic
branching fraction and the lifetime are $\pm 0.37\times 10^{-3}$ and 
$\pm 0.18\times 10^{-3}$, respectively.
The dominant systematic uncertainty, of theoretical origin, comes from the
determination of the kinetic energy of the $b$-quark inside the $b$-hadron
as explained in Appendix \ref{appendixC}.

\mysubsection{Exclusive $\Vcb$ determination}
\label{sec:vcbexclu}

In the exclusive method, the value of \vcb\ is extracted by studying
the decay rate for the process \btods\ as a function of the recoil kinematics
of the \dsp\ meson. The decay rate is parameterized as a function
of the variable $w$, defined as the product of the four-velocities of the 
\dsp\ and the $\Bdb$ mesons.
This variable is related to the square of
the four-momentum transfer from the $\Bdb$ to the $\ell^-{\overline \nu}_\ell$
system, $q^2$, by:
\begin{equation}
w = \frac{m_{\rm D^{*+}}^2+m_{\rm B^0_d}^2-q^2}{2m_{\rm B^0_d}
m_{\rm D^{*+}}},
\end{equation}
and its values range from $1.0$, when the \dsp\ is produced at rest in the \bbar\ rest 
frame, to about $1.50$.  Using HQET, the differential partial width for this
decay is given by: 
\begin{eqnarray}
\label{eq:decayw}
{\frac{{\rm d} \Gamma}{{\rm d} w}}&=
&{\cal K}(w){\cal F}_{D^*}^2(w) \Vcb^2
\end{eqnarray}
where ${\cal K}(w)$ is a known phase space term  
and ${\cal F}_{D^*}(w)$ is the hadronic form factor for
the decay.  Although the shape of this
form factor is not known, its magnitude
at zero recoil, $w=1$, can be estimated using HQET.  
In the heavy quark limit ($m_{\rm b}\rightarrow \infty$), \fw\ 
coincides with the Isgur-Wise function \cite{neub,neuc} which is
normalised to unity at the point of zero recoil.
Corrections to \fone\ have been calculated to take into account
the effects of finite quark masses and QCD corrections \cite{luke}.  
Calculations of this correction yield \fone\ $=0.88 \pm 0.05$ 
(Appendix \ref{appendixC}).
  Since the phase space factor ${\cal K}(w)$ tends to zero
as $w\rightarrow 1$, the decay rate vanishes at
$w=1$ and the 
 accuracy of the extrapolation relies on achieving a reasonably
constant reconstruction efficiency in the region close to $w=1$.  
The  unknown function ${\cal F}_{D^{*}} (w)$ is approximated with an expansion
around $w=1$ due to Caprini, Lellouch and Neubert (CLN) \cite{CLN}:
\begin{equation}
{\cal F}_{D^{*}}(w)= {\cal F}_{D^{*}}(1) \times 
\left[ 1 - 8 \rho^2 z + (53 \rho^2 - 15)z^2
- (231 \rho^2 -91) z^3 \right],
\label{eq:clnextr}
\end{equation}
where $\rho^2$ is the slope parameter at zero recoil and 
$\large z= \frac{\sqrt{w+1} - \sqrt{2}}{\sqrt{w+1} + \sqrt{2}}$.
The ratio between the axial and vector form factors is included in
${\cal K}(w)$.
Theoretical predictions restrict 
values of $\rho^2$ to be in the range: $-0.14<\rho^2<1.54$.
An alternative parametrization, obtained earlier, can be found in
\cite{ref:grinstein}.

\mysubsubsection{Sources of systematic uncertainties}

The systematic 
uncertainties in the determination of 
\vcb\ using the semileptonic decay \btods\ can be
grouped into the following categories:
\begin{itemize}
\item {\it normalisation}: $\Bdb$ meson production rate, 
${\rm D}^{(*)}$ branching fraction to 
the tagged final states (including topological BR), 
$\Bdb$ lifetime (this is needed to obtain the 
$\Bdb \rightarrow \Dstarp   \ell^-\overline{\nu_{\ell}}$
decay partial 
width), and the $b \rightarrow {\rm B}$ fragmentation function
(which influences the reconstruction efficiency).


\item {\it background from physical processes:} comprising \Btau, \Bxc\ 
(followed by the 
semileptonic decay $\tau/\mathrm{X}_{\overline{c}} \rightarrow \ell^-
\overline{\nu_{\ell}} {\rm X}$) 
and, particularly, the intermediate production of excited charm mesons
$\Dstarstar$
which subsequently decay to a $\Dstarp$;

\item {\it detector specific items}: selection efficiency 
(lepton identification, tracking, vertexing),
non physics background 
(combinatorial, hadron mis-identification), resolution, 
fitting, etc. 
This last set is treated as uncorrelated among experiments, and 
therefore will be ignored in the following discussion.
\end{itemize}

\mysubsubsection{Normalisation}
The  $\Bdb$  production rate at LEP is given by the product:
\begin{eqnarray}
\frac{\Gamma(\Zz \rightarrow b \overline{b})}{\Gamma(\Zz \rightarrow \mathrm{hadrons})} 
~ {\rm BR}(b \rightarrow \Bdb) ~=~ \Rb ~ \fd 
\end{eqnarray}
where $\Rb$\ is taken from
\cite{HFLEPEW} (Table \ref{tab:gensys}) and 
$\fd$\ 
from Section \ref{sec:results}. 

The values for the 
 charm meson decay branching fractions are taken from the PDG \cite{PDG98}. 
Correlations among some of them (e.g. 
$\Do \rightarrow {\rm K}^-\pi^+$ with $\Do \rightarrow {\rm K} n\pi$, etc.)
are included. Analyses based on the inclusive 
reconstruction
of pions from $\Dstarp$ cascade decays
 may be affected by the knowledge of the topological 
$\Do$\ branching fractions
\footnote{A topological branching fraction corresponds to the branching 
fraction into a fixed number of charged particles emitted in the final 
state.}; they are taken from MARKIII measurements
 \cite{ref:mark3}.

The $\Bdb$ lifetime determined in
Section \ref{sec:taubd} is used.

Knowledge of the $\Bdb$\ fragmentation function is necessary  in order to 
compute the fraction of $\Bdb$\ which were not reconstructed
because they did not have enough energy to be detected.
$\Bdb$\ 
hadrons produced in $e^+ e^-$ annihilations carry on average 
a large fraction, $<x_E>$, 
 of the beam energy (Table \ref{tab:gensys}); 
consequently only a small fraction
 of them are outside the selection acceptance. 

\mysubsubsection{Physics background}
Charged ${\rm D}^{*}$ mesons can be accompanied by a lepton of opposite
charge in two other $b$-hadron decay processes:
\begin{itemize}
\item
$\overline{{\rm B}} \rightarrow \Dstarp \tau^- \overline{\nu_{\tau}}$ with 
$\tau \rightarrow \ell^-\overline{\nu_{\ell}} \nu_{\tau}$,
\item
$\overline{{\rm B}} \rightarrow \Dstarp \overline{{\rm D}} {\rm X}$ with 
$\overline{{\rm D}} \rightarrow \ell^-\overline{\nu_{\ell}} {\rm Y}$.
\end{itemize}
Their respective branching fractions have been evaluated in Section 
\ref{sec:ajous1}.

The production rates of the different $\Dstarstar$
states (see Sections \ref{sec:dssb} and \ref{sec:dssc}) and the variation of 
their corresponding form factors as a function of $w$, have also
to be considered.
Published results on $\Vcb$
are based on the old Isgur-Wise model \cite{isgw}, which
predicts a sizeable $\Dstarstar$ 
background rate near the end-point spectrum.
As a consequence, in this model the error on the overall amount 
of $\Dstarstar$
has a large effect on $\Vcb$\ while having a negligible contribution
on the slope parameter $\rho^2$. However, HQET predicts that,
in the infinite charm mass limit, the rate near $w=1$ is suppressed by a 
further factor $(w^2-1)$ when compared with the signal \cite{Pene,Babar}. 
In this case, the $\Dstarstar$ rate uncertainty would have a large 
effect on the 
slope with only a  small influence on $\Vcb$ \cite{DELPHI_vcb}.
However, models in this extreme case fail to predict the 
ratio ${\rm R}^{**}$ (Equation~(\ref{eq:Rss}))
between the production rates of the two narrow states.

\begin{figure}[t]
\centerline{\epsfxsize 14.0truecm \epsfbox{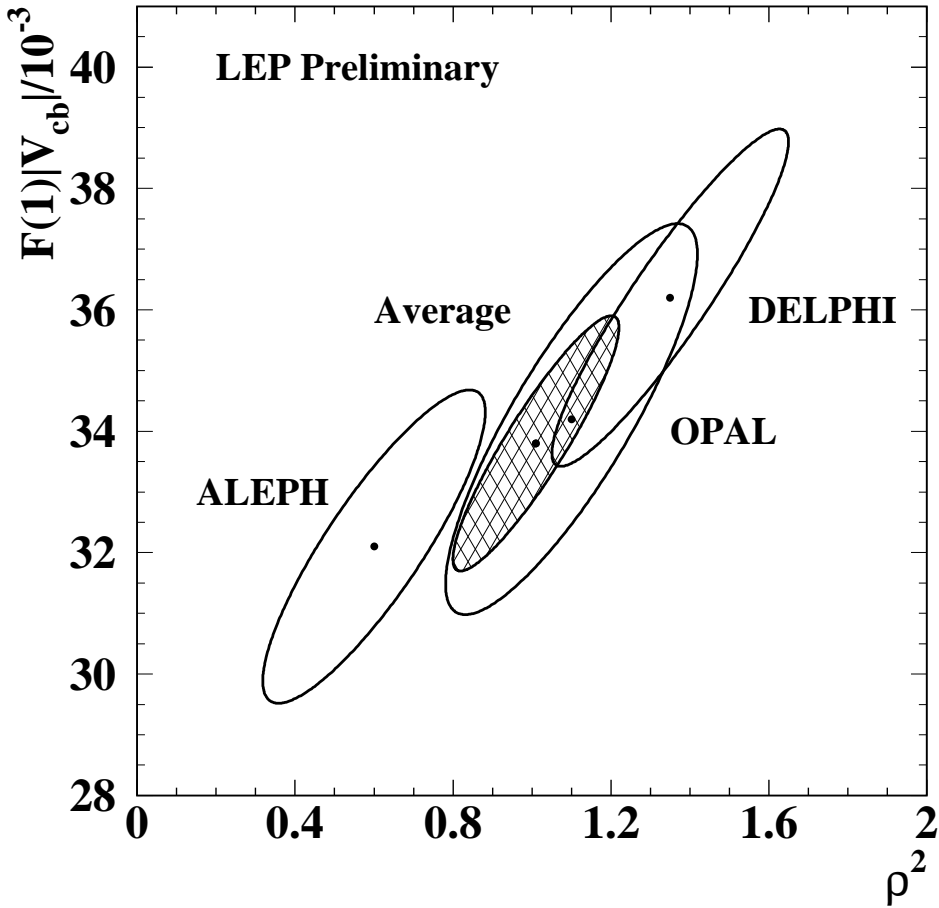}}
\caption{{\it Corrected measurements and LEP average of the
quantities \fvcb\ and $\rho^2$ using the
exclusive method.
The error ellipses, centred on the different measurements,
correspond to contours at the 39 $\%$ C.L. and include systematics.
The hatched ellipse corresponds to the average of the three measurements.
}\label{f:vcbell}}

\end{figure}

A treatment which accounts for ${\cal O}(1/m_c)$ corrections 
is proposed in \cite{ref:ligeti}. 
Several possible approximations of the form factors are provided,
depending on five different expansion schemes
and on three different input parameters. To be conservative, at the
present stage of the analysis, each proposed scheme was
tested in turn and the input parameters were varied over
their full range. The quoted central value
corresponds to the arithmetic average of the values 
obtained with the two extreme models.
The systematic error due to the
modelling of the $\rm D^{**}$ background was computed as half the difference
between the two extreme results.

\mysubsubsection {Corrections applied to the measurements}

Since the three LEP measurements have been performed using different 
methods and inputs, they must be put on the same footing before being 
averaged. ALEPH and DELPHI measurements have been updated to use 
the CLN \cite{CLN} extrapolation method (\ref{eq:clnextr}) and the
Ligeti \cite{ref:ligeti} \btodss\ model.
Since OPAL uses the Caprini-Neubert 
extrapolation method and the JETSET \btodss\ model, corrections to the OPAL
measurement have been estimated using the ALEPH analysis, which is similar. 

Corrections for changing to the standard input parameters, listed in Table \ref{t:sys},
have been calculated as for the $\dmd$ measurement (see Section \ref{sec:method}).
The central value of each analysis is adjusted according to the 
difference between the used and desired parameter values and the 
associated systematic error. The systematic error 
itself is then scaled to reflect the desired uncertainty on the input
parameter.
Table~\ref{t:correxp} lists the corrected results for the three experiments.

\begin{figure}[htb]
\begin{center}
\begin{tabular}{cc}
\mbox{\epsfxsize8.0cm \epsfysize6.0cm\epsffile{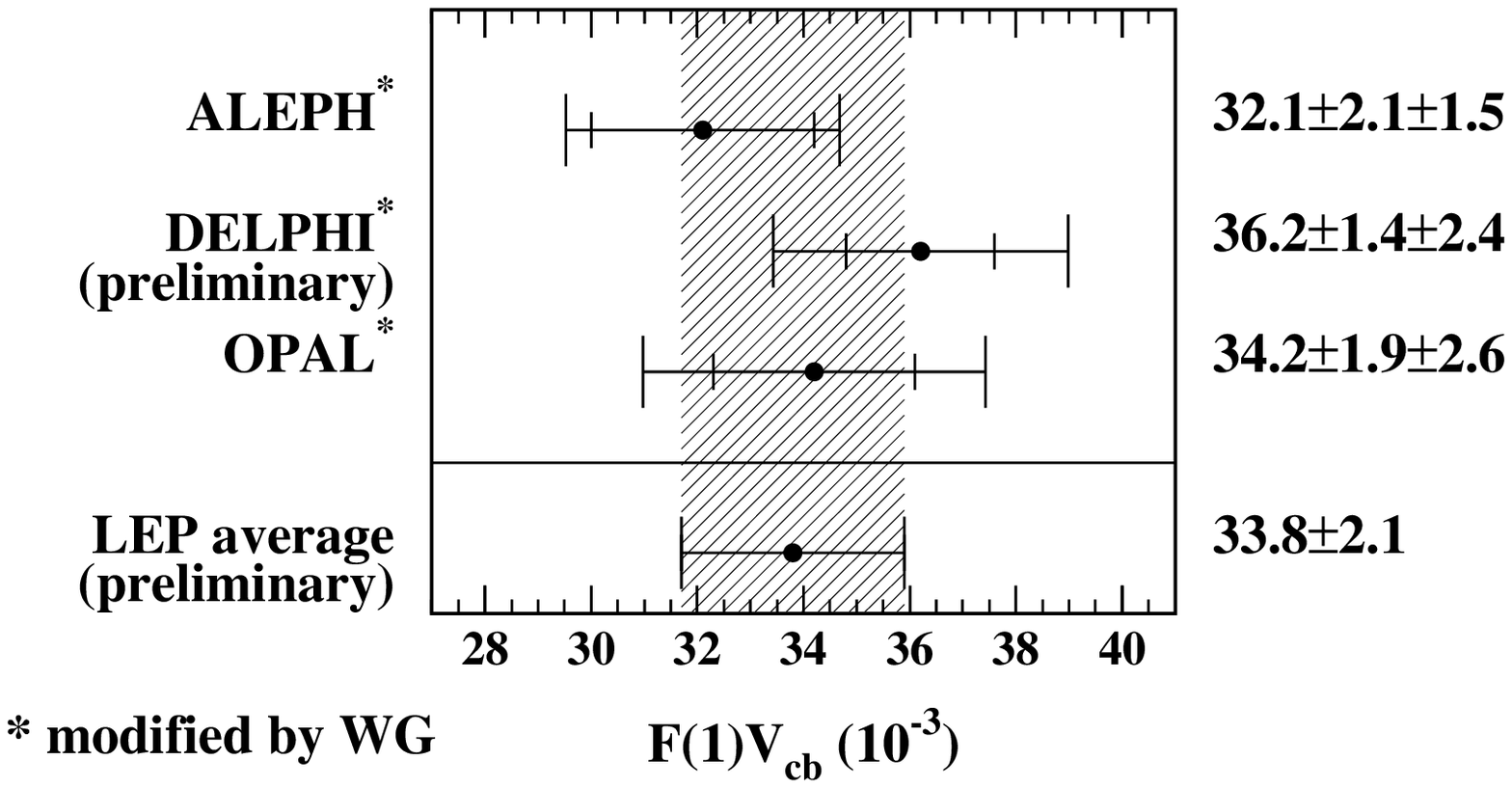}} &
\mbox{\epsfxsize8.0cm \epsfysize6.0cm\epsffile{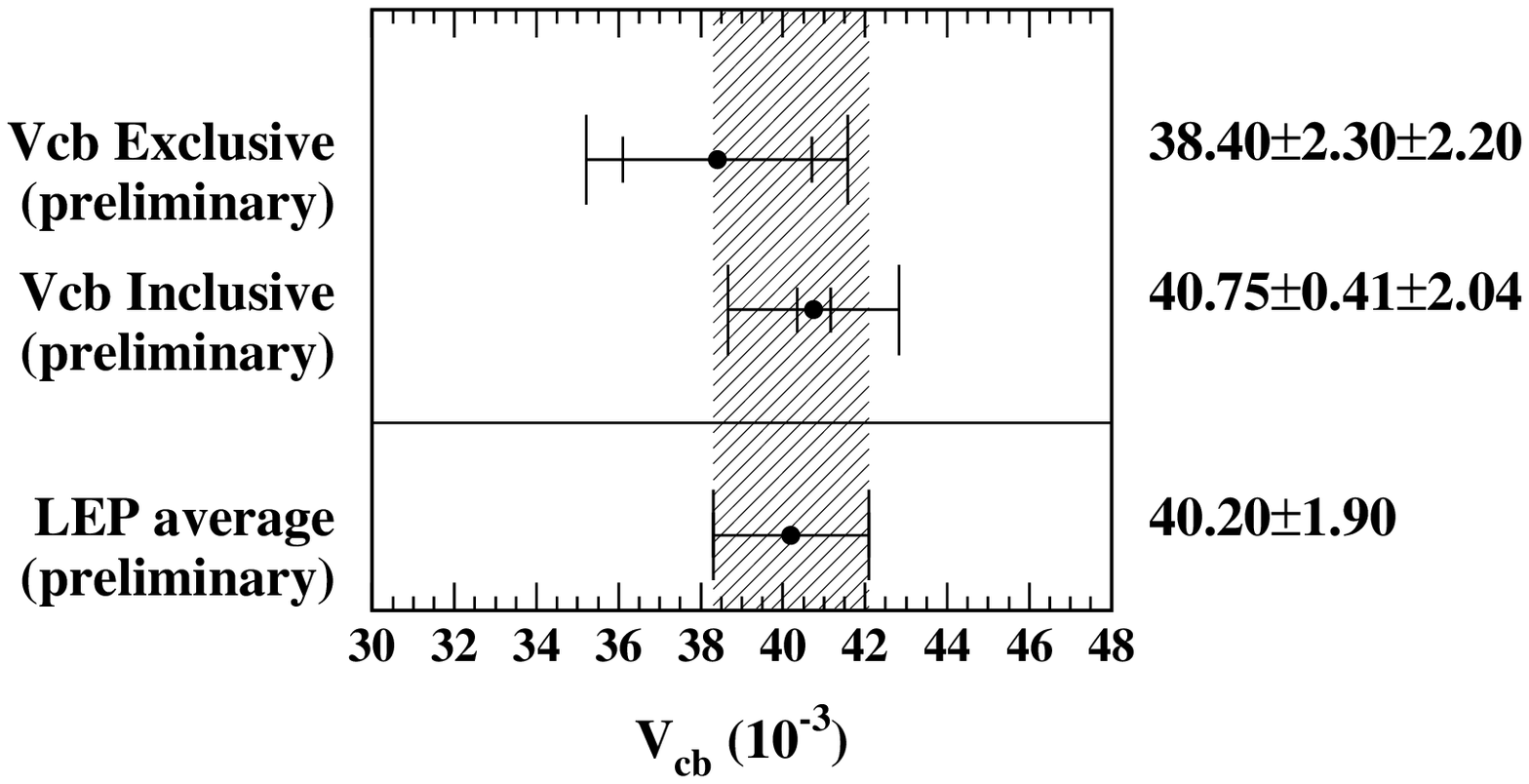}}\\
\end{tabular}
\caption{{\it Left: Corrected \fvcb\ values and LEP average
 using the exclusive
method.  The values shown in the plot have been adjusted by the 
working group and
are those used to perform the average. The original values
can be found in the experimental papers.
Right: \vcb\ LEP average.}\label{f:vcbav}}

\end{center}
\end{figure}


\begin{table}
\begin{center}
\begin{tabular}{|l|lc|} \hline
Experiment   & ${\cal F}_{D^{*}}(1)|V_{\rm cb}|~(\times 10^3)$& $ \rho^2$ \\ \hline
 ALEPH ~~\cite{ALEPH_vcb}      &  32.1 $\pm$  2.1 $\pm$ 1.5 & 0.60 $\pm$ 0.25 $\pm$ 0.13 \\
 DELPHI~\cite{DELPHI_vcb}      &  36.2 $\pm$  1.4 $\pm$ 2.4 & 1.35 $\pm$ 0.13 $\pm$ 0.27 \\
 OPAL~~~\cite{OPAL_vcb}        &  34.2 $\pm$  1.9 $\pm$ 2.6 & 1.10 $\pm$ 0.24 $\pm$ 0.21 \\ \hline
\end{tabular}
\end{center}
\caption{{\it Experimental results on \fvcb\ and $\rho^2$ corrected for common 
inputs.} \label{t:correxp} }

\end{table}

\mysubsubsection{Exclusive $\Vcb$ average}
The combination method for $\mbox{\fvcb}$ and $\rho^2$ is the same as 
the method used for
$\dmd$
(see Section \ref{sec:method}) generalized to the combination of two or more correlated
parameters.
The LEP average (see Figure \ref{f:vcbav}-right)
gives:
\begin{eqnarray}
\mbox{\fvcb} & = & \favcb \\
  \rho^2 & = & \rhobb 
\end{eqnarray}
The parameters used are listed in Table~\ref{tab:gensys}. The
uncertainty on the variation with $w$ of the  
 \btodss\  form factors is taken to be fully correlated between experiments.
The dominant systematic uncertainties on \fvcb\ 
are listed in Table~\ref{t:sys}.
The largest comes from the \btodss\ 
contribution.
\begin{table}
\begin{center}
\begin{tabular}{|l|c|c|}\hline
Source &  $\sigma({\cal F}_{D^{*}}(1)|V_{\rm cb}|)/{\cal F}_{D^{*}}(1)|V_{\rm cb}|~(\%)$& $\sigma(\rho^2)$ \\  \hline
{\bf    BR's} & & \\
$\mathrm{BR}(\mathrm D\rightarrow \mathrm K n \pi)$ 
                        & 1.0 & 0.01  \\
$\mathrm{BR}(\mathrm D^{*+}\rightarrow\mathrm D^0\pi^+)$  & 1.0 & -\\
$\Gamma _{{\mathrm b }{\overline{ \mathrm b}}}/\Gamma_{\mathrm{ had}}$
                        & 0.19 & -\\
$\fd$          
                        & 1.2 &  -  \\ \hline
{\bf         Background} & & \\
$\rm B^- \rightarrow D^{*+} \pi^- \ell^-\overline{\nu_{\ell}}$  & 4.5 &  0.16 \\
$\rm \overline{B} \rightarrow D^{*+} X_{\overline{c}}(\rightarrow \ell^- X)$  & 0.20 &  -   \\ 
$\rm \overline{B} \rightarrow D^{*+} \tau^- \overline{\nu_{\tau}}$ & 0.17 &  -   \\ \hline
{\bf Detector  }        & 2.1 & 0.10 \\ \hline
{\bf   Other inputs}    &     &       \\ 
Fragmentation           & 0.9 &  -    \\ 
 $\Bdb$  lifetime    & 1.2 &  -    \\ \hline
{\bf Total syst. }      & 5.5 &  0.19  \\  \hline  
{\bf Statistical }      & 2.7 &  0.09  \\  \hline  
\end{tabular}                  
\end{center}
\caption{{\it Dominant systematic uncertainties on \fvcb~ and $\rho^2$ expressed,
respectively, as relative (in $\%$) and absolute values.} \label{t:sys} }

\end{table}
The confidence level of the fit is 15\%. The error ellipses of the 
corrected measurements and of the LEP average are shown on 
Figure~\ref{f:vcbell}.

The theoretical estimate, \fone=$0.88\pm0.05$ (see Appendix 
\ref{appendixC}), is used to determine:
\begin{equation}
\Vcb^{excl.} =  \vcbexc .
\end{equation}

\mysubsection{Overall $\Vcb$ average}
\label{sec:vcbav}

The combined \vcb\ average (see Figure \ref{f:vcbav}-right)
 can be extracted taking into account
correlations between the inclusive and exclusive methods. 
The most important source of correlations comes from theoretical
uncertainties in the evaluation of $\mu_{\pi}$,
the average momentum of the $b$-quark inside the $b$-hadron
 (Appendix \ref{appendixC}). 
In the determination of experimental
systematic uncertainties, theoretical uncertainties in the modelling of 
$b \rightarrow \ell$ decays and the
exact amount of $b \rightarrow \Dstarstar$ decays are taken as fully correlated. 
If these correlations were neglected, the central value for the 
$\Vcb$ average would change by only 0.3\%.
Uncertainties from lepton identification and background
also contribute, but to a much lesser extent.
All other sources provide negligible contributions to the
correlated error.
The various contributions
to the uncertainty are categorised in Table \ref{corr}. 
The combined value is:
\begin{equation}
\Vcb = \vcbavg
\end{equation}
where, within the total error of 1.9, 1.5 comes from uncorrelated sources and
1.2 from correlated sources.

\begin{table}[ht!]
\begin{center}
\begin{tabular}{|c||c|c||c|c|} \hline
source & \multicolumn{2}{|c|} {correlated} & \multicolumn{2}{|c|} {uncorrelated} \\\hline
         & $\Vcb$ incl. &  $\Vcb$ excl.     & $\Vcb$ incl. &  $\Vcb$ excl. \\\hline
theory               & 2.4\%     &   2.7\%        & 4.4\%     &   5.0\% \\ 
exp. syst.           & 0.6\%     &   4.6\%        & 0.7\%     &   3.0\% \\ 
stat.                &           &                & 0.4\%     &   2.7\% \\ \hline
total                & 2.5\%     &   5.3\%        & 4.5\%     &   6.4\% \\ \hline
\end{tabular}   
\parbox{15cm}{\caption {\it {Contributions to correlated and
    uncorrelated errors on $\Vcb^{incl.}$ and $\Vcb^{excl.}$,
expressed as relative errors.
\label{corr}}}}
\end{center}
\end{table}

\mysection{Average of LEP $\Vub$ measurements}
\label{sec:vub}

The first LEP combined determinations of 
BR($b \rightarrow \ell^-\overline{\nu_{\ell}}{\rm X}_u$)($\ell^-= e^-$ or $\mu^-$) and the derivation of 
$\Vub$ have been obtained by combining the results reported 
by the 
ALEPH~\cite{aleph_vub},  
DELPHI~\cite{delphi_vub} and L3~\cite{l3_vub} 
Collaborations\footnote{The present members of the $\Vub$ working group are:
D. Abbaneo, M. Battaglia, P. Henrard, S. Mele, E. Piotto, Ph. Rosnet and 
Ch. Schwick.}. 
The three analyses rely on different techniques to measure the inclusive 
yield of $b \rightarrow u$ transitions in semileptonic $b$-hadron decays. 
The experimental details can be found in the original publications.
All three experiments reported evidence for the 
$b \rightarrow  \ell^-\overline{\nu_{\ell}}{\rm X}_u$ transition and measured 
its rate
BR($b \rightarrow  \ell^-\overline{\nu_{\ell}}{\rm X}_u$). 
DELPHI fitted
$\Vub/\Vcb$ directly 
to the fraction of candidate $b \rightarrow \ell^-\overline{\nu_{\ell}}{\rm X}_u$ decays in the 
selected data sample. For this averaging, the corresponding value of
BR($b \rightarrow \ell^-\overline{\nu_{\ell}}{\rm X}_u$) has been derived by using
the value of $\Vcb$ obtained in Section \ref{sec:vcbav}
and the relationship between 
$\Vub$ and  BR($b \rightarrow \ell^-\overline{\nu_{\ell}}{\rm X}_u$) 
given below.

In order to average these results, the sources of systematic uncertainties 
have been divided into two categories. The first contains (a) uncorrelated
systematics due to experimental systematics, such as lepton 
identification, $b$-tagging, vertexing efficiency and energy 
resolution, and (b) uncorrelated systematics from signal modelling and 
background description. 
The second contains correlated systematic uncertainties deriving 
from the simulation of signal $b \rightarrow u$ and background 
$b \rightarrow c$ transitions.
The contributions from the statistical,
experimental, and uncorrelated and correlated modelling uncertainties are 
summarised in Table~\ref{tab:brslu}. 

\begin{table}[ht!]
\begin{center}
\begin{tabular}{|l|c c c c c|}
\hline
Experiment & BR & (stat.) & (exp.) & 
(uncorrelated) & (correlated) \\
\hline 
ALEPH~\cite{aleph_vub} & 1.73 & $\pm$ 0.48 & $\pm$ 0.29 &
$\pm$ 0.29 ($^{\pm 0.29~b\rightarrow c}_{\pm 0.04~b\rightarrow u}$) & 
$\pm$ 0.47 ($^{\pm 0.43~b\rightarrow c}_{\pm 0.19~b\rightarrow u}$) \\
 & & & & &\\ 
DELPHI~\cite{delphi_vub} & 1.57 & $\pm$ 0.35 & $\pm$ 0.37 &
$\pm$ 0.17 ($^{\pm 0.12~b\rightarrow c}_{\pm 0.12~b\rightarrow u}$) 
& $\pm$ 0.42 ($^{\pm 0.34~b\rightarrow c}_{\pm 0.20~b\rightarrow u}$) \\
& & & & &\\
L3~\cite{l3_vub} & 3.30 & $\pm$ 1.00 & $\pm$ 0.80 & 
$\pm$ 0.68 ($^{\pm 0.68~b\rightarrow c}_{~~~-~~~b\rightarrow u}$)
& $\pm$ 1.40 ($^{\pm 1.29~b\rightarrow c}_{\pm 0.54~b\rightarrow u}$) \\
\hline
\end{tabular}
\end{center}
\caption[]{{\it The results for 
10$^3 \times${\rm BR}($b \rightarrow  \ell^-\overline{\nu_{\ell}}{\rm X}_u$) from the 
LEP experiments with the statistical, experimental, 
model uncorrelated, and model correlated uncertainties.} \label{tab:brslu}} 
\end{table}

The correlated systematics 
are summarised in Table~\ref{tab:ebrslu}.
Differences in the analysis techniques adopted by the three 
experiments are reflected by the different sizes of the systematics 
uncertainties estimated from each common source. Important common
systematics arise from the D topological branching fractions and the 
rate of ${\rm D} \rightarrow {\rm K}^0$ decays. 
${\rm D}$ decays 
represent a potential source of background for $b \rightarrow u$
decays because both are characterized 
by a small hadronic mass and a low charged 
multiplicity. The sensitivity to the topological branching fractions is 
reduced in the DELPHI analysis by applying a rescaling of the mass 
${\rm M_X}$ of the hadronic system, based on the reconstructed 
$\ell^- \overline{ \nu_{\ell}} {\rm X}$ 
mass, and by the use of identified kaons for separating signal from background 
events. This explains the different sensitivity of the ALEPH and 
DELPHI analyses to these two important sources of systematic uncertainty.
ALEPH and L3 are sensitive to the uncertainties in the $b$ fragmentation
function due to the kinematical variables used for discriminating 
$b \rightarrow \ell^- \overline{\nu_{\ell}} {\rm X}_u$ from 
$b \rightarrow \ell^- \overline{\nu_{\ell}} {\rm X}_c$ 
decays. 
The DELPHI result is sensitive to the assumed production rates of
$b$ hadron species due to the use of kaon 
anti-tagging to reject $b \rightarrow c$, thus rejecting also $\Bs$ and 
$\Lb$ decays; and it is sensitive
to the contribution of ${\rm D}^{(*)} \pi$ and 
${\rm D}^{**}$ 
states in semileptonic decays because the 
resulting difference in the vertex topology is also used for discriminating 
$b \rightarrow u$ from $b \rightarrow c$ decays. 


\begin{table}[hb!]
\begin{center}
\begin{tabular}{|l|c|c|c|}
\hline
Source & ALEPH & DELPHI & L3 \\
\hline 
$b$  species                                 & 0.01  & 0.12  &  - \\
$b$ fragmentation                            & 0.22  & 0.03 & 0.32  \\
$b \rightarrow \ell$ model                   & 0.11 & 0.08 & 1.24\\
$c \rightarrow \ell$ model                   & 0.14 & 0.13 & 0.12\\
D topological BR's                           & 0.31 & 0.06 & -\\
BR(D$ \rightarrow {\rm K}^0$)                & 0.08 & 0.19  & -\\
$\Dstarstar$, ${\rm D}^{(*)} \pi$ production & 0.04 & 0.19 & -\\
\hline
$b \rightarrow u$ inclusive model            & 0.18  & 0.08 & 0.25\\
$b \rightarrow u$ exclusive model            & 0.05  & 0.18  & 0.20\\
$\Lambda_b$ decay model                      & 0.04 & -    & 0.44\\
\hline
\end{tabular}
\end{center}
\caption[]{{\it Correlated sources of systematic uncertainties (in units of
10$^{-3}$) entering in the measurement of 
{\rm BR}($b \rightarrow \ell^-\overline{\nu_{\ell}}{\rm X}_u$).} \label{tab:ebrslu}}
\end{table}

The  systematic uncertainties on the $b \rightarrow u$ signal have been 
grouped into {\sl inclusive model} and {\sl exclusive model} classes, 
which are assumed to be fully
correlated. The first corresponds to the uncertainty in the modelling of
the kinematics of the $b$-quark in the heavy hadron. It has been 
estimated from 
the spreads of the results obtained with 
the ACCMM~\cite{accmm}, the 
Dikeman-Shifman-Uraltsev~\cite{dsu}, and the parton~\cite{parton} models 
in the ALEPH and DELPHI analyses. In the case of the L3 analysis,
the uncertainties in the single pion and in the lepton energy spectra
were evaluated  
from the discrepancies between the model of Ref~\cite{burdman} and the 
ISGW~\cite{isgw} model respectively to the JETSET~7.4~\cite{jetset} 
prediction. 
The {\sl exclusive model} uncertainty corresponds to the modelling of the 
hadronic final state in the $b \rightarrow \ell^-\overline{\nu_{\ell}}
{\rm X}_u$ 
decay.
These uncertainties have been estimated by replacing the parton shower 
fragmentation model in JETSET~\cite{jetset} with the fully exclusive 
ISGW2~\cite{isgw2} 
model by ALEPH and DELPHI, and by propagating a 100\% uncertainty 
on the ${\rm B} \rightarrow \pi \ell^-\overline{\nu_{\ell}}$ rate by L3. 
ALEPH and L3 have also taken into account the uncertainty from the 
modelling of the charmless semi-leptonic decay of $b$-baryons. This has not
been considered by DELPHI as they remove decays containing identified
protons and kaons, thus suppressing the contribution of $b$-baryons as 
mentioned above. In addition to these sources, 
ALEPH has estimated a $b \rightarrow u$ uncertainty from the 
energy cut-off value $\Lambda$ for the hybrid model adopted~\cite{hybrid}; 
DELPHI allowed for the $b$-quark pole mass $m_b$ and the 
expectation value of the kinetic energy operator $<p^2_b>$ uncertainties
which have been assumed to be uncorrelated.

\begin{figure}[h!] 
\begin{center}
\begin{tabular}{cc}
\mbox{\epsfxsize8.0cm \epsfysize10.0cm\epsffile{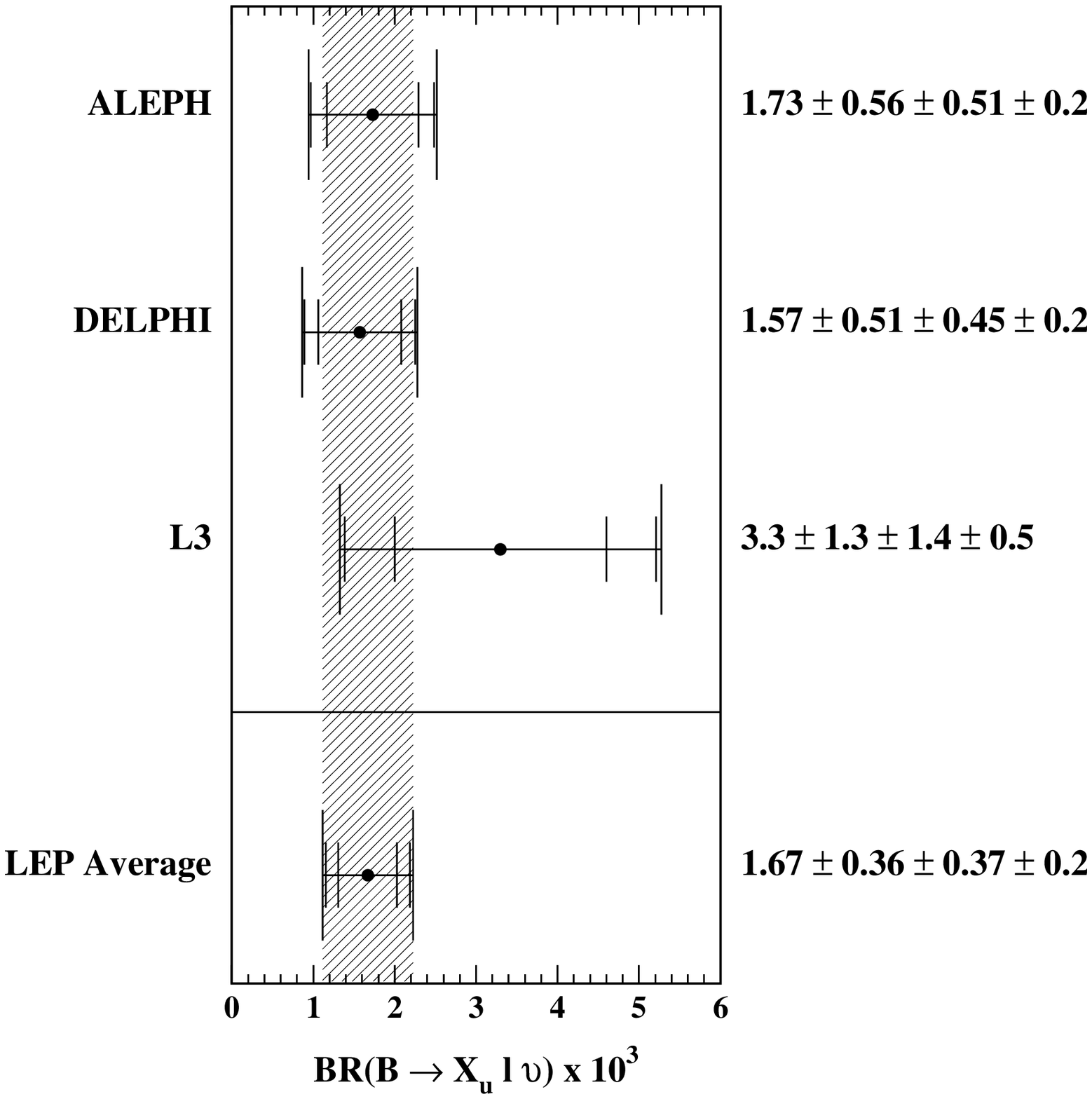}} &
\mbox{\epsfxsize8.0cm \epsfysize10.0cm\epsffile{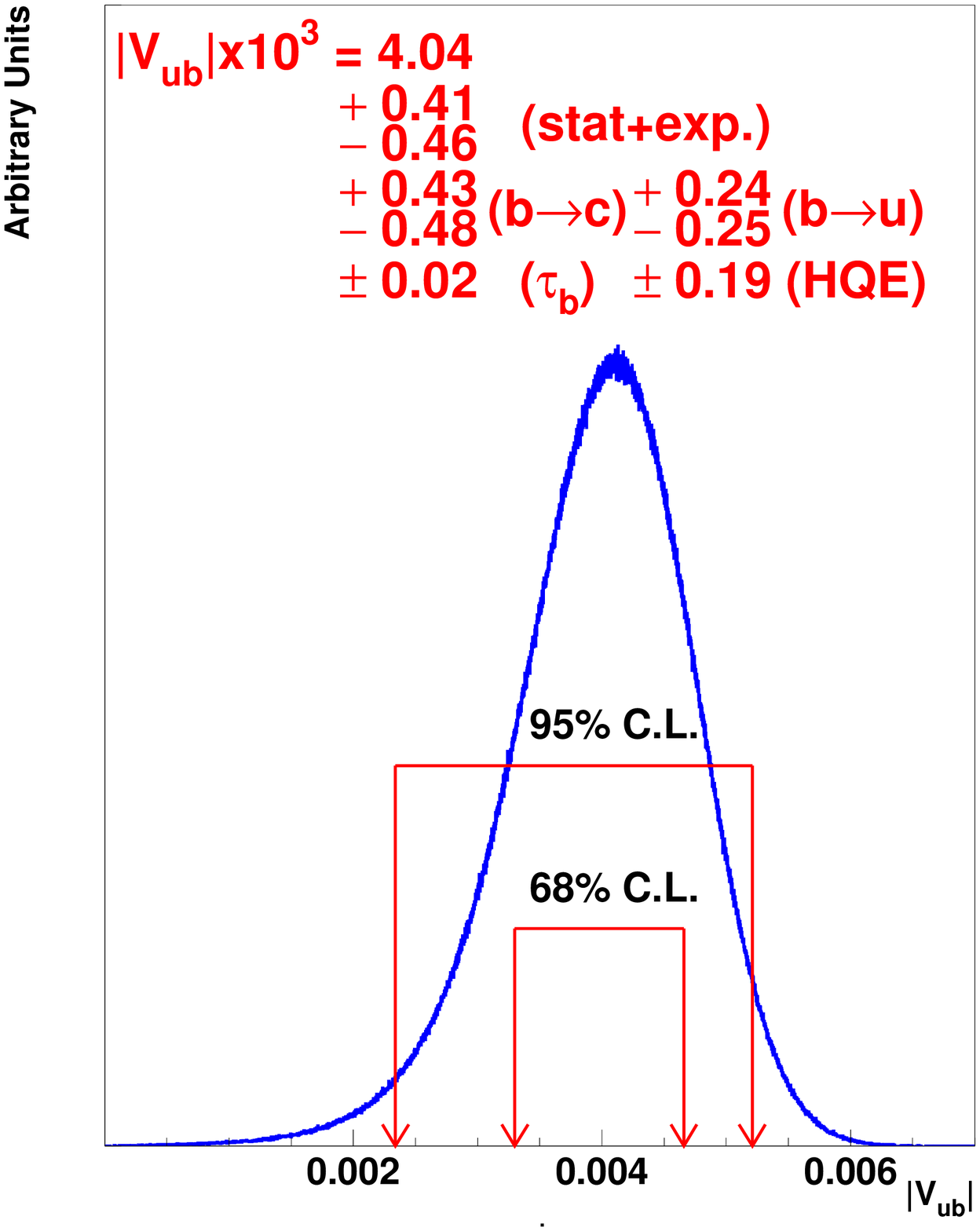}} \\
\end{tabular}
\caption{{\it Left: the determinations of 
${\rm BR}(b \rightarrow \ell^-\overline{\nu_{\ell}}{\rm X}_u $) by 
ALEPH, DELPHI and L3 and the resulting LEP average.
Right: the probability density function for $\Vub$ corresponding to 
the LEP average value of ${\rm BR}(b \rightarrow \ell^-\overline{\nu_{\ell}}{\rm X}_u$)
with the value of 
the median and two confidence intervals indicated. 
Unphysical negative entries have been
discarded and the probability density function renormalised accordingly.}
\label{fig:brslu}}
\end{center}
\end{figure}

The three measurements of BR($b \rightarrow  \ell^-\overline{\nu_{\ell}}{\rm X}_u$) 
have been averaged 
using the Best Linear Unbiased Estimate (B.L.U.E.) technique~\cite{blue}.
Using the inputs from Table~\ref{tab:brslu} and Table~\ref{tab:ebrslu}, the 
LEP average value for BR($b \rightarrow \ell^-\nu {\rm X}_u$) was found to be:
\begin{eqnarray} 
{\rm BR}(b \rightarrow \ell^-\overline{\nu_{\ell}}{\rm X}_u)& = &(1.67 \pm 0.36 {\rm (stat.+exp.)} \pm 0.37 (b \rightarrow c) 
\pm 0.20 (b \rightarrow u)) \times 10^{-3} \nonumber \\
 & =& (1.67 \pm 0.55) \times 10^{-3} 
\end{eqnarray} 
with a confidence level for the combination of 0.70 
(see Figure~\ref{fig:brslu}).

The value of the $\Vub$ element has been extracted using the 
following relationship derived in the context of (OPE)~\cite{uraltsev,hoang}:
\begin{eqnarray}
\Vub & =& 0.00445~ \sqrt{\frac{ \mathrm{BR}(b \rightarrow {\rm X}_u \ell^-\overline{\nu_{\ell}})}
{0.002} \frac{1.55 \mathrm{ps}}{\tau_b}} \times \nonumber \\
 & & (1 \pm 0.010 \mathrm{(pert)} \pm 0.030 (1/m_b^3) \pm 0.035 
(m_b))
\end{eqnarray}
by assuming $m_b$ = (4.58 $\pm$ 0.06)~$\GeV/c^2~$ (Appendix \ref{appendixC}).

From the LEP average of  
${\rm BR}(b \rightarrow \ell^-\overline{\nu_{\ell}}{\rm X}_u $)
obtained above, 
the $b$-lifetime value, $\tau_b$, obtained in Section \ref{sec:Atau},
and the quoted $\Vub$ uncertainty coming from OPE ($\pm$4.7 \%
relative), a probability density function for $\Vub$ has been calculated.
The resulting distribution is shown in Figure~\ref{fig:brslu}-right,
where all errors are convoluted together
 assuming that they are Gaussian in 
${\rm BR}(b \rightarrow \ell^-\overline{\nu_{\ell}}{\rm X}_u$), 
with the exception of the OPE error assumed 
to be Gaussian in $\Vub$.
The small part of 
this function in the negative, unphysical region, corresponding
to only 0.14\%, has been discarded and the 
probability density function renormalised accordingly.
The median of this function has been chosen as the best estimate of
$\Vub$, and $\pm 34.135\%$ and $\pm 47.725\%$ of the integral of 
the probability density function around this value have been used to define
the 1$\sigma$ and 2$\sigma$ confidence regions, denoted hereafter as the 
68\% and 95\% confidence levels, obtaining:
\begin{equation}
\Vub = (4.04^{+0.62}_{-0.74}) \times 10^{-3}~{\rm at~ the~ 68\%~ C.L.}
\end{equation}
and
\begin{equation}
\Vub = (4.04^{+1.17}_{-1.71}) \times 10^{-3}~{\rm at~ the~95\%~ C.L.}
\end{equation}

All the uncertainties have been included in these estimates. 
The application of the above procedure for each error source separately 
yields the following detailed result for the 68\% confidence level:

\begin{eqnarray}
\Vub &= &(4.04 ^{+0.41}_{-0.46}~{\rm (stat.+det.)} 
      ^{+0.43}_{-0.48}~(b \rightarrow c~{\rm syst.)} 
      ^{+0.24}_{-0.25}~(b \rightarrow u~{\rm syst.)}  \nonumber \\
    & &  \pm 0.02~(\tau_b) 
      \pm 0.19~{\rm (OPE)}) \times 10^{-3}.
\end{eqnarray}

\mysection{Summary of all results}
\label{sec:conclusion}

This paper provides precise combined results (see Table \ref{tab:conclusion}),
from measurements
submitted to the 1999 Summer Conferences,
on parameters which govern
production and decay properties of $b$-hadrons emitted in
high energy $b$-quark jets.

At the $\Zz$ pole, the polarization of $b$-baryons is 
found to be 
significantly different from zero
but it is reduced as compared to the initial $b$-quark polarization of $-0.94$.
Production of $\Sigma_b^{(\ast)}$ baryons has been invoked 
\cite{ref:popolb} to explain
this difference.

The accuracy on $\Bdb$ and $\Bm$ lifetimes is better than 2$\%$. The $\Bm$
lifetime is significantly longer than the $\Bdb$ lifetime, in agreement with 
the original expectations based on the OPE and parton-hadron duality 
\cite{ref:bigilife}. But the clear difference between measured and expected 
lifetimes of $b$-baryons remains to be explained. It may point to a failure
of parton-hadron duality in inclusive B decays, or to a failure of quark
models used to evaluate the mean values of operators contributing to 
$b$-baryon decays.
\begin{table}[thb!]
  \begin{center}
    \begin{tabular}{|l| c |}
      \hline
      {\bf {\boldmath $b$}-hadron lifetimes} &  CDF-LEP-SLD, Section \ref{sec:Atau}\\
      \hline
      $\tau(\Bd)$ & $(\taubd)$~ps \\
      $\tau(\Bp)$ & $(\taubp)$~ps \\
      $\tau(\Bs)$ & $(\taubs)$~ps \\
      $\tau(\Lb)$ & $(\taulb)$~ps \\
      $\tau(\Xi_b)$ & $(\tauxib)$~ps \\
      $\tau(b-{\rm baryon})$ & $(\taubbar)$~ps \\
      $\tau_b$ & $(\taubav)$~ps \\
      $\frac{\tau(\Bm)}{\tau(\Bdb)}$ & $ \taubpovertaubd$\\

      \hline
      {\bf {\boldmath $b$}-hadron production rates} & CDF-LEP, Section \ref{sec:boscill}\\
      \hline
      $\fs$ & $\fbs$ \\
      $\fb$ & $\fbar$ \\
      $\fd = \fu$ & $\fbd$ \\
      $\rho(\fs,\fb)$ & $\rhosbar$ \\
      $\rho(\fd,\fs)$ & $\rhosd$ \\
      $\rho(\fs,\fb)$ & $\rhobard$ \\
      \hline
     {\bf {\boldmath ${\rm B}^0 - \overline{{\rm B}^0}$} oscillations} &  ARGUS-CDF-CLEO-LEP-SLD, Section \ref{sec:boscill}\\
      \hline
      $\dmd$ & $(\dmdx)$ \ips \\
      $\chi_d$& $\chix$ \\
      $\dms$ & $> \dmslim$ \ips {\rm at~ the}~95$\%$ C.L. \\
      \hline
    {\bf  Limit on {\boldmath $\Delta \Gamma_{\Bs} $}} &  CDF-LEP-SLD, Section \ref{sec:deltag}\\
      \hline
      $\Delta \Gamma_{\Bs} /\Gamma_{\Bs}  $ & $<0.31  $ at the~95$\%$ C.L.\\
      \hline
     {\bf  Measurement of {\boldmath $\Vcb$}} &  LEP, Section \ref{sec:vcb}\\
      \hline
      ${\rm BR}(b \rightarrow \ell X) $ & $(10.58 \pm 0.07 \pm 0.17)\% $ \\
      $\mbox{\fvcb}$ & $\favcb$ \\
      $\rho^2$  & $\rhobb$ \\ 
      $\Vcb^{incl.}$ & $ \vcbinc$ \\
      $\Vcb^{excl.}$ & $ \vcbexc$ \\
      $\Vcb$ & $ \vcbavg$ \\
      \hline
    {\bf  Measurement of {\boldmath $\Vub$}} &  LEP, Section \ref{sec:vub}\\
      \hline
      $\Vub$ & $ (4.04^{+0.62}_{-0.74}) \times 10^{-3}$ \\
      \hline
    {\bf  {\boldmath $\Lb$} polarization in {\boldmath $\Zz$} decays} &  LEP, Section \ref{sec:systgen}\\
      \hline
      ${\cal P}(\Lb)$ & $ -0.45^{+0.17}_{-0.15} \pm 0.08$ \\
      \hline
    \end{tabular}
  \end{center}
    \caption{{\it Summary of the results obtained 
on $b$-hadron production rates
and decay properties using data available by mid 1999. More details
on systematic uncertainties can be found in the corresponding sections.}
    \label{tab:conclusion}}
\end{table}

The accuracy on $\Bdb$ and $\Bm$ production rates in $b$-quark jets has 
improved by a factor two as compared to published values \cite{PDG98}.
In conjunction with the better determination of the $\Bdb$ lifetime,
this  measurement has given a more precise determination of $\Vcb$
from $\Bdb \rightarrow \Dstarp \ell^- \overline{\nu_{\ell}}$ decays.
The $\Bsb$ fraction in jets is close to 10$\%$ with an uncertainty which 
has been reduced by 30$\%$. Its determination is an important input to 
searches for $\Bs$-$\Bsb$ oscillations. The accuracy on the $b$-baryon
rate has improved by about a factor two because of new measurements
using spectator baryons.

The mass difference between mass eigenstates of the $\Bd$-$\Bdb$ system
is measured with a relative precision better than 4$\%$.
The corresponding quantity for the $\Bs$-$\Bsb$ system is still
unmeasured, in spite of impressive progress in the 
sensitivity reached by the experiments.

A first combined result on the decay width difference between 
mass eigenstates of
the $\Bs$-$\Bsb$ system has been obtained.

Because of the very accurate measurements obtained on the inclusive
$b$-lifetime and semileptonic branching fraction, the accuracy
of $\Vcb$ determined from inclusive semileptonic decays is entirely limited by
theoretical uncertainties. An experimental control of these uncertainties
is thus needed.
The measurement of $\Vcb$ from exclusive 
$\Bdb \rightarrow \Dstarp {\rm X} \ell^- \overline{\nu_{\ell}}$
decays is limited by uncertainties related to $\Dstarstar$ 
production and decay mechanisms.
Present experimental results on production characteristics of these states
in $b$-hadron semileptonic decays have been detailed in Section 
\ref{sec:systgen} of the present report.

The fact that the theoretical uncertainties
on the determinations of $\Vcb$ from inclusive and exclusive
measurements are largely uncorrelated has been used in evaluating a 
global average.

The new measurement of $\Vub$ from LEP experiments has reached an
accuracy similar to previous determinations 
at the $\Upsilon(4S)$ but with a better sensitivity over a larger
fraction of the phase space than in measurements
using the lepton energy end-point region. 

New measurements from LEP and SLD are still expected in the year 2000 and 
will improve present determinations of $b$-hadron production and decay
properties, which already provide stringent constraints on the shape of the 
CKM unitarity triangle just before the era of B-factories.

\section*{Acknowledgements}
We would like to thank the CERN accelerator divisions for the efficient 
operation of the LEP accelerator, and their close collaboration with the 
four experiments. We would like to thank members of the CDF and SLD 
Collaborations for making results available to us in advance of the 
conferences and for useful discussions concerning their combination.
Useful contacts with members of the CLEO Collaboration are also 
acknowledged. We thank W. Venus for a careful reading of the manuscript
and for helpful comments.
Finally results on $\Vcb$ and $\Vub$ have been obtained after discussions
with several theorists to understand the meaning and the importance of 
theoretical uncertainties. Among them we would like to thank in particular:
M. Beneke, I.I. Bigi, G. Buchalla, F. Defazio, A. Hoang, L. Lellouch,
Z. Ligeti
and N. Uraltsev.

\newpage
\clearpage
\appendix
\mysection{Production rates of the ${\rm D}_1$ and ${\rm D}^*_2$ mesons 
in semileptonic $b$-decays}
\label{appendixA}

\begin{table}[ht!]
\begin{center}
\begin{tabular}{|l||c|c|c|} \hline
 Experiment    & Channel & Value $\times 10^3$ & Ref. \\   \hline
ALEPH        & ${\rm BR}(b \rightarrow \overline{\rm B}) {\rm BR}({\rm \overline{B}} \rightarrow {\rm D}^{+}_1 \ell^-\overline{\nu_{\ell}}){\rm BR}({\rm D}^{+}_1 \rightarrow \Dstaro \pi^+)$ & $ 2.06^{+0.55~ +0.29}_{-0.51~-0.40} $ & \cite{DssALEPH} \\
ALEPH        & ${\rm BR}(b \rightarrow \overline{\rm B}) {\rm BR}({\rm \overline{B}} \rightarrow {\rm D}^{0}_1 \ell^-\overline{\nu_{\ell}}){\rm BR}({\rm D}^{0}_1 \rightarrow \Dstarp \pi^-)$ & $ 1.68^{+0.40~ +0.28}_{-0.36~-0.29} $ & \cite{DssALEPH} \\
ALEPH        & ${\rm BR}(b \rightarrow \overline{\rm B}) {\rm BR}({\rm \overline{B}} \rightarrow {\rm D}^{0}_1 \ell^-\overline{\nu_{\ell}}){\rm BR}({\rm D}^{0}_1 \rightarrow \Dstarp \pi^-)$ & $ 3.62^{+1.78}_{-1.48}\pm 0.77 $ & \cite{DssALEPH} \\
CLEO     & ${\rm BR}({\rm B^-} \rightarrow {\rm D}^{0}_1 \ell^-\overline{\nu_{\ell}}){\rm BR}({\rm D}^{0}_1 \rightarrow \Dstarp \pi^-)$ & $ 3.73 \pm 0.85 \pm 0.52 \pm 0.24 $ &\cite{ref:cleodss}  \\
\hline

\end{tabular}
\caption[]{{\it Measured values of the ${\rm D}_1$ production rate in 
$b$-semileptonic decays.
The third systematic uncertainty, quoted in the CLEO analysis,
comes from the variation of the detection efficiency when changing the 
parameters of the signal form factors in the ISGW2
model \cite{isgw2}.} \label{tab:darate}}

\end{center}
\end{table}

\begin{table}[ht!]
\begin{center}
\begin{tabular}{|l||c|c|c|} \hline
 Experiment    & Channel & Value $\times 10^4$ & Ref. \\   \hline
ALEPH        & ${\rm BR}(b \rightarrow \overline{\rm B}) {\rm BR}({\rm \overline{B}} \rightarrow {\rm D}^{*+}_2 \ell^-\overline{\nu_{\ell}}){\rm BR}({\rm D}^{*+}_2 \rightarrow \Do \pi^+)$ & $ 3.1^{+2.4~ +0.4}_{-2.2~-0.6} $ & \cite{DssALEPH} \\
ALEPH        & ${\rm BR}(b \rightarrow \overline{\rm B}) {\rm BR}({\rm \overline{B}} \rightarrow {\rm D}^{*0}_2 \ell^-\overline{\nu_{\ell}}){\rm BR}({\rm D}^{*0}_2 \rightarrow \Dstarp \pi^-)$ & $ 5.1^{+3.0~ +0.8}_{-2.6~-0.9} $ & \cite{DssALEPH} \\
ALEPH        & ${\rm BR}(b \rightarrow \overline{\rm B}) {\rm BR}({\rm \overline{B}} \rightarrow {\rm D}^{*0}_2 \ell^-\overline{\nu_{\ell}}){\rm BR}({\rm D}^{*0}_2 \rightarrow \Dp \pi^-)$ & $ 3.8^{+2.4~ +0.8}_{-1.9~-0.8} $ & \cite{DssALEPH} \\
CLEO     & ${\rm BR}({\rm B^-} \rightarrow {\rm D}^{*0}_2 \ell^-\overline{\nu_{\ell}}){\rm BR}({\rm D}^{*0}_2 \rightarrow \Dstarp \pi^-)$ & $ 5.9 \pm 6.6 \pm 1.0 \pm 0.4 $ &\cite{ref:cleodss}  \\
\hline

\end{tabular}
\caption[]{{\it Measured values of the ${\rm D}^{*}_2$ production rate in 
$b$-semileptonic decays. In the original publication from ALEPH, 
as these values differ from zero only by one
or two standard deviations, only upper limits
were quoted. The third systematic uncertainty, quoted in the CLEO analysis,
comes from the variation of the detection efficiency when changing the 
parameters of the signal form factors in the ISGW2
model \cite{isgw2}.} \label{tab:dbstar}}

\end{center}
\end{table}

\mysection{$\Lb$ polarization measurements}
\label{appendixAb}
\begin{table}[ht!]
\begin{center}
\begin{tabular}{|l||c|c|} \hline
 Experiment    & Value & Ref. \\   \hline
ALEPH        & $ -0.23^{+0.24~ +0.08}_{-0.20~-0.07} $ & \cite{ref:alpol} \\
DELPHI       & $ -0.49^{+0.32~ +0.17}_{-0.30~-0.17} $ & \cite{ref:delpol} \\
OPAL         & $ -0.56^{+0.20~ +0.09}_{-0.13~-0.09} $ & \cite{ref:oppol} \\

\hline
\end{tabular}
\caption[]{{\it Measured values of the $\Lb$ polarization in 
$\Zz$ decays.}\label{tab:polar} }
\end{center}
\end{table}

\newpage
\clearpage

\mysection{Measurements used in the evaluation of 
$b$-hadron lifetimes}
\label{appendixB}
In the Tables below, new measurements available since Winter 1999 are labelled with ``(n)'' and preliminary results with ``(p)''.

\begin{table}[ht!]
\begin{center}
\begin{tabular}{|l||c|c|c|c|} \hline
 Experiment    & Method                & Data set        & $\tau_{B^0}$ (ps)                & Reference \\   \hline
ALEPH (n, p)         & D$^{(*)} \ell$         &  91--95        & 1.524$\pm 0.053^{+0.035}_{-0.032} $ & \cite{ALEB01}  \\
ALEPH          &  Excl. rec.            &  91--94        & 1.25$^{+0.15}_{-0.13} \pm 0.05$ & \cite{ALEB0}  \\
ALEPH          &Partial rec. $\pi^+ \pi^-$&  91--94       & 1.49$^{+0.17+0.08}_{-0.15-0.06}$ & \cite{ALEB0} \\
CDF            & D$^{(*)} \ell$        & 92--95         & 1.474$\pm 0.039^{+0.52}_{-0.51}$  & \cite{CDFB02}  \\
CDF            & Excl. (${\rm J}/\psi K$)    & 92--95         & 1.58$\pm$ 0.09 $\pm$ 0.02        & \cite{CDFB01}  \\
DELPHI         & D$^{(*)} \ell$        &  91--93        & 1.61$^{+0.14}_{-0.13} \pm 0.08$  & \cite{DELB01}  \\
DELPHI         & Charge sec. vtx.      &  91--93        & 1.63 $\pm$ 0.14  $\pm$ 0.13      & \cite{DELB02} \\
DELPHI         & Inclusive D$^* \ell$  &  91--93       & 1.532$\pm$  0.041 $\pm$0.040      & \cite{DELB03} \\
L3             & Charge sec. vtx.      &  94--95        & 1.52$\pm$  0.06  $\pm$ 0.04      & \cite{L3B01}  \\ 
L3 (p)            & Inclusive D$^* \ell$  &  94           & 1.74$\pm$  0.12  $\pm$ 0.04      & \cite{L3B02}  \\ 
OPAL           & D$^{(*)} \ell$        &  91--93        & 1.53$\pm$  0.12  $\pm$ 0.08      & \cite{OPAB0}  \\ 
OPAL (n, p)          & Charge sec. vtx.      &  93--95        & 1.523$\pm$ 0.057  $\pm$ 0.053    & \cite{OPAB1}  \\ 
SLD            & Charge sec. vtx.$\ell$&  93--95        & 1.56$^{+0.14}_{-0.13} \pm$ 0.10 & \cite{SLDB01}  \\ 
SLD  (n, p)          & Charge sec. vtx.      &  93--98        & 1.565$\pm$ 0.021 $\pm$ 0.043    & \cite{SLDB02}  \\ \hline
Average        &                       &               & 1.562 $\pm$ 0.029                 &    \\   \hline

\end{tabular}
\caption[]{{\it Measurements of the $\Bd$ lifetime.}}
\end{center}
\end{table}
 
\begin{table}[ht!]
\begin{center}
\begin{tabular}{|l||c|c|c|c|} \hline
 Experiment    & Method            & Data set      & $\tau_{B^+}$(ps)                 & Reference \\   \hline
ALEPH (n, p)         & D$^{(*)} \ell$    &  91-95        & 1.646$\pm 0.056^{+0.036}_{-0.034} $ & \cite{ALEB01}  \\
ALEPH          & Excl. rec.        &  91-94        & 1.58$^{+0.21+0.04}_{-0.18-0.03}$ & \cite{ALEB0} \\
CDF            & D$^{(*)} \ell$    & 92-95         & 1.637$\pm 0.058^{+0.45}_{-0.43}$  & \cite{CDFB02}  \\
CDF            & Excl. (${\rm J}/\psi K$)& 92-95         & 1.68 $\pm$ 0.07 $\pm$ 0.02       & \cite{CDFB01}  \\
DELPHI         & D$^{(*)} \ell$    &  91-93        & 1.61 $\pm$ 0.16 $\pm$ 0.12       & \cite{DELB01}$^a$  \\
DELPHI         & Charge sec. vtx.  &  91-93        & 1.72 $\pm$ 0.08 $\pm$ 0.06       & \cite{DELB02}$^a$ \\
L3             & Charge sec. vtx.  &  94-95        & 1.66$\pm$  0.06  $\pm$ 0.03      & \cite{L3B01}  \\ 
OPAL           & D$^{(*)} \ell$    &  91-93        & 1.52 $\pm$ 0.14 $\pm 0.09$       & \cite{OPAB0}  \\ 
OPAL (n, p)          & Charge sec. vtx.  &  93-95        & 1.643$\pm$ 0.037  $\pm$ 0.025    & \cite{OPAB1}  \\ 
SLD            & Charge sec. vtx. $\ell$&  93-95   & 1.61$^{+0.13}_{-0.12} \pm$ 0.07 & \cite{SLDB01}  \\
SLD  (n, p)          & Charge sec. vtx. &  93-98         & 1.623$\pm$ 0.020 $\pm$ 0.034    & \cite{SLDB02}  \\ \hline
Average        &                   &               & 1.656 $\pm$ 0.025                 &    \\   \hline

\end{tabular}
\caption[]{{\it Measurements of the $\Bp$ lifetime. \\
           a) The combined DELPHI result quoted in \cite{DELB02} is (1.70 $\pm$ 0.09) ps.} \label{tab:bp}}
\end{center}
\end{table}

\begin{table}[ht!]
\begin{center}
\begin{tabular}{|l||c|c|c|c|} \hline
 Experiment    & Method              & Data set     & $\tau_{B_s}$ (ps)                 & Reference \\   \hline
ALEPH          & D$_s \ell$          & 91-95        & 1.54$^{+0.14}_{-0.13} \pm$ 0.04   & \cite{ALEBS1}  \\
ALEPH          & D$_s h$             & 91-95        & 1.47$\pm$ 0.14 $\pm$ 0.08         & \cite{ALEBS2}  \\
CDF (p)            & D$_s \ell$          & 92-96        & 1.36$ \pm 0.09 ^{+0.06}_{-0.05} $ & \cite{CDFBS}  \\
CDF            & Excl. ${\rm J}/\psi \phi$ & 92-95        & 1.34$^{+0.23}_{-0.19} \pm$ 0.05   & \cite{CDFB01}  \\
DELPHI (n, p)        & D$_s \ell$          & 91-95        & 1.42$^{+0.14}_{-0.13} \pm 0.03$   & \cite{DELBS0} \\
DELPHI (n, p)        & D$_s h$             & 91-95        & 1.49$^{+0.16}_{-0.15}$ $ ^{+0.07}_{-0.08}$   & \cite{DELBS0}\\
DELPHI         & D$_s$ inclus.       & 91-94        & 1.60$\pm 0.26^{+0.13}_{-0.15}$    & \cite{DELBS2}\\
OPAL           & D$_s \ell$          & 90-95        & 1.50$^{+0.16}_{-0.15} \pm$ 0.04   & \cite{OPABS1}  \\ 
OPAL           & D$_s$ inclus.       & 90-95        & 1.72$^{+0.20+0.18}_{-0.19-0.17}$  & \cite{OPABS2}  \\ \hline
Average        &                     &              &  1.464 $\pm$ 0.057                  &    \\   \hline

\end{tabular}
\caption[]{{\it Measurements of the $\Bs$ lifetime.} \label{tab:bs} } 
\end{center}
\end{table}
\begin{table}[bth]
\begin{center}
\begin{tabular}{|l||c|c|c|c|} \hline
Experiment&Method               &Data set&$\tau_{\Lambda_{\rm{b}}}$ (ps)    & Reference \\\hline
ALEPH     &$\Lambda   \ell$        &  91-95 & 1.20$^{+0.08}_{-0.08} \pm$ 0.06  & \cite{ALELAM}\\
DELPHI    &$\Lambda \ell \pi$ vtx  &  91-94 & 1.16$\pm 0.20 \pm 0.08$          &  \cite{DELLAM0}$^a$\\
DELPHI    &$\Lambda   \mu$ i.p.    &  91-94 & 1.10$^{+0.19}_{-0.17} \pm 0.09 $ &  \cite{DELLAM1}$^a$ \\
DELPHI    &p$\ell$                 &  91-94 & 1.19$ \pm 0.14 \pm$ 0.07      &  \cite{DELLAM0}$^a$\\
OPAL      &$\Lambda   \ell$ i.p.&  90-94 & 1.21$^{+0.15}_{-0.13} \pm$ 0.10 &  \cite{OPALAM1}$^b$  \\ 
OPAL      &$\Lambda   \ell$ vtx.&  90-94 & 1.15$\pm 0.12 \pm$ 0.06          & \cite{OPALAM1}$^b$ \\ 
\hline
Avg. above 6   &                   &     & $1.170^{+0.066}_{-0.054}$     &      \\   \hline
ALEPH     &$\Lambda_c \ell$        &  91-95 & 1.18$^{+0.13}_{-0.12} \pm$ 0.03  &  \cite{ALELAM}\\
ALEPH     &$\Lambda \ell^- \ell^+$ &  91-95 & 1.30$^{+0.26}_{-0.21} \pm$ 0.04  &  \cite{ALELAM}\\
CDF       &$\Lambda_c \ell$        &  91-95 & 1.32$\pm 0.15        \pm$ 0.06   &  \cite{CDFLAM}\\
DELPHI    &$\Lambda_c   \ell$      &  91-94 & 1.11$^{+0.19}_{-0.18} \pm 0.05$  &  \cite{DELLAM0}$^a$\\
OPAL      &$\Lambda_c \ell \& \Lambda \ell^- \ell^+$ &  90-95 & 1.29$^{+0.24}_{-0.22} \pm$ 0.06  &  \cite{OPALAM2}\\ \hline
Avg. above 5   & $ \tau_{\Lambda_b}$  &        & $1.229^{+0.081}_{-0.079}$                             &      \\   \hline
Avg. above 11   &                   &     & $1.208 \pm 0.051$     &      \\   \hline
ALEPH     &$\Xi \ell$              &  90-95 & 1.35$^{+0.37+0.15}_{-0.28-0.17}$ &  \cite{ALELAM1}\\
DELPHI    &$\Xi \ell$           &  91-93 & 1.5 $^{+0.7}_{-0.4} \pm$ 0.3     &  \cite{DELLAM2} \\
\hline
Avg. above 2   &   $ \tau_{\Xi_b}$ &     & $1.39^{+0.34}_{-0.28}$     &      \\   \hline
\end{tabular}
\caption[]{{\it Measurements of the $b$-baryon lifetime.\\
           a) The combined DELPHI result quoted in \cite{DELLAM0} is (1.14$\pm 0.08 $ $\pm$ 0.04) ps. \\
           b) The combined OPAL result quoted in \cite{OPALAM1} is (1.16 $\pm$ 0.11 $\pm$ 0.06) ps. }\\
 \label{tab:lam}} 
\end{center}
\end{table}

\begin{table}[ht!]
\begin{center}
\begin{tabular}{|l||c|c|c|c|} \hline
Experiment & Method            &Data set& $\tau_{B}$ (ps)                 & Reference \\   \hline
ALEPH      & Lepton i.p. (3D)  & 91-93  & 1.533 $\pm$ 0.013 $\pm$      0.022 & \cite{ALEIN1}  \\
L3         & Lepton i.p. (2D)  & 91-94  & 1.544 $\pm$ 0.016 $\pm$      0.021 & \cite{L3IN1}$^b$ \\
OPAL       & Lepton i.p. (2D)  & 90-91  & 1.523 $\pm$ 0.034 $\pm$      0.038 & \cite{OPAIN1}  \\ \hline
Average set 1&                 &        & 1.537 $\pm$ 0.020                  &    \\   \hline
ALEPH      & Dipole            &   91   & 1.511 $\pm$ 0.022 $\pm$      0.078 & \cite{ALEIN2}  \\
ALEPH (p)     & Sec. vert.        &  91-95 & 1.601 $\pm$ 0.004 $\pm$      0.032 & \cite{ALEIN22}  \\
DELPHI     & All track i.p.(2D)& 91-92  & 1.542 $\pm$ 0.021 $\pm$      0.045 & \cite{DELIN0}$^a$ \\
DELPHI     & Sec. vert.        & 91-93  & 1.582 $\pm$ 0.011 $\pm$      0.027 & \cite{DELIN}$^a$ \\
L3         & Sec. vert. + i.p. & 91-94  & 1.556 $\pm$ 0.010 $\pm$      0.017 & \cite{L3IN1}$^b$ \\
OPAL       & Sec. vert.        & 91-94  & 1.611 $\pm$ 0.010 $\pm$      0.027 & \cite{OPAIN2}  \\ 
SLD        & Sec. vert.        &  93    & 1.564 $\pm$ 0.030 $\pm$      0.036 & \cite{SLDIN}  \\ \hline
Average set 2 &                 &         & 1.577 $\pm$ 0.016                  &    \\   \hline
Average sets 1-2   &                   &        & 1.564 $\pm$ 0.014                       &    \\   \hline
CDF        & ${\rm J}/\psi$ vert.    &  92-95 & 1.533 $\pm$ 0.015 $^{+0.035}_{-0.031}$  & \cite{CDFB01}  \\ \hline
\end{tabular}
\caption[]{{\it Measurements of the average $b$-hadron lifetime.\\
            a) The combined DELPHI result quoted in \cite{DELIN} is (1.575 $\pm$ 0.010 $\pm$ 0.026) ps.\\
            b) The combined L3 result quoted in \cite{L3IN1} is (1.549 $\pm$ 0.009 $\pm$ 0.015) ps.} \label{tab:bhad} }

\end{center}
\end{table}

\begin{table}[ht!]
\begin{center}
\begin{tabular}{|l||c|c|c|c|} \hline
 Experiment    & Method            & Data set      & Ratio $\tau_+ /\tau_0$           & Reference \\   \hline
ALEPH (n, p)         & D$^{(*)} \ell$    &  91-95        & 1.080$\pm 0.062 \pm 0.018$        & \cite{ALEB01}  \\
ALEPH          & Excl. rec.       &  91-94         & 1.27$^{+0.23+0.03}_{-0.19-0.02}$  & \cite{ALEB0}  \\
CDF            & D$^{(*)} \ell$   & 92-95         & 1.110$\pm 0.056^{+0.033}_{-0.030}$  & \cite{CDFB02}  \\
CDF            & Excl. (${\rm J}/\psi K$)& 92-95         & 1.06 $\pm$ 0.07 $\pm$ 0.02          & \cite{CDFB01}  \\
DELPHI         & D$^{(*)} \ell$    &  91-93        & 1.00$^{+0.17}_{-0.15} \pm$ 0.10     & \cite{DELB01} \\
DELPHI         & Charge sec. vtx.  &  91-93        & 1.06$^{+0.13}_{-0.11} \pm 0.10$     & \cite{DELB02} \\
L3             & Charge sec. vtx.  &  94-95        & 1.09$\pm$  0.07  $\pm$ 0.03         & \cite{L3B01}  \\ 
OPAL           & D$^{(*)} \ell$    &  91-93        & 0.99$ \pm 0.14^{+0.05}_{-0.04}$     & \cite{OPAB0}  \\ 
OPAL (n)          & Charge sec. vtx.  &  93-95        & 1.079$\pm$ 0.064  $\pm$ 0.041       & \cite{OPAB1}  \\ 
SLD            & Charge sec. vtx. $\ell$&  93-95   & 1.03$^{+0.16}_{-0.14} \pm$ 0.09     & \cite{SLDB01}  \\ 
SLD (n, p)         & Charge sec. vtx.  &  93-98        & 1.037$^{+0.025}_{-0.024} \pm$ 0.024 & \cite{SLDB02}  \\ \hline
Average        &                   &               & 1.065 $\pm$ 0.023                     &    \\   \hline

\end{tabular}
\caption[]{{\it Measurements of the ratio  
$\tau_{\Bp} /\tau_{\Bd}$.} \label{tab:rat}  }
\end{center}
\end{table}

\newpage
\clearpage

\mysection{Measurements of $b$-hadron production rates}
\label{appendixAc}

\mysubsection{$\Bs$ production rate}
\begin{table}[ht!]
\begin{center}
\begin{tabular}{|l|r@{\,$\pm$\,}l|c|}
\hline
    \multicolumn{1}{|c|}{Quantity}      &  
    \multicolumn{2}{|c|}{Value}         &   
    Ref. \\
\hline
    {\hspace{1mm}}
$\fs~ {\rm BR}(\Bs \rightarrow \Dsm \ell^+ \nu_{\ell} {\rm X})~{\rm BR}(\Dsm \rightarrow \phi \pi^-)  $    {\hspace{1mm}} 
    &   (2.87  &  $ 0.32 \pm {}^{0.25}_{0.41})$ \,$\times 10^{-4}$  &  \cite{A:tbs} \\ 
    {\hspace{1mm}}
$\fs~ {\rm BR}(\Bs \rightarrow \Dsm \ell^+ \nu_{\ell} {\rm X})~{\rm BR}(\Dsm \rightarrow \phi \pi^-) $     {\hspace{1mm}} 
    &   (4.2  &  $ 1.9 )$ \,$\times 10^{-4}$  &  \cite{A:tbs2} \\ 
    {\hspace{1mm}}
$\fs~ {\rm BR}(\Bs \rightarrow \Dsm \ell^+ \nu_{\ell} {\rm X})~{\rm BR}(\Dsm \rightarrow \phi \pi^-)$      {\hspace{1mm}} 
    &   (3.9  &  $ 1.1 \pm 0.8)$ \,$\times 10^{-4}$  &  \cite{A:tbs3} \\ 
\hline
Average of the three measurements   {\hspace{1mm}} 
    &   (3.00  &  $ ^{0.38}_{0.41})$ \,$\times 10^{-4}$  &   \\ 
\hline
    {\hspace{1mm}} $\fs/(\fu+\fd)~{\rm BR}(\Dsm \rightarrow \phi \pi^-)$       {\hspace{1mm}} 
    &   (7.7  &  $ 1.5)$\,$\times 10^{-3}$   &  \cite{A:flamcdf} \\ 
\hline
\end{tabular}
\caption []{{\it Inputs used in the calculation of
the $\Bs$ production rate.} \label{tab:sfrac} } 
\end{center}

\end{table}

All published results have been multiplied by the branching fraction
for the decay $\Dsm \rightarrow \phi \pi^-$ to be independent of the assumed
central value and uncertainty of this quantity; quoted uncertainties
have been reevaluated accordingly.

The ALEPH measurement \cite{ref:anotused}
of $\fs~ {\rm BR}(\Bs \rightarrow \Dsm {\rm X})~
 {\rm BR}(\Dsm \rightarrow \phi \pi^-)=(3.1 \pm 0.7 \pm 0.6)\times 10^{-3}$
has not been used to obtain an additional measurement on $\fs$ because of the model dependence attached to the evaluation of 
${\rm BR}(\Bs \rightarrow \Dsm {\rm X})$. Instead, the value of $\fs$, quoted
in Table \ref{tab:rates}, can be used to extract, from this measurement, the
inclusive branching fraction for $\Dsm$ production in $\Bs$ decays:
\begin{equation}
{\rm BR}(\Bs \rightarrow \Dsm {\rm X})~{\rm BR}(\Dsm \rightarrow \phi \pi^-)=
(3.1 \pm 0.7 \pm 0.7)\times 10^{-2}
\end{equation}
and, using the value for ${\rm BR}(\Dsm \rightarrow \phi \pi^-)$
 given in Table \ref{tab:gensys}:
\begin{equation}
{\rm BR}(\Bs \rightarrow \Dsm {\rm X})=0.86 \pm 0.19 \pm 0.29.
\end{equation}

\mysubsection{$b$-baryon production rate}
\begin{table}[ht!]
\begin{center}
\begin{tabular}{|l|r@{\,$\pm$\,}l|c|}
\hline
    \multicolumn{1}{|c|}{Quantity}      &  
    \multicolumn{2}{|c|}{Value}         &   
    Ref. \\
\hline
    {\hspace{1mm}}
     $\prodlb~{\rm BR}(\Lc \rightarrow {\rm p} {\rm K}^- \pi^+)$     {\hspace{1mm}}
    &   (3.78  & $ 0.31 \pm 0.23)$ \, $\times 10^{-4}$  &  \cite{AD:prodlamb} \\ 
    {\hspace{1mm}}
     $\prodlb~{\rm BR}(\Lc \rightarrow {\rm p} {\rm K}^- \pi^+)$     {\hspace{1mm}}
    &   (5.19  & $ 1.14 \pm {}^{1.19}_{0.66})$ \, $\times 10^{-4}$  &  \cite{AD:prodlamb2} \\ 
\hline
Average of the two measurements   {\hspace{1mm}} 
    &   (3.90  &  $ 0.42)$ \,$\times 10^{-4}$  &   \\ 
\hline
    {\hspace{1mm}} $\fb/(\fu+\fd)~{\rm BR}(\Lc \rightarrow {\rm p} {\rm K}^- \pi^+)$      {\hspace{1mm}} 
    &   (5.9  &  $ 1.4)\times 10^{-3}$   &  \cite{A:flamcdf} \\ 
\hline
    {\hspace{1mm}}     $\prodxb$     {\hspace{1mm}} 
    &  ( 5.4   & $ 1.1 \pm 0.8 )$ $\,\times \ 10^{-4}$  &  \cite{AD:cascb} \\ 
    {\hspace{1mm}}     $\prodxb$     {\hspace{1mm}} 
    &  ( 5.9   & $ 2.1 \pm 1.0 )$ $\,\times \ 10^{-4}$  &  \cite{AD:cascb2} \\ 
\hline
Average of the two measurements   {\hspace{1mm}} 
    &   (5.5  &  $ 1.2)$ \,$\times 10^{-4}$  &   \\ 
\hline
    {\hspace{1mm}} $\fb$      {\hspace{1mm}} 
    &   0.102  &  $ 0.007 \pm 0.027 $   &  \cite{A:flamdir} \\ 
\hline
\end{tabular}
\caption []{{\it Inputs used in the calculation of
the $b$-baryon production rate.} \label{tab:bfrac}} 
\end{center}

\end{table}

All published results which are using the $\Lc$ baryon
have been multiplied by the branching fraction
for the decay $\Lc \rightarrow {\rm p} {\rm K}^- \pi^+$ 
to be independent of the assumed
central value and uncertainty of this quantity; quoted uncertainties
have been reevaluated accordingly.

\mysubsection{$\Bp$ production rate}

\begin{table}[ht!]
\begin{center}
\begin{tabular}{|l|r@{\,$\pm$\,}l|c|}
\hline
    \multicolumn{1}{|c|}{Quantity}      &  
    \multicolumn{2}{|c|}{Value}         &   
    Ref. \\
\hline
    {\hspace{1mm}} $\fu$      {\hspace{1mm}} 
    &   0.414  &  $ 0.016 $   &  \cite{ref:delphibplus} \\ 
\hline
\end{tabular}
\caption [] {{\it Direct measurement of the $\Bp$ 
production rate.} \label{tab:bpfrac}} 
\end{center}

\end{table}

This value is obtained from the measurement of the production rate
of charged weakly decaying $b$-hadrons. A small correction has been applied
to account for $\Xi_b^-$ production, as given in 
Table \ref{tab:bfrac}.
\newpage
\clearpage

\mysection{Theoretical uncertainties relevant to the measurements of
$\Vub$ and $\Vcb$}
\label{appendixC}

At beginning of June 1999 a workshop entitled
``Informal Workshop on the Derivation of $\Vcb$
and $\Vub$: Experimental Status and 
Theory Uncertainties''\footnote{Participants at this workshop were:
D. Abbaneo, P. Ball, E. Barberio, M. Battaglia, M. Beneke, 
I.I. Bigi, 
G. Buchalla, M. Calvi, O. Cooke, F. Defazio, L. di Ciaccio, R. Fleischer,
 P. Gagnon, P. Henrard, A. Hoang, L. Lellouch, J. Lu, S. Mele,
E. Piotto, Ph. Rosnet, P. Roudeau, D. Rousseau, Ch. Schwick and F. Simonetto.}
was held at CERN.

As, at that time, there was not a general 
consensus on the values for theoretical 
errors to be used to evaluate $\Vcb$ and $\Vub$~\cite{Babar}, 
the main aim of this workshop has been to scrutinize
the uncertainties attached to the different parameters and to define
a common set of values to be adopted for the derivation of $\Vub$ and $\Vcb$.
The various contributing systematic uncertainties have been 
added in quadrature. This procedure is justified when several independent 
sources of uncertainties contribute, independently of their exact
distributions, unless a single source clearly dominates. A linear sum
can be justified if uncertainties
are fully and positively correlated, which is not really the case of the
presently considered quantities.
In addition, theoretical errors correspond usually to 
a confidence level larger than 68$\%$, which is the confidence level assumed
when these uncertainties are taken as standard deviations of Gaussian
distributions.

The procedure adopted in the end which is explained in the conclusions,
is not entirely satisfactory;
it is simply pragmatic in the absence, at present, of a direct experimental 
control,
or of other evaluations with different theoretical techniques, of the 
main sources of systematic theoretical uncertainties.
More details can be found in \cite{ref:bigi0}. 

\mysubsection{Measurement of $\Vub$ using the decay 
$b \rightarrow  \ell^-\overline{\nu_{\ell}}{\rm X}_{u}$}
\label{sec:A}

Based on studies developed independently by two 
groups \cite{uraltsev,hoang}, 
a relative  theoretical uncertainty of 5\% has been evaluated for the 
extraction of $\Vub$ from the measured inclusive charmless semileptonic rate 
${\rm BR}(b\rightarrow \ell^-\overline{\nu_{\ell}}{\rm X}_{u})$. The central values
of the two analyses also 
agree\footnote{In practice the initial central value quoted in \cite{uraltsev}
has been corrected and the value obtained 
in \cite{hoang} is
$\Vub~=~0.00443 \left ( \frac{{\rm BR}(b \rightarrow  \ell
\overline{\nu_{\ell}}{\rm X}_u)}{0.002}  \right )^{1/2}
\left ( \frac{1.55~{\rm ps}}{\tau_b} \right )^{1/2}
 \times \left (1 \pm 0.020_{pert.} \pm 0.030_{m_b} \right ) $.},
and the following relationship 
has been adopted in the extraction of $\Vub$:

\begin{equation}
\begin{array}{cc}
\Vub=
& 0.00445 \left ( \frac{{\rm BR}(b \rightarrow \ell^-
\overline{\nu_{\ell}} {\rm X}_u)}{0.002}  \right )^{1/2}
\left ( \frac{1.55~{\rm ps}}{\tau_B} \right )^{1/2}\\
  & \times \left (1 \pm 0.010_{pert.} \pm 0.030_{1/m_b^3} \pm 0.035_{m_b} \right ) \\
\end{array}
\label{eq:vub}
\end{equation}

There have been extensive discussions concerning the uncertainty to be 
attributed to the value of $m_b$, the $b$-quark mass. 
The analyses reported in~\cite{ref:mass1,ref:mass2}
independently indicate a range of $\pm 0.060~\GeV/c^2$ or less to be 
assigned to 
$m_b$. However during the workshop this narrow mass range raised concerns
 from a third independent analysis described in ~\cite{ref:mass3}.  
As a result of this ongoing discussion, the value  
$m_b(1~\GeV)=(4.58 \pm 0.060) \GeV/c^2$, where the 
error is meant to define the 68\% confidence level, has been adopted here
for the 
derivations of $\Vub$ and $\Vcb$\footnote{As explained in 
Section \ref{sec:C}, in the inclusive determination of $|V_{cb}|$, 
all theoretical uncertainties have in addition been multiplied by two, 
in a rather arbitrary way.}.

$m_b(\mu)$ has been defined in a way which leads to a linear dependence
on $\mu$:
\begin{equation}
\frac{dm_b(\mu)}{d\mu}~=~-c_m~\frac{\alpha_s(\mu)}{\pi}~+~....
\label{eq:mb}
\end{equation}
The number $c_m$ specifies the concrete definition within this general class;
its value is equal to $\frac{16}{9}$. 
More details on the definition of $m_b(\mu)$ can be found in 
\cite{ref:cza}.

There has been a claim~\cite{ref:vub2} that a larger fractional uncertainty 
of $^{+13}_{-10}\%$, as well as a 7.5\% lower central value for $\Vub$,
has 
to be used. From the arguments which have been developed at the 
workshop, it has been concluded that the results of the 
analyses~\cite{uraltsev,hoang} are based on well defined 
theoretical grounds and that they provide a reliable estimate of the
theoretical uncertainties.

Recent measurements performed at LEP, based on the inclusive selection
of $b \rightarrow \ell^-\overline{\nu_{\ell}}{\rm X}_u $ decays, are sensitive to
a large fraction of the mass distribution of the hadronic system
and to the full range of the lepton energy spectrum.
This mass distribution $M_X$ has been studied by theorists \cite{ref:mass} for
the $b \rightarrow \ell^-\overline{\nu_{\ell}}{\rm X}_u $ transition.
The conclusion was\footnote{In practice experimental 
analyses take into account the expected
mass distribution of the hadronic system, and the variation of the
detection efficiencies as a function of this mass, to estimate
this systematic uncertainty.}
that the additional theoretical uncertainty on $\Vub$ 
from $m_b$ and $\mu_{\pi}^2$ is less than 10\% if the experimental inclusive
technique is sensitive up to at least 
$M_X \simeq 1.5$~$\GeV/c^2$.

It has thus been concluded that the theoretical errors attached to the 
determination of
$\Vub$ using the present experimental technique are under control.

\mysubsection{Measurement of $\Vcb$ using the decay 
$\Bdb \rightarrow \Dstarp \ell^-\overline{\nu_{\ell}}$}
\label{sec:B}

It has been proposed to use the parametrization given in
\cite{ref:vcb1} to account for the dependence in $w$ and to 
extract the value at $w=1$ of the differential decay rate.
The quantity ${\cal F}_{D^{\ast}}(1) \Vcb$ is then obtained.

In order to measure $\Vcb$,
the value at $w=1$ of the form factor is taken from theory by evaluating
different corrections \cite{ref:bigi1} which have to be applied to the naive 
expectation of  ${\cal F}_{D^{\ast}}(1)=1$:
\begin{equation}
\left | {\cal F}_{D^{\ast}}(1) \right |^2~=~\xi_A(\mu) -\Delta^A_{1/m^2} 
 -\Delta^A_{1/m^3} -\sum_{0<\epsilon_i<\mu} \left | {\cal F}_i \right |^2
\end{equation}

Expressions for the different terms contributing to this equation
can be found in \cite{ref:bigi1}. $\xi_A(\mu)$ results
from perturbative QCD expansion at the scale $\mu$. The other quantities
are of non-perturbative QCD origin and are obtained in the O.P.E.
formalism. It has to be noticed that these terms correspond
to an expansion in $1/m_Q$ where the heavy quark can have
the $b$ or the $c$ flavours. In the evaluation of related
uncertainties, the charm quark mass determination is thus
essential.

The central value of ${\cal F}_{D^{\ast}}(1)$
has been recently lowered by 2$\%$
by evaluating higher order perturbative corrections \cite{ref:ural1}.

The central value of the $b$-quark mass and its uncertainty are the
same as used in Section \ref{sec:A}.

The charm quark mass is obtained from the mass difference 
$m_b-m_c$, inferred from the measured values of the 
spin-averaged beauty and charm meson masses
\footnote{$\left < M_B \right >=(M_B+3M_{B^{\ast}})/4$ and a similar expression
for $\left < M_C \right >$.}:
\begin{equation}
\begin{array}{cc}
m_b~-~m_c~=~
&  \left < M_B \right > -\left < M_C \right >~+~
\mu_{\pi}^2 \left ( \frac{1}{2m_c}- \frac{1}{2m_b} \right )~+~
{\cal O} \left ( 1/m^2_{c,b} \right ) \\
 ~~~~~~~~~~=~ &  (~3.50~+~0.040 \frac{\mu_{\pi}^2-0.5}{0.1}~+~\Delta M_2~ )~(\GeV/c^2)\\
\end{array}
\label{eq:mc}
\end{equation}
$\mu_{\pi}^2$ is the average of the square of the heavy quark momentum 
inside the B hadron. $\Delta M_2$ includes all terms of order
$m_Q^{-n}$, with $n \geq 2$ and its absolute value is expected
to be smaller than 15 $\MeV/c^2$.
The main difference between the theoretical analyses given in
\cite{ref:bigi1} and \cite{ref:neu1} comes from the evaluation of the 
 uncertainties on non-perturbative corrections. A different
approach can be found in \cite{ref:grin}.

The detailed balance of uncertainties given in \cite{ref:bigi1}, 
which has been revised
for consistency
with the values adopted at present for the different parameters, is
the following:
\begin{equation}
{\cal F}_{D^{\ast}}(1)~=~0.880 -0.024 \frac{\mu^2_{\pi}-0.5~ \GeV^2}{0.1~\GeV^2}
\pm 0.035_{excit.} \pm 0.010_{pert.} \pm 0.025_{1/m^3}
\label{eq:fdstar}
\end{equation}

Uncertainties originating from non-perturbative QCD dominate
over the perturbative contribution ($pert.$).
The hypotheses or the results on which
their contributions were based
are reviewed in the following. The label $excit.$ denotes the
contribution of excited charmed final states.

\mysubsubsection{$\mu^2_{\pi}$} 
 
The value $\mu^2_{\pi}=(0.5 \pm 0.1)~\GeV^2$ has been used and it gives
$\pm0.024$ uncertainty on ${\cal F}_{D^{\ast}}(1)$.
This evaluation has been justified considering that, using
QCD sum rules, the following value has been obtained~\cite{ref:ball1}:
\begin{equation}
\mu^2_{\pi}=(0.5 \pm 0.1)~\GeV^2
\end{equation}
and that a model-independent lower bound has been established~\cite{ref:bigi2}:
\begin{equation}
\mu^2_{\pi}>\mu^2_G \simeq \frac{3}{4}(M^2_{B^\ast}-M^2_B)
\approx 0.4 ~\GeV^2
\end{equation}
$\mu^2_G$ is the expectation value of the chromomagnetic operator.

\mysubsubsection{High mass excitations}

In QCD sum rules, used in
\cite{ref:bigi1,ref:neu1}, the effect expected
from high mass hadronic states has been introduced in a rather
arbitrary way. To be conservative,
in \cite{ref:bigi1}, it has been assumed that their effect,
on the correction terms which behave as $m_Q^{-2}$, can vary between
0 and 100$\%$ of $\Delta^A_{1/m^2}$. Such a variation corresponds
to $\pm 0.035$ on ${\cal F}_{D^{\ast}}(1)$.

\mysubsubsection{Higher order non-perturbative corrections}

The value of $\pm 0.025$ uncertainty related to the contribution from
terms of order at least $m_Q^{-3}$ is
obtained in \cite{ref:bigi1}.

\mysubsubsection{Adopted value}
Combining the uncertainties quoted in 
(\ref{eq:fdstar})
gives:
\begin{equation}
{\cal F}_{D^{\ast}}(1)~=~0.88 ~\pm 0.05.
\end{equation}
%
%
%
%

\mysubsubsection{Expected improvements in the control of theoretical errors}

 The uncertainty attached to ``excited states'' is expected to be decreased
if better measurements of 
$\Dstarstar$
 production rates and
hadronic mass distributions are 
obtained\footnote{$\Dstarstar$ denotes all decay modes which are not
$\overline{{\rm B}} \rightarrow {\rm D} \ell^-\overline{\nu_{\ell}}$ and
$\overline{{\rm B}} \rightarrow \Dstar \ell^-\overline{\nu_{\ell}}$.}.

 The uncertainty related to $\mu^2_{\pi}$ 
is expected to decrease by measuring
moments of the lepton momentum in the B rest frame. 

\mysubsubsection{Form factors for $\Dstarstar$ production near to  $w=1$}

It has been realized that the $w$-dependence of the form factors used
to describe $\Dstarstar$ production in $b$-hadron semileptonic decays is
important for evaluating the systematic uncertainty on $\Vcb$
related to the fraction of $\Dstar$ mesons coming from $\Dstarstar$ decays.
It has been proposed to use the model described in \cite{ref:ligeti},
which predicts a rate for the ratio of 2$^+$ over 1$^+$ narrow states
more in agreement with the experimental results given in Section 
\ref{sec:systgen}. 
Additional information on $\Dstarstar$ production which may bring 
constraints 
on such models would be welcome.

\mysubsection{Measurement of $\Vcb$ using the inclusive semileptonic decay 
$b \rightarrow \ell^-X$ rate}
\label{sec:C}

The expression relating $\Vcb$ to the inclusive semileptonic
branching fraction can be found in \cite{ref:bigi1}:

\begin{equation}
\begin{array}{cc}
\Vcb=
& 0.0411 \left ( \frac{{\rm BR}(b \rightarrow \ell^-
\overline{\nu_{\ell}}{\rm X}_c)}{0.105}  \right )^{1/2}
\left ( \frac{1.55~{\rm ps}}{\tau_b} \right )^{1/2}\\
  & \times \left ( 1 -0.012 \frac{\mu_{\pi}^2-0.5 \GeV^2}{0.1 \GeV^2} \right )\\
  & \times \left (1 \pm 0.015_{pert.} \pm 0.010_{m_b} \pm 0.012_{1/m_Q^3}\right ) \\
\end{array}
\label{eq:huit}
\end{equation}
The central value of reference \cite{ref:bigi1} has been lowered by 2$\%$ 
to account for a different choice of the $b$-quark mass. 
A very similar result for the central value and the  uncertainties 
can be found in \cite{hoang}.

\mysubsubsection{Uncertainties related to quark masses}
It has been assumed that $m_b(\mu)$ is known with an uncertainty of
$\pm$60~$\MeV/c^2$, as discussed in Section~\ref{sec:A}, and that the difference 
between the 
$b$- and $c$-quark masses
is known with an uncertainty of $\pm$40~$\MeV/c^2$, related to the 
error on $\mu_{\pi}^2$. These variations induce $\pm 0.010$ 
and $\pm 0.012$ uncertainties
on $\Vcb$, respectively.


\mysubsubsection{Adopted value}

Adding in quadrature the quoted uncertainties
in Equation (\ref{eq:huit}) leads to a $\pm$2.5$\%$ relative 
uncertainty on $\Vcb$. 
Since these errors have not been cross-checked by other theoretical
approaches or experimental measurements and
because most of these uncertainties depend on the evaluation
of $m_c$ (or $\mu_{\pi}^2$) which is less under control
than $m_b$,
it has been decided to inflate the total error by an arbitrary 
factor of two, leading to 
a theoretical uncertainty of $\pm$5$\%$ (see reference \cite{ref:bigi0} for 
a more detailed discussion).
This implies, in practice, that uncertainties of
$\pm 120~\MeV/c^2$, $\pm 80~\MeV/c^2$, and $\pm 0.2~\GeV^2$ have been
attached to the values of $m_b$, $m_b-m_c$,
and $\mu_{\pi}^2$ respectively.



\mysubsection{Common sources of theoretical errors for the two determinations of
$\Vcb$}

Theoretical uncertainties attached to the exclusive and inclusive
 measurements of $\Vcb$
are largely uncorrelated. When evaluating
the average, only the uncertainties related
to $\mu_{\pi}^2$ and $m_b$ have been considered fully correlated.

\mysubsection{Sources of theoretical errors entering into the measurement
of the ratio $\frac{\Vub}{\Vcb}$}

In the evaluation of the ratio $\frac{\Vub}{\Vcb}$, several common experimental
and model systematics have a reduced effect.
 
For the  ratio $\frac{|{\rm V}_{ub}|_{incl}}{|{\rm V}_{cb}|_{incl}}$, uncertainties
originating from the determination of
 $m_b$ have to be considered as
fully correlated. Leading effects of order ${\cal O}(1/m^3)$ are expected
to be uncorrelated between $\Vub$ and $\Vcb$ determinations 
\cite{ref:bigi0}. This is also expected for perturbative uncertainties.

With the conventions used in the previous sections, the relative errors
on the ratio $\frac{|{\rm V}_{ub}|_{incl}}{|{\rm V}_{cb}|_{incl}}$ are:

\begin{equation}
\pm 0.015(m_b)~\pm 0.032(pert.)~\pm 0.024(\mu_{\pi}^2) \pm 0.038(1/m^3)
~=~\pm 0.06
\label{eq:rincl}
\end{equation}

For the ratio $\frac{|{\rm V}_{ub}|_{incl}}{|{\rm V}_{cb}|_{excl}}$, all uncertainties
can be considered as uncorrelated giving:
\begin{equation}
\pm 0.035(m_b)~\pm 0.014(pert.)~\pm 0.024(\mu_{\pi}^2) \pm 0.039(1/m^3)
\pm 0.035(excit.)~=~\pm 0.07
\label{eq:rexcl}
\end{equation}

In practice, the ratio $\frac{|{\rm V}_{ub}|}{|{\rm V}_{cb}|}$ will be obtained using the average
of the two measurements of $\Vcb$ and taking into account common systematics
with $\Vub$.

\mysubsection{Conclusions and summary}
The conclusions of the workshop which are given in this Appendix
have left aside
 several subtleties concerning
the exact meaning of the parameters entering into the different expressions
for $\Vub(incl.)$, $\Vcb(incl.)$ and $\Vcb(excl.)$.
The interested reader has therefore to consult the documents quoted in 
the references
for more information.

Two groups (at least) have obtained consistent results on central values 
and uncertainties for $\Vub(incl.)$ and $\Vcb(incl.)$.

Uncertainties on $\Vcb(incl.)$ have been enlarged, in a rather arbitrary way,
by a factor two to have some margin because of possible additional 
contributions (reliability of $1/m_c$ expansion, need for other techniques
to evaluate $m_b$,...). 

Dedicated experimental studies on $\Dstarstar$ production and on the 
distributions of moments of the lepton momentum in the B rest frame
are needed to evaluate with  better accuracy and greater confidence the most
important 
sources of systematic errors contributing in the theoretical expressions.


The following central values and theoretical uncertainties have been used 
in the measurements of $\Vub$ and $\Vcb$ obtained from combined
LEP analyses:
\begin{itemize}
\item{inclusive measurement of $\Vub$:}
\begin{equation}
\begin{array}{cc}
\Vub=
& 0.00445 \left ( \frac{{\rm BR}(b \rightarrow \ell^-
\overline{\nu_{\ell}}{\rm X}_u)}{0.002}  \right )^{1/2}
\left ( \frac{1.55~{\rm ps}}{\tau_b} \right )^{1/2}\\
  & \times \left (1 \pm 0.010_{pert.} \pm 0.030_{1/m_b^3} \pm 0.035_{m_b} 
\right ) \\
\end{array}
\end{equation}

\item{inclusive measurement of $\Vcb$:}
\begin{equation}
\begin{array}{cc}
\Vcb=
& 0.0411 \left ( \frac{{\rm BR}(b \rightarrow \ell^-
\overline{\nu_{\ell}}{\rm X}_c)}{0.105}  \right )^{1/2}
\left ( \frac{1.55~{\rm ps}}{\tau_b} \right )^{1/2}\\
  & \times \left ( 1 -0.024 \frac{\mu_{\pi}^2-0.5 \GeV^2}{0.2 \GeV^2} \right )\\
  & \times \left (1 \pm 0.030_{pert.} \pm 0.020_{m_b} \pm 0.024_{1/m_Q^3}
\right ) \\
\end{array}
\end{equation}

\item{exclusive measurement of $\Vcb$:}
\begin{equation}
{\cal F}_{D^{\ast}}(1)~=~0.880 -0.024 \frac{\mu^2_{\pi}-0.5~ \GeV^2}{0.1~\GeV^2}
\pm 0.035_{excit.} \pm 0.010_{pert.} \pm 0.025_{1/m^3}
\end{equation}
\end{itemize}

All uncertainties have been added in quadrature.

The correlated uncertainty between $\Vcb(incl.)$ and $\Vcb(excl.)$ measurements
stem from the effect of $\mu^2_{\pi}$.

The correlated uncertainty between $\Vub(incl.)$ and $\Vcb(incl.)$ measurements
is limited to the effect of $m_b$.
Hence no correlated uncertainty
between $\Vub(incl.)$ and $\Vcb(excl.)$ measurements is assumed.


\begin{thebibliography}{ref99}
\bibitem{dnewdsh} The value quoted for the measured lifetime differs
slightly from the one quoted in Table \ref{tab:bs} because it corresponds
to the present status of the analysis in which the information
on $\dgs$ has been obtained.
\end{thebibliography}
\end{document}